\documentclass[%
 reprint,
 amsmath,amssymb,
 aps,
rmp,
]{revtex4-2}
\usepackage{float}%
\usepackage{color}
\usepackage{graphicx}
\usepackage{dcolumn}
\usepackage{bm}
\usepackage{booktabs}
\usepackage{hyperref}
\usepackage{wasysym} 

\usepackage{citehack}
\setcitestyle{numbers,square,comma,sort&compress}

\def\bk{{\mathbf{k}}}

\def\bK{{\mathbf{K}}}

\def\bq{{\mathbf{q}}}

\def\bG{{\mathbf{G}}}

\begin{document}

\title{A Sport and a Pastime: Model Design and Computation in Quantum Many-Body Systems}

\author{Gaopei Pan}
\author{Weilun Jiang}%
\affiliation{%
	Beijing National Laboratory for Condensed Matter Physics and Institute of Physics, Chinese Academy of Sciences, Beijing 100190, China
}%

\affiliation{%
School of Physical Sciences, University of Chinese Academy of Sciences, Beijing 100049, China}%

\author{Zi Yang Meng}
 \email{zymeng@hku.hk}
\affiliation{%
Department of Physics and HKU-UCAS Joint Institute of Theoretical and Computational Physics, The University of Hong Kong, Pokfulam Road, Hong Kong SAR, China}%


\date{\today}

\begin{abstract}
We summarize the recent developments in the model design and computation for a few representative quantum many-body systems, encompassing quantum critical metals beyond the Hertz-Millis-Moriya framework with pseudogap and superconductivity, SYK non-Fermi-liquid with self-tuned quantum criticality and fluctuation induced superconductivity, and the flat-band quantum Moir\'e lattice models in continuum where the interplay of quantum geometry of flat-band wave function and the long-range Coulomb interactions gives rise to novel insulating phases at integer fillings and superconductivity away from them. Although the narrative choreography seems simple, we show how important the appropriate model design and their tailor-made algorithmic developments -- in other words, the scientific imagination inspired by the corresponding fast experimental developments in the aforementioned systems -- compel us to invent and discover new knowledge and insights in the sport and pastime of quantum many-body research.
\end{abstract}

\maketitle

\tableofcontents

\section{Introduction}
\label{sec:I}
In the 200 pages short novel "A Sport and a Pastime"~\cite{salterA1967} -- generally regarded as a modern classic -- the writer and the great stylist James Salter, has successfully established the standard not only for fiction, but for the principal organ of literature --  the imagination.

Set in provincial France in the 1960s, through the wanderings of a young American middle-class college drop-out Philip Dean and an even younger, small-town French girl, Anne-Marie, the novel reveals the country's "secret life $\cdots$ into which one can not penetrate". It is "the life of photograph albums, uncles, names of dogs that have died". The green, bourgeois and rural France would be inaccessible, remote, and lifeless to Philip Dean without his affair with the beautiful, though cheap, Anne-Marie. In a way she becomes Philip, and together they become the person who illuminates travel, the person the readers always dream of meeting, and without whom the museums are tedious, the roads empty, the rivers are dry and the food and drinks a necessity. Anne-Marie provides Dean, twenty-four and a dropout from Yale, a perspective, rescuing him from the lost ranks of student sightseers and together they provide the readers a personal reference, in which the architechture, mountains, great rivers, villages and green scenery can be easily understood from an emotional as well as numerical scale~\cite{dowieA1988}.

The book, as has been praised as "what appears at first to be a short, tragic novel about a love affair in provincial France is in fact an ambitious, refractive inquiry into the nature and meaning of storytelling, and the reasons artists are compelled to invent. That such a feat occurs across a mere 200 pages is breathtaking, and though its narrative choreography seems simple, the novel is anything but minor."~\cite{hallBeautiful2017}

The same inquiry and the same compulsion that forces scientists to invent and discover, emotionally as well as numerically, also apply to our pursuit in the model design and computation for quantum many-body systems. This short review, in this sense like the "A Sport and a Pastime" of James Slater, offers the readers a personal reference of our reflection upon the actively-going-on research efforts in quantum critical metals, SYK non-Fermi-liquid and quantum Moir\'e lattice models in recent years. 

Our interests in the quantum critical metals lie in the rich history of the Hertz-Millis-Moriya(HMM) theory~\cite{hertzQuantum1976,millisEffect1993,moriyaSpin1978}, where the dynamic properties of the itinerant ferromagnetic or antiferromagnetic quantum critical point (QCP) are the focus. The extension of the theory to study fermionic properties~\cite{leeGauge1989,altshulerLow1994,oganesyanQuantum2001,abanovQuantum2003,rechQuantum2006}, predicts that
fermions near such QCPs
are overdamped, with fermionic self-energy scaling as $\Sigma\propto\omega_n^{2/3}$ for ferromagnetic QCP and $\Sigma\propto\omega_n^{1/2}$ for antiferromagnetic ones, where $\omega_n$ is the fermionic Matsubara frequency.
The fact that power in frequency is less than $1$ implies that the system is a non-Fermi-liquid (nFL) in the quantum critical region spanned by temperature, energy and other control parameter axes.
Within the one-loop framework,
these conclusions and scaling exponents are universal for all itinerant QCPs.
When higher order contributions are taken into account, additional phenomena may appear,
e.g. first order behavior, spiral phases, and low-frequency scaling violations ~\cite{kirkpatrickNature2003,maslovNonanalytic2009,millisNonanalytic2006,conduitInhomogeneous2009,metlitskiQuantum2010,metlitskiInstabilities2010,holderAnomalous2015,greenQuantum2018,schliefExact2017}, as well as
superconductivity.
In particular,
if the bosonic order parameter (OP) is not conserved, higher order processes
modify the damping of the bosons in the long-wavelength limit, and change the value of dynamic exponent $z$, for example in the ferromagnetic case from $z=3$ to $z=2$~\cite{mineevMagnetic2013,chubukovNonLandau2014}.

These fundamental discussions are not only for the curiosities of theorists, the quantum critical nFL behaviors have been observed in a variety of materials~\cite{lohneysenFermi2007}, such as the Kondo lattice materials UGe$_2$~\cite{huxleyMagnetic2003}, URhGe~\cite{levyMagnetic2005}, UCoGe~\cite{stockAnisotropic2011}, YbNi$_4$P$_2$~\cite{steppkeFerromagnetic2013} and more recently CeRh$_6$Ge$_4$~\cite{shenStrange2020,wuAnisotropic2021}, where in the latter a pressure-induced ferromagnetic quantum critical point (QCP) with the characteristic nFL specific heat and resistivity are reported. These experimental progress poses a series of theoretical questions on the origin and characterization of these nFL behaviors. In particular, it is of crucial importance to understand the fundamental principles that govern these QCPs and to identify the universal properties that are enforced by these principles.

In Sec.~\ref{sec:II}, we will review our model design and quantum Monte Carlo (QMC) simulation technique developments upon the antiferromagnetic~\cite{liuItinerant2018,liuElective2019,liuItinerant2019} and ferromagnetic QCP nFLs with both conserved~\cite{xuNonFermi2017,xuRevealing2019,xuIdentification2020}, and more importantly, non-conserved OP such as the quantum rotor model~\cite{jiangSolving2022}, where the transformation from $z=3$ to $z=2$ QCP scaling behavior is observed at weak coupling~\cite{liuDynamical2022} and the pseudogap and superconductivity induced by the quantum critical fluctuations are observed at stronger coupling~\cite{jiangMonte2022}. The analysis procedure the authors in Ref~\cite{xuIdentification2020} developed to unambiguously reveal the nFL fermion self-energy and the bosonic dynamic susceptibilities will be presented in details. From here, few immediate directions and open questions, both in theoretical and numerical developments and their implications for the experiments, will be discussed.

The itinerant QCP models, require the coupling between the critical bosonic modes with the Fermi surface (FS) such that inside the quantum critical region there emerges nFL from the quantum critical fluctuations, but in the ordered or disordered phases of the bosons, the fermions are essentially in a FL state with different FS geometry and systems are essential non-interacting. In Sec.~\ref{sec:III}, we focus on another type of nFL systems, the spin-1/2 Yukawa-Sachdev-Ye-Kitaev (Yukawa-SYK) model~\cite{sachdevGapless1993, kitaev2015simple,wangQuantum2020,panYukawa2021,wangPhase2021}. Unlike the itinerant QCP models in Sec.~\ref{sec:II} where the systems live in 2 spatial dimensions, the Yukawa-SYK models live in infinite dimensions and have been shown to “self-tune” to quantum criticality within large-$N$ approximation~\cite{wangQuantum2020}, that is, independent of the bosonic bare mass, the system becomes critical due to the strong mutual feedback between the bosonic and fermionic sectors.  This fact renders Yukawa-SKY model more convenient to study the nFL behaviors. We therefore implemented the lattice models where the bosons are Yukawa coupled to the fermions and reveal the SYK type of nFL with Green's function power-law in both frequency and imaginary time axes in the self-tuned quantum criticality with QMC simulations~\cite{panYukawa2021}, and we further discovered the  fluctuation mediated pairing in the Yukawa-SYK model~\cite{wangPhase2021}. Few representative recent developments in model design and numerical solutions for the related SYK nFL systems are also discussed.

\begin{figure*}[!htp]
	\centering
	\includegraphics[width=\linewidth]{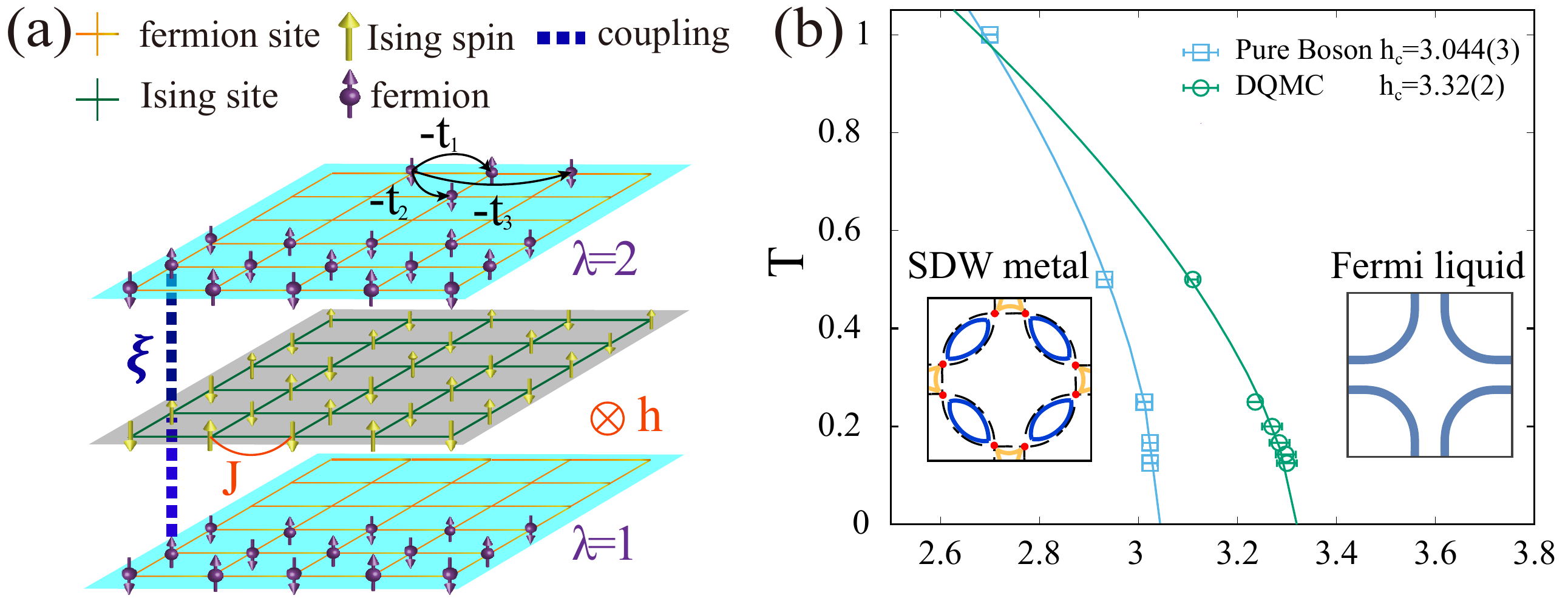}
	\caption{ (a) Lattice model for antiferromagnetic quantum critical metal. Fermions reside on two layers ($\lambda$ = 1,2) with intralayer nearest-, second-, and third-neighbor hoppings $t_1$, $t_2$, and $t_3$. The middle layer is composed of Ising spins, subject to nearest-neighbor antiferromagnetic Ising coupling $J$ and a transverse magnetic field $h$. Between the layers, an on-site Ising coupling $\xi$ is introduced between fermion and Ising spins. (b) Phase diagram of model. The light blue line marks the phase boundaries of the pure bosonic model $H_{\text{Ising}}$, with a QCP (light blue circle) at $h_c=3.044(3)$ with 3D Ising universality. After coupling with fermions, the QCP shifts to higher values. The green solid circle is the QCP obtained with DQMC $(h_c=3.32(2))$.  The system sizes in the DQMC simulations are upto $28\times28\times200$. The figure is adapted from Ref.~\cite{liuItinerant2019}.}
	\label{fig:fig1}
\end{figure*}

Another way to generate the all-to-all strong interaction such as those in the SYK models is via the truly long-ranged Coulomb interactions. Usually in the condensed matter materials, the Coulomb interaction is screened and becomes short-ranged, but in the quantum Moir\'e materials, such as twisted bilayer graphene (TBG) and twisted metal transition metal dichalcogenides (TMD), due to the perfect 2D setting with flat-bands, these systems are bestowed with the quantum geometry of wavefunctions -- manifested in the distribution of Berry curvature in the flat bands -- and strong long-range Coulomb electron interactions, and they exhibit rich phase diagram of correlated insulating and superconducting phases thanks to the high tunability by twisting angles, gating and tailored design of the dielectric environment~\cite{tramblyLocalization2010,tramblyNumerical2012,bistritzerMoire2011,lopesContinuum2012,lopesGraphene2007,caoUnconventional2018,shenCorrelated2020,xieSpectroscopic2019,KhalafCharged2021,KevinStrongly2020,pierceUnconventional2021,caoCorrelated2018,luSuperconductors2019,liaoValence2019,liaoCorrelated2021,moriyamaObservation2019,chenTunable2020,rozhkovElectronic2016,ChatterjeeSkyrmion2020,kerelskyMaximized2019,rozenEntropic2021,tomarkenElectronic2019,soejimaEfficient2020,liuSpectroscopy2021,liaoCorrelation2021,KhalafSoftmodes2020,ZondinerCascade2020,saitoPomeranchuk2021,GhiottoCriticality2021,SchindlerTrion2022,WangTMD2020,Parkchern2021,anInteraction2020,huangGiant2020}. In Sec.~\ref{sec:IV}, we introduce the model design and numerical methodology developments in the quantum Moir\'e lattice models. In particular, the momentum-space quantum Monte Carlo method developed by us~\cite{zhangMomentum2021,daiQuantum2022} and its applications in the study of ground state phase diagram~\cite{zhangSuperconductivity2021}, the dynamic properties of single-particle and collective excitations~\cite{panDynamical2022}, as well as the possible pairing mechanism~\cite{zhangSuperconductivity2021} and the {\it Sign bound theory} for the flat-band correlated Hamiltonians~\cite{zhangSign2021,panSign2022,ouyangProjection2021}, are thoroughly discussed. Our model design and computation also reveal important symmetry-breaking patterns in the ground state~\cite{xuKekule2018,liaoValence2019,liaoCorrelated2021,liaoCorrelation2021}  and universal thermodynamic and dynamic properties of the correlated flat-band systems that deeply rooted in the collective excitations unique to the Moir\'e materials~\cite{liaoCorrelation2021,chenRealization2021,linExciton2022,panDynamical2022}. From these efforts, the mysteries about the correlated insulating and superconducting states, with their topological and multi-degrees of freedom such as spin, valley, layer and bands characteristics, are gradually being addressed in a controlled manner.

Finally, in Sec.~\ref{sec:V} we point out several immediate directions along the main content of this review, some of which are being actively pursued by us and other members of the community, and it is with these collective efforts that we firmly believe the sport and pastime of the model design and computation for quantum many-body systems shall expand and prevail.

\section{Lattice model and simulations for quantum critical metals}
\label{sec:II}
	
\subsection{Antiferromagnetic and ferromagnetic quantum critical metal models}
The lattice models of such quantum critical metals have generic form. For example, in the Ref.~\cite{liuItinerant2019}, the authors design a square lattice AFM model with two fermion layers and
one Ising spin layer in between, which is shown in Fig.~\ref{fig:fig1}(a). The Hamiltonian is given as:
\begin{equation}
	H = H_{\text{f}} + H_{\text{Ising}} + H_{\text{Int}}
	\label{eq:A1}
\end{equation}
where,
\begin{eqnarray}
		H_{\text{f}}  &= & -t_1\sum_{\left \langle ij \right \rangle \sigma \lambda }c_{i \sigma \lambda, }^\dagger c_{j \sigma \lambda }-t_2\sum_{\left \langle \left \langle ij \right \rangle \right \rangle  \sigma \lambda }c_{i\sigma \lambda }^\dagger c_{j \sigma\lambda}\nonumber \\
		&  &-t_3\sum_{\left \langle \left \langle \left \langle ij \right \rangle \right \rangle \right \rangle \sigma \lambda }c_{i \sigma\lambda }^\dagger c_{j \sigma \lambda}+h.c.-\mu\sum_{i \sigma \lambda }n_{i \sigma \lambda } \nonumber\\
		H_{\text{Ising}} &= & J\sum_{\left \langle ij \right \rangle }s_i^z s_j^z-h\sum_{i}s_i^x \nonumber\\
		H_{\text{Int}}&= & -\xi \sum_{i}s_i^z\left(\sigma_{i,1}^z + \sigma_{i,2}^z \right).
		\label{eq:A2}
\end{eqnarray}
$H_{\text{f}}$ is the free fermion Hamiltonian with $\sigma=\uparrow, \downarrow$ and $\lambda =1,2$ the spin and layer indices. Intralayer fermions are subject to nearest, second and third neighbor hoppings $t_{1}=1.0$, $t_{2}=-0.32$ and $t_{3}=0.128$, as well as the chemical potential $\mu=-1.11856$ (which means the electron density is $\left\langle n_{i\lambda}\right\rangle \sim 0.8$ here).  $H_{\text{Ising}}$ is the bosonic Hamiltonian realized in the form of a transverse field quantum Ising model with antiferromagnetic nearest neighbor coupling $J=1$. $H_{\text{Int}}$ is the coupling term between the Ising spin and the fermion via an inter-layer onsite Ising coupling $\xi=1$ and $\sigma^{z}_{i\lambda}=\frac{1}{2}(c^{\dagger}_{i\uparrow \lambda}c_{i\uparrow\lambda}-c^{\dagger}_{i\downarrow\lambda}c_{i\downarrow\lambda})$ is the fermion spin along $z$. The parameters are chosen according to Ref. \cite{sachdevDensity2014} such that there are four big Fermi pockets and four pairs of hot spots (hot spot number $N_{\text{h.s.}} = 8\times 2 = 16$ where the factor 2 comes from two fermion layers) when it is deep in the AFM phase of Ising spins.

In a series of works Ref.~\cite{liuItinerant2019,liuItinerant2018,liuElective2019}, the authors have established the nFL self-energy $\omega_n^{1/2}$, bosonc susceptibility $\chi^{-1}(q,\Omega_m)\sim (q^2 + \Omega_m)^{1-\eta}$, the anomalous dimension $\eta \sim \frac{1}{N_{h.s.}}$ roughly consistent with the HMM theory and its higher order perturbative renormalization group analyses~\cite{abanovQuantum2003,metlitskiQuantum2010_2}. However, as we will explain in the  Sec.~\ref{sec:IIc} and ~\ref{sec:IId} below, to numerically obtain these results require careful analysis and we have also seen signatures beyond the one-loop perturbative RG calculations. Whether the other fixed points, proposed for such type of AFM QCP with $z=1$~\cite{schliefExact2017} and their possible numerical investigations~\cite{luntsNon2022} can be verified, still remains open question. 

Compared with the model of the antiferromagnetic Ising quantum critical metal, the ferromagnetic quantum critical metals we studied consist of three similar terms. Here, we introduce two similar ferromagnetic models, with quantum Ising spin~\cite{xuNonFermi2017,xuIdentification2020,xuRevealing2019} and quantum rotor~\cite{liuDynamical2022,jiangMonte2022} coupled to fermions, respectively. The free fermion part remains two layers to avoid sign problem, two spin flavors offering degenerate FS geometry. After certain unitary transformation without changing the values of weights, the matrix elements of the two layers are complex conjugate to each other, that is, the corresponding weights are complex conjugate to each other, and the total weight is a positive real number. There is no sign problem here.  The bosonic degrees of freedom are of $Z_2$ and O(2) symmetries, corresponding to the Ising and XY cases, with the latter replaces the original separate rotation of fermions and the Ising spins to the joint rotation of fermions and rotors and renders the OP non-conserved. The coupling term $H_{\text{Int}}$ is onsite, leading to criticality on the entire FS. The Hamiltonian of Ising coupled fermions follows Eq.~\eqref{eq:A2}, where we set $J<0$. For rotor coupled fermions, we have $H= H_{\text{f}} + H_{\text{rotor}} +H_{\text{Int}}$ with
\begin{eqnarray}
	H_{\text{f}}  &= & -t_1\sum_{\left \langle ij \right \rangle \sigma \lambda }c_{i \sigma \lambda, }^\dagger c_{j \sigma \lambda }-t_2\sum_{\left \langle \left \langle ij \right \rangle \right \rangle  \sigma \lambda }c_{i\sigma \lambda }^\dagger c_{j \sigma\lambda}\nonumber \\
	&  &-t_3\sum_{\left \langle \left \langle \left \langle ij \right \rangle \right \rangle \right \rangle \sigma \lambda }c_{i \sigma\lambda }^\dagger c_{j \sigma \lambda}+h.c.-\mu\sum_{i \sigma \lambda }n_{i \sigma \lambda } \nonumber\\
	H_{\text{rotor}} &=& \frac{U}{2}\sum\limits_i  L_i^2  - t_\text{b}\sum\limits_{\langle i,j \rangle } \cos \left({{ \theta }_i} - {{ \theta }_j}\right) \nonumber\\
	H_{\text{Int}} &=& - \frac{K}{2}\sum\limits_{i\lambda } {\left( c_{i\lambda }^\dagger {\sigma^x} c_{i\lambda } \cdot \cos {\theta_i} + c_{i\lambda }^\dagger {\sigma ^y}c_{i\lambda} \cdot \sin{\theta_i}\right)}.
\label{eq:eqA3}
\end{eqnarray}
Likewise in Eq.~\eqref{eq:A2}, $H_{\text{f}}$ contains fermions from two layers, labeled $\lambda$, and two spin flavors, labeled $\sigma$. While for XY case, $H_{\text{rotor}}$ describes the dynamics of quantum rotors, where $L_i$ is the angular momentum and $\theta_i\in[0,2\pi)$ is the angle . $H_{\text{Int}}$ couples the rotor and fermions with on-site coupling strength $K$. The expression in brakets is equivalent to $c_{i\lambda}^{\dagger} \vec{\sigma} c_{i\lambda} \cdot \vec{S}_{i} $, written in fermionic bilinears representations. 

\begin{figure}[!htp]
	\centering
	\includegraphics[width=\columnwidth]{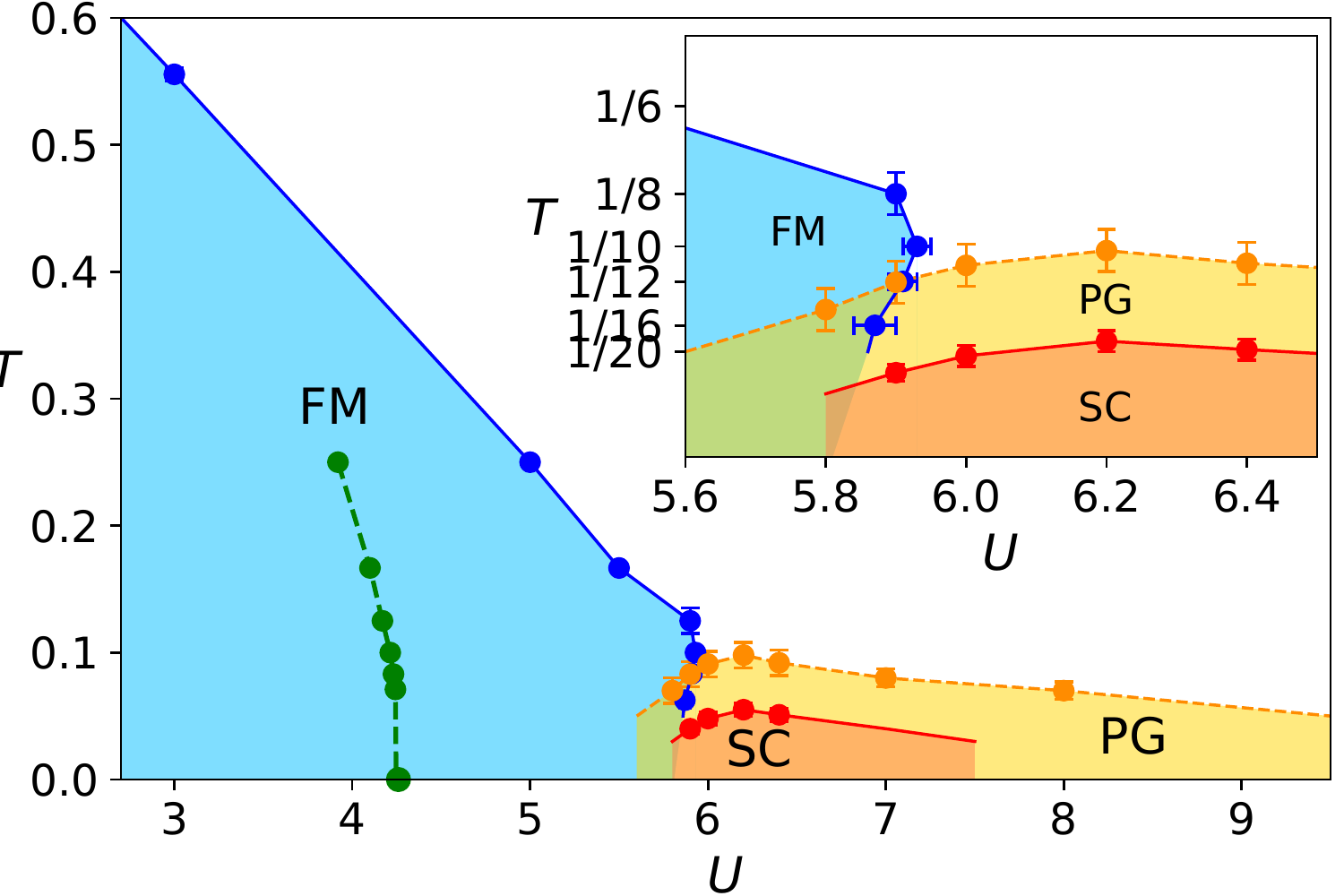}
	\small
	\caption{The $T$-$U$ phase diagram of coupling strength $K=4$ rotor coupled fermions model. Blue region represents ferromagnetic/superfluid phase of rotors. Blue dots with solid line ( green dots with dashed line ) are the phase boundary of coupled(bare) rotor model, which are determined by the superfluid data collapse. Yellow and red regions denote pseudogap and superconducting phase of fermions, respectively. The upper boundary is estimated by identifying the onset of the gap of fermionic spectrum function. While superconducting upper boundary is determined by the pairing susceptibility data collapse. Superconducting fluctuation is enhanced, since $K$ is large, and so-called pseudogap region appears. On the other hand for bosonic part, the phase boundary bends over at low temperature, giving rise to the re-entrance phenomenon of superfluid phase. The figure is adapted from Ref.~\cite{jiangMonte2022}.}
	\label{fig:fig_phase1}
\end{figure}

Without the coupling term, the bare rotor model exhibits quasi-long-range ferromagnetic order at small $U/t_{\text{b}}$. Conversely, at large $U/t_{\text{b}}$, the rotors degenerate to individual degree of freedom on each lattice site, which represents the disordered phase. The two phases are bridged via a Kosterlitz-Thouless transition. The quantum Monte Carlo simulation of the quantum rotor model has been discussed thoroughly in Ref.~\cite{jiangSolving2022}. When tuning on the coupling strength, gapless excitations near quantum critical points drives fermions to develop effective interactions, and changes rotors dynamics as well. In our studies~\cite{liuDynamical2022,jiangMonte2022}, we fix the FS geometry by assigning $t_1=1, t_2=0.2, t_3=0, \mu=0$, to avoid nesting. Besides, we set $t_{\text{b}}=1$, and change $U$ and temperature $T$ to obtain the phase diagram. Note that the different choice of $K$ leads to totally distinct physics. Here, we consider two values, $K=1,4$, giving rise to the nFL behavior and pseudogap, superconductivity behavior at intermediate temperature scales. We will focus on these results in Sec.~\ref{sec:IIB} and ~\ref{sec:IIc}.

\begin{figure}[!htp]
	\centering
	\includegraphics[width=\columnwidth]{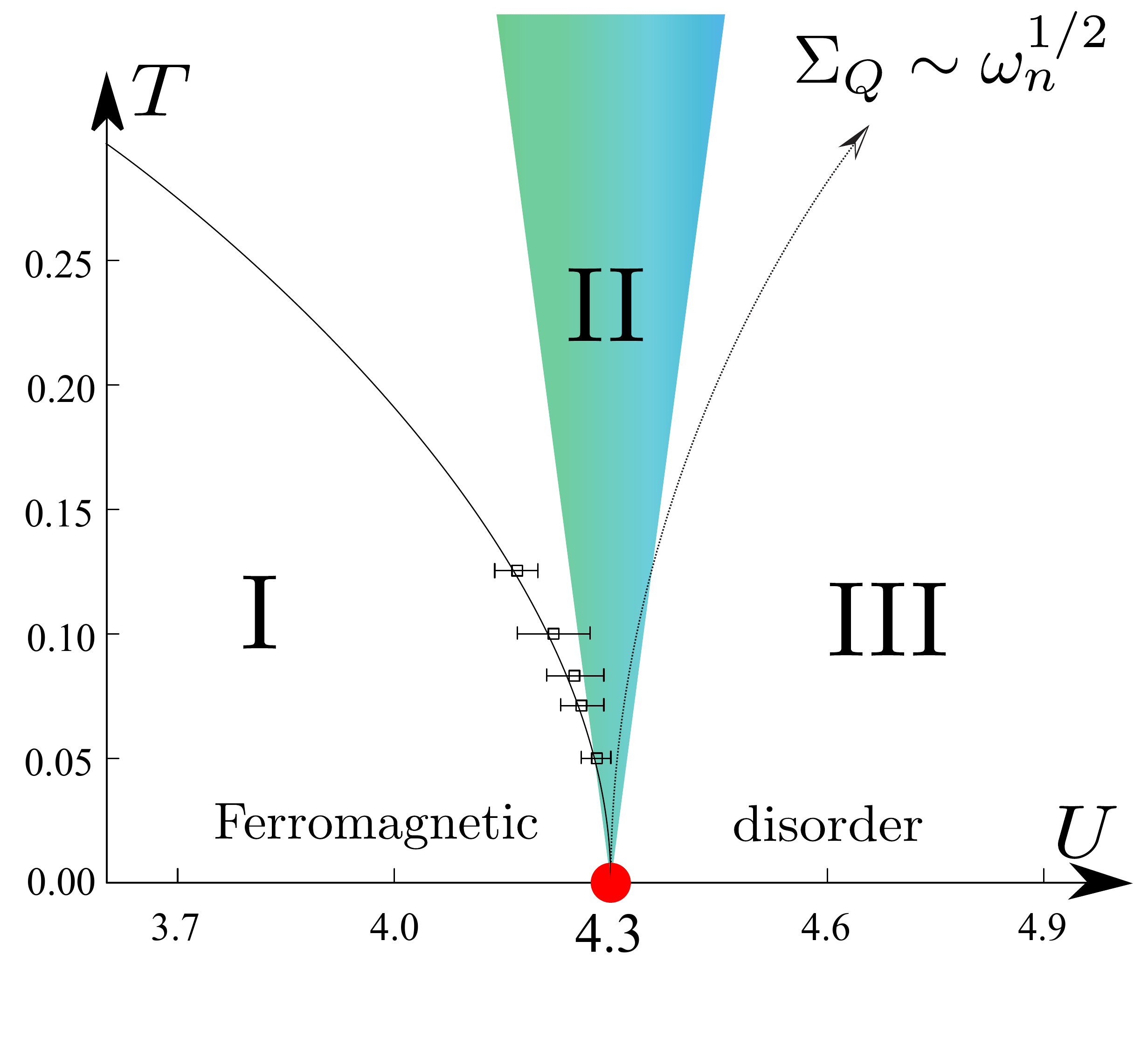}
	\small
	\caption{The $T$-$U$ phase diagram of $K=1$ rotor coupled fermions model. Compared with $K=4$ in Fig.~\ref{fig:fig_phase1}, the superconducting fluctuations is suppressed in the temperature range we studied, which is convenient for us to explore nFL behavior. The  putative QCP is shown as red dots at $U=4.3$. Region \uppercase\expandafter{\romannumeral1}, \uppercase\expandafter{\romannumeral3} denote ferromagnetic and disorder phase for bosonic part. The phase boundary is determined by the susceptibility data collapse at fixed temperature. nFL labeled region \uppercase\expandafter{\romannumeral2} appears near QCP, where the quantum part of self energy satisfies $\sim \omega_n^{1/2}$. The figure is adapted from Ref.~\cite{liuDynamical2022}.}
	\label{fig:fig_phase2}
\end{figure}

\subsection{Pseudogap and superconductivity near ferromagnetic critical point}
\label{sec:IIB}
First, we show the phase diagram at $K=1,4$ in Fig.~\ref{fig:fig_phase1} and Fig.~\ref{fig:fig_phase2}. As described in Eq.~\eqref{eq:eqA3}, we set large coupling strength $K=4$ to produce effective pairing for fermions. At $K=1$, the superconducting fluctuations is not detected until $T=0.05$. Previous studies of spin-fermion model also observed similar superconducting dome near quantum critical point Ref.~\cite{schattnerCompeting2016,gerlachQuantum2017,LiNature2017,bauerHierarchy2020}. However, in our study, we identify a pseudogap phase that never observed in such a spin-fermion model. To carefully study the properties of these phase diagrams, primarily, we measure correlation functions of Cooper pairs in various pairing channels, including s-wave and p-wave, intra-layer singlet and triplet, spin-singlet and triplet. We find that the dominant pairing channel is the on-site, s-wave, layer-singlet, spin-triplet, with total spin $S=0$, written as $\Delta(\mathbf{r})=\frac{1}{\sqrt{2}} ( c_{\mathbf{r}1\uparrow} c_{\mathbf{r}2\downarrow} - c_{\mathbf{r}2\uparrow} c_{\mathbf{r}1\downarrow}) = \frac{1}{\sqrt{2}} (c_{\mathbf{r}1\uparrow} c_{\mathbf{r}2\downarrow} + c_{\mathbf{r}1\downarrow} c_{\mathbf{r}2\uparrow})$. 

\begin{figure}[!htp]
	\centering
	\includegraphics[width=\columnwidth]{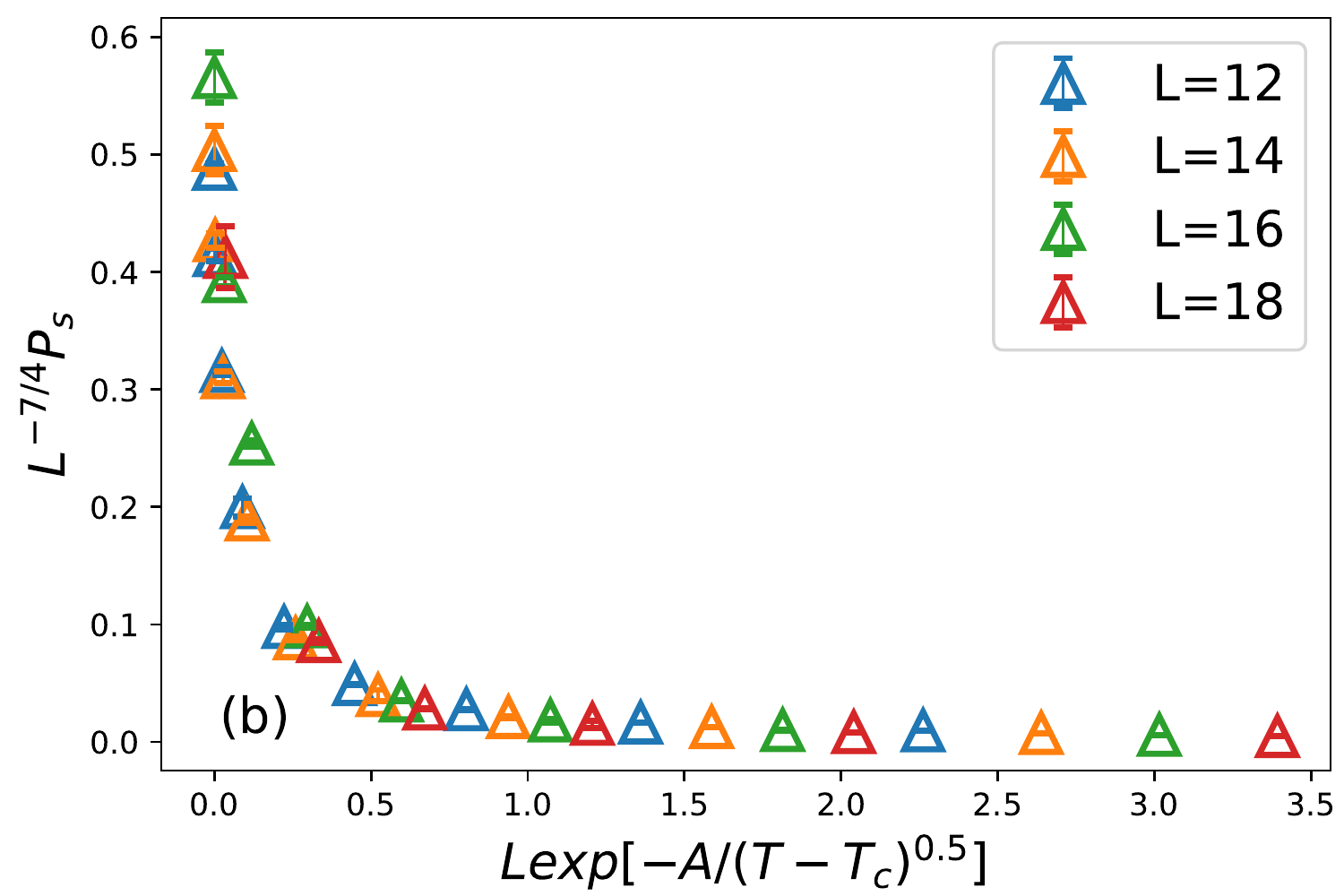}
	\small
	\caption{Data collapse of pairing susceptibility following KT scaling behavior, $P_s = L^{2-\eta_c} f(L e^{-A/(T-T_c)^{1/2}})$, where $\eta_c=0.25$, $f$ is a universal function. The best fit gives $A=0.075, T_c=0.048$. The figure is adapted from Ref.~\cite{jiangMonte2022}.}
	\label{fig:fig_ps}
\end{figure}

It is feasible to calculate the paring susceptibility using dominant paring channels $\Delta$ by defining $P_\text{s}= \frac{1}{L^2}\int_0^{\beta} \sum_i (\Delta^{\dagger}(\mathbf{r}_i,\tau) \Delta(\mathbf{0},0))$. $P_s$ reflects the correlation between cooper pairs, and there is quasi-long-range order at two dimensional cases. Utilizing the Kosterlitz-Thouless form $P_\text{s} = L^{2-\eta_c} f(L\cdot \exp(-\frac{A}{(T-T_\text{c})^{1/2}}))$ for $T > T_\text{c}$, with $\eta_{c}=1/4$, the authors in Ref~\cite{jiangMonte2022} perform the data collapse to determine the superconducting transition temperatures for various $U$. Fig.~\ref{fig:fig_ps} shows the fitting results at $U=6.0$. $T_c$ at various $U$ is plotted as red dots in Fig.~\ref{fig:fig_phase1}, whose maximum locates at $T_c \sim 0.048$ and $U \sim 6.2$.  

\begin{figure}[!htp]
	\centering
	\includegraphics[width=\columnwidth]{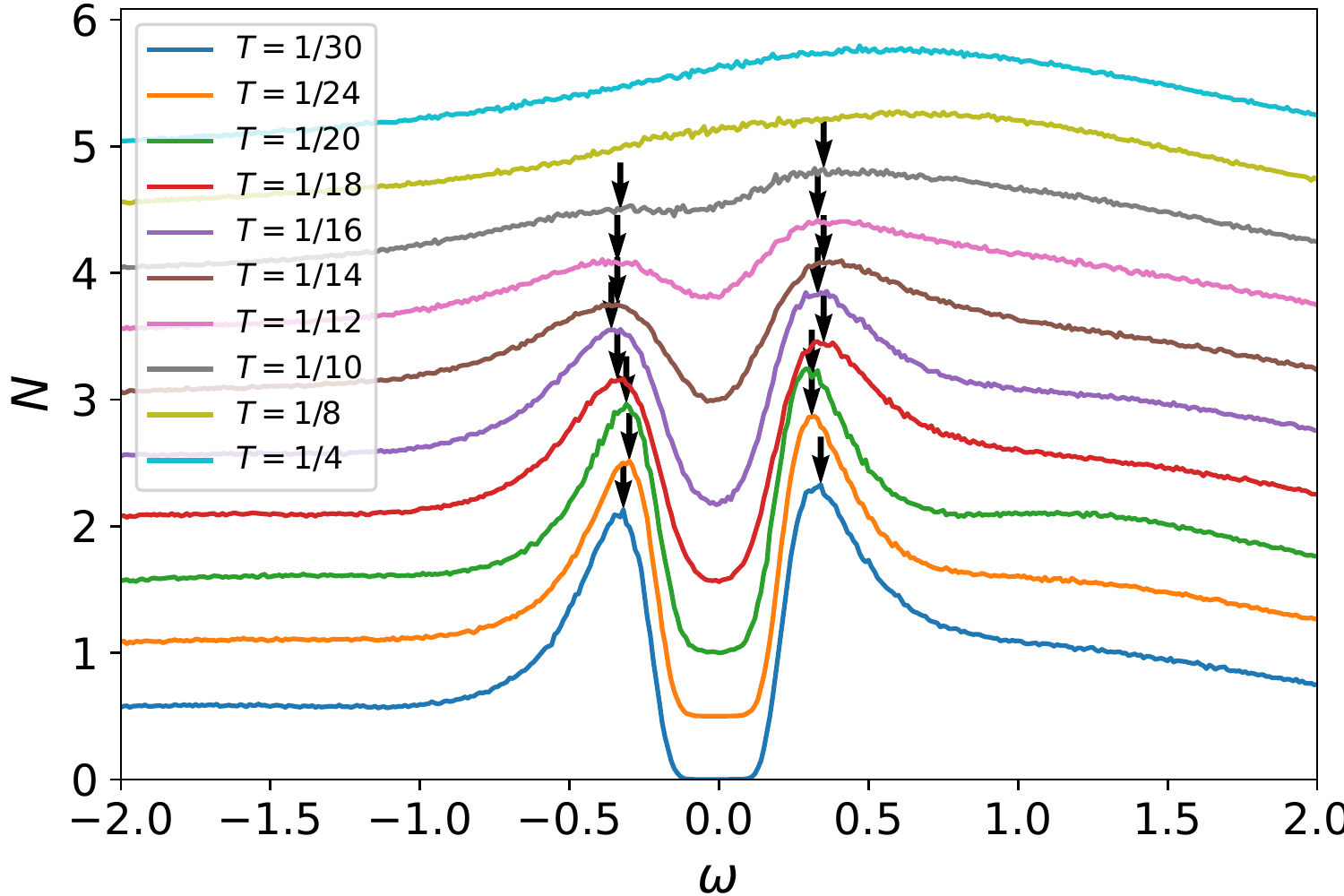}
	\small
	\caption{Local density of states at $U=6.0$, $L=12$ for various temperature. The fermionic state goes through nFL, pseudogap, and superconducting state, corresponding temperature ranges $T>0.1$, $0.1>T>0.048$, $0.048>T$. At nFL state, the spectrum has no minimum near $\omega=0$, indicating high-temperature continuum behavior. The onset of pseudogap behavior is identified by existing local minimum near $\omega=0$, corresponding $T=0.1$ in the spectrum. Decreasing the temperature, the gap exhibits gap-filling evolution, instead gap closing for conventional superconductor. At lowest temperature, full gap appears, as in the superconducting phase. The figure is adapted from Ref.~\cite{jiangMonte2022}.}
	\label{fig:fig_spec}
\end{figure}

Subsequently, we focus on the temperature scales higher than $T_c$. Former studies showed nFL behavior ~\cite{xuNonFermi2017,liuItinerant2019}, which is similar to the cuprate phase diagram. The most direct way is to examine the single particle density of states. Nonetheless, in our model, due to strong bosonic fluctuation, we assume there exists a pseudogap region surrounding the superconducting phase. To clarify this in a most direct way, the authors in Ref~\cite{jiangMonte2022} compute the single particle density of states in QMC simulation, imitating the angle-resolved photoemission spectroscopy (ARPES) experiments results ~\cite{vishikPhotoemission2018,sobotaAngle2021}. They calculate the local Green's function along imaginary axis and perform the analytic continuation~\cite{sunDynamical2018,zhou2020amplitude,yan2021topological,wangFractionalized2021} and finally obtain the results in real frequency shown in Fig.~\ref{fig:fig_spec}. Remarkably, when focusing on the onset of the full-gap near $\omega=0$, one find the consistent temperature approximately at $T=0.05$, comparable with that of $T_c$.

It must be stressed that the pseudogap region here is not directly comparable with the pseudogap phenomenon in cuprate high temperature superconductors, since we are working in a ferromagnetic QCP regime rather an antiferromagnetic one. Besides, the cooper pairs in this model possess s-wave symmetry, instead of p-wave symmetry near nematic QCP, or d-wave near antiferromagnetic QCP, which results from the isotropic FS, in contrast with the nodal-antinodal structure. Generally, the first appearance of the high-temperature gap at the antinodal momentum is identified as the onset of pseudogap behavior. In view of this, we determine the upper boundary by observation of density of states curve to find whether there exists a minimum near fermi energy. In Fig.~\ref{fig:fig_ps} at $U=6.0$, we find $T_{\text{PG}} \sim 0.1$. Tuning $U$, we obtain one crossover line shown as the yellow dashed line in the phase diagram, whose maximum is also located near QCP. 

Another way to estimate the upper boundary of the pseudogap region comes from the Eliashberg equation. Within this theory, one solves the set of self-consistent equations for fermionic self-energy and bosonic propagator, with the latter as the input information and obtained by the numerical simulation. We further map the whole model with the $\gamma$-model of theoretical analysis and find $T_{\text{PG}} = 0.08$. Note here, $\gamma$-model are theoretical quantum-critical models for which the dynamical four-fermion interaction $V(q,\omega_n) = \bar g^{\gamma} /|\omega_n|^{\gamma}$, where $\bar g$ represents effective four-fermion interactions. Our model corresponds to $\gamma = 1/3$ model~\cite{wuInterplay2020}, where the results of the Eliashberg equation gives $T_{\text{PG}} = 4.4 \bar g$. We calculate $\bar g$ with respect to the simulation results of self-energies, and finally get $T_{\text{PG}} = 0.08$, which is in good agreement with the measurement of $T_{\text{PG}} = 0.1$ at $U=6.0$.

Between $T_{\text{PG}}$ and $T_c$ is the region, which we regard as the pseudogap phase. In contrast with the conventional superconductors, the temperature evolution of the fermion spectral gap follows a gap-filling regime, instead of a gap-closing regime. The density of state remains finite at $\omega=0$.

\begin{figure}[!htp]
	\centering
	\includegraphics[width=\columnwidth]{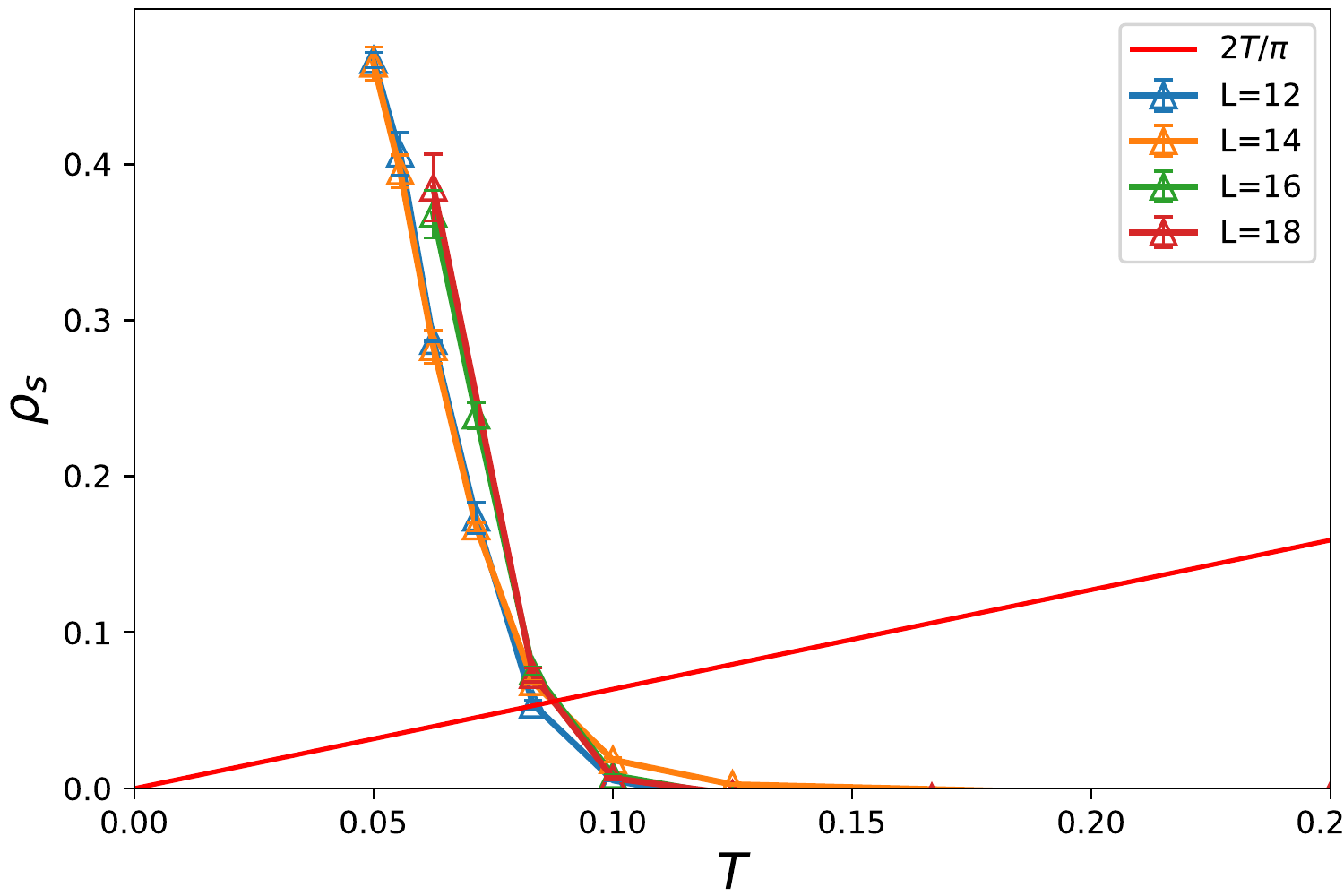}
	\small
	\caption{Superfluid density $\rho_s$ versus temperature at $U = 6$ for system sizes various system size. The onset temperature of superconducting fluctuation is approximated by the crossover temperature for curve of $\rho_s(L \rightarrow \infty)$ and linear function with slope $2/\pi$. For studied system size, we estimate such temperature is at the scale of
	$T \sim 0.1$, consistent with the onset of pseudogap in the phase diagram Fig.~\ref{fig:fig_phase1}. The figure is adapted from Ref.~\cite{jiangMonte2022}.}
	\label{fig:fig_sf}
\end{figure}

To further study the temperature scales of the pseudogap region, we calculate the superfluid density $\rho_s$. 
In such model, $\rho_s$ is used for determining the approximate onset temperature of pseudogap, as it reaches the universal coefficient $\frac{2\pi}{T}$, shown in Fig.~\ref{fig:fig_sf} and we found the temperature scale of $T\sim 0.1$, consistent with the $T_{\text{PG}}$ discussed above.

To sum up, we identify a pseudogap region above $T_c$ and determine its phase boundary primarily by observing single particle spectrum structure, assisted with other observables.

\subsection{nFL behavior and fermionic self-energy analysis}
\label{sec:IIc}

\begin{figure}[!htp]
	\centering
	\includegraphics[width=\columnwidth]{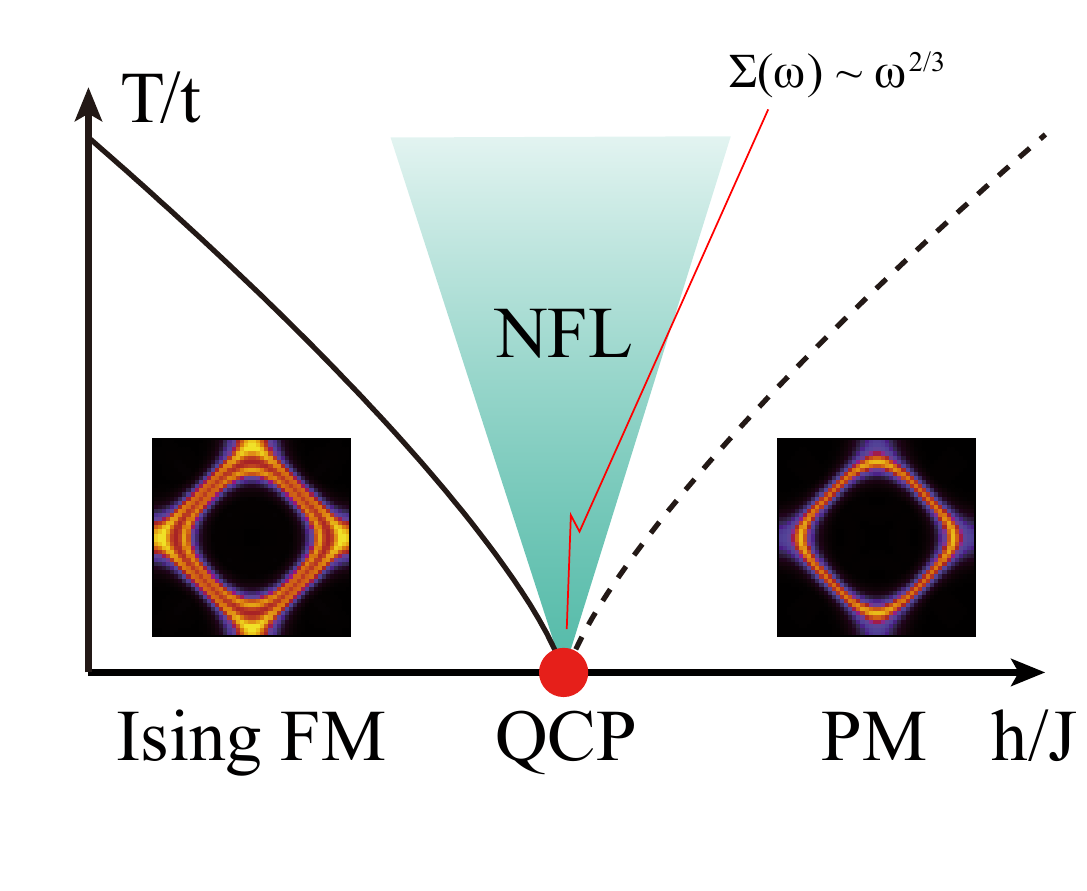}
	\small
	\caption{The $T$-$h$ phase diagram of Ising coupled fermions model, which is similar to the phase diagram in Fig.~\ref{fig:fig_phase2}, where both nFL state, ferromagnetic and disorder phase exist. The difference comes from the nFL state, where in this case, the quantum part of self energy satisfies $\sim \omega_n^{1/2}$. The figure is adapted from Ref.~\cite{xuNonFermi2017}.}
	\label{fig:fig_phase3}
\end{figure}

Above $T_c$, the fermions near QCP exhibit incoherent properties, which is regarded as the nFL state. To construct this state in a lattice model, reminiscent of prior chosen coupling strength $K$, we expect it is more convenient to study the nFL in the small $K$ regime, where the superconducting fluctuation is greatly suppressed. We take $K=1$ rotor coupled fermions model in Eq.~\eqref{eq:eqA3}, accompanied by Ising coupled fermion model in Eq.~\eqref{eq:A2} as two examples, whose phase diagrams are shown in Fig.~\ref{fig:fig_phase2} and Fig.~\ref{fig:fig_phase3}. 

\begin{figure}[!htp]
	\centering
	\includegraphics[width=\columnwidth]{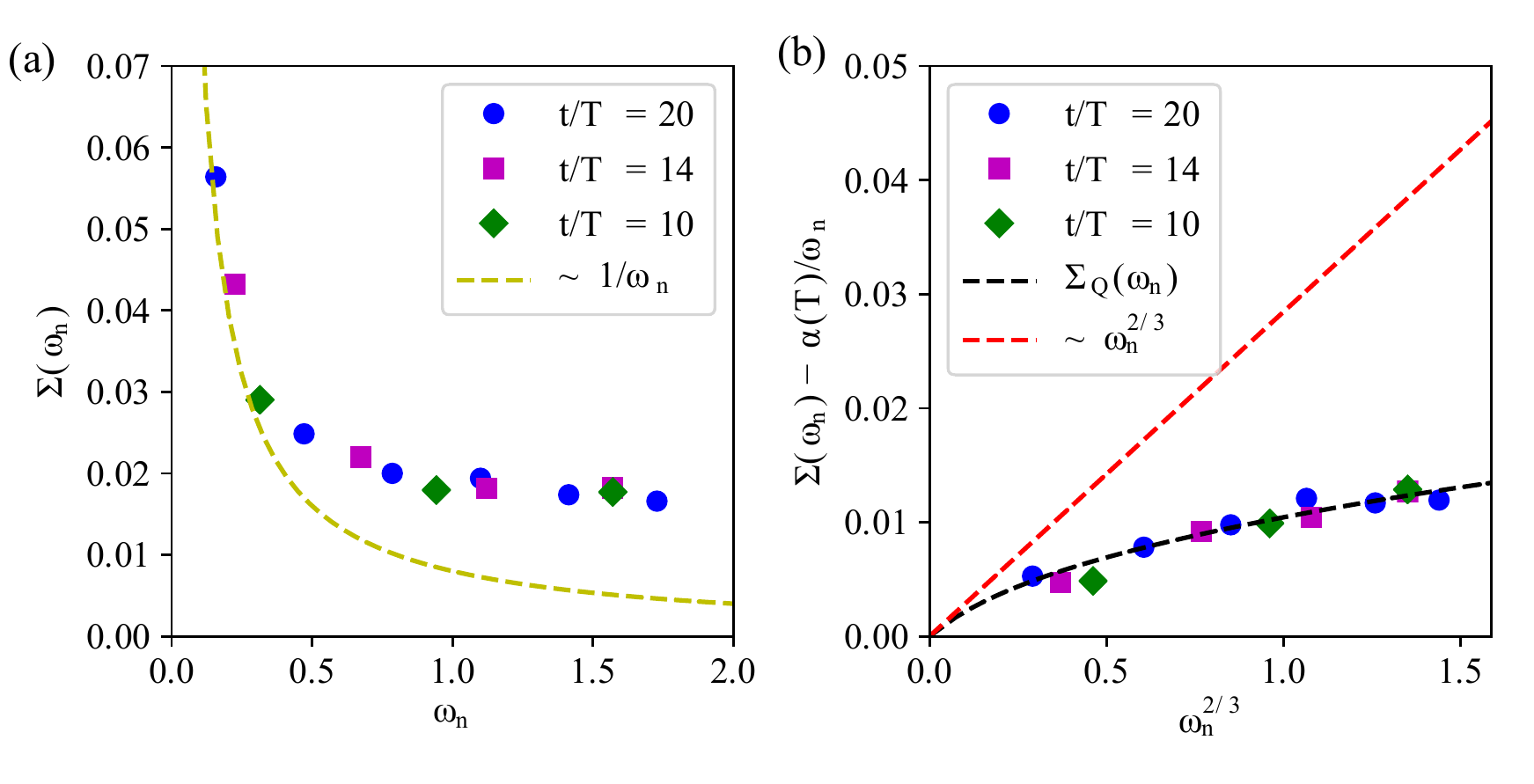}
	\small
	\caption{Self energy analysis of Ising coupled fermions model. (a) shows the self energy versus Matsubara frequencies $\omega_n$ at various temperature, where clear $\frac{1}{\omega_n}$ behavior(shown in dashed line) is observed at small $\omega_n$. (b) Subtracting contributions of thermal part, $y$-axis reviews quantum part contribution. $x$-axis is set to be $\omega^{2/3}$, and red dashed line is guided to the eyes as $\Sigma_Q \sim \omega^{2/3}$. Numerica data of solid dots fall on the black dashed line, which asymptotically approaches $\sim \omega^{2/3}$. The figure is adapted from Ref.~\cite{xuIdentification2020}.}
	\label{fig:fig_se}
\end{figure}

In early studies, the ferromagnetic critical point was explored by integrating out fermions departure from FS in the effective action and obtaining the effective Lagrangian for bosonic
degrees of freedom, known as HMM theory~\cite{hertzQuantum1976,millisEffect1993,moriyaSpin1978}. The theory predicts nFL states near QCP, and the change of critical exponents and university class from the pure bosonic theory. Hertz's calculation found the dynamic critical point near ferromagnetic QCP is $z=3$ . To verify the nFL behavior, an obvious judgement is the quasi-particle weight, which approaches zero as temperature goes down. Numerical studies in Ising fluctuation coupled with free fermions clearly show this feature ~\cite{hertzQuantum1976}. Next, we focus on the fermionic self-energy. We plot the imaginary part of self-energy Im$\Sigma$ versus fermionic Matsubara frequency $\omega_n=(2n+1)\pi T$ in Fig.~\ref{fig:fig_se}. It is well known in a FL state, the self energy is linearly proportional to $\omega_n$, contributing to the quasi-particle weight. Here, we find at small $\omega$, Im$\Sigma$ diverges approaching zero frequency. To explain this, we use the Eliashberg equation to calculate the explicit expression of $\Sigma$. In the first place, we have the following relations in such coupling strength
\begin{equation}
	\omega_{\text{F}} \ll \Sigma \ll \pi T, \bar{g}, \omega_b \ll E_{\text{F}}
	\label{eq:B1}
\end{equation}
where $\omega_{\text{F}}$/$\omega_b$ is fermionic/bosonic crossover frequency scale, where the self-energies change its power law behavior. $\omega_{\text{F}} \ll \omega_b$ guarantees the studied Matsubara frequencies $\omega_n$ is much smaller than $\omega_b$. Therefore, shown as following, one can expand $\Sigma$ with the power of $\frac{\omega_n}{\omega_b}$. Besides, $\bar{g}$ is the effective fermion-boson coupling, and satisfies $\bar{g} \ll E_{\text{F}}$, which provides
data for $\Sigma(\omega_n)$ for a substantial number of Matsubara points in the range $\omega_n \gg \Sigma(\omega_n)$. For this reason, vertex corrections that contribute to $\Sigma$ can be
neglected in this regime. These conditions simplify the solving process of Eliashberg equations~\cite{abrikosovMethods2012}. 

Our way to deal with the divergence is to separate the contribution of quantum part and thermodynamics part, indicated as $\Sigma(\omega_n) = \Sigma_{\text{T}}(\omega_n) + \Sigma_{\text{Q}}(\omega_n)$. The so-called thermal part $\Sigma_{\text{T}}(\omega_n)$ is the contribution from static thermal fluctuations and possess the form $\Sigma(\omega_n) = \frac{\alpha(T)}{\omega_n}$. Simply speaking, the finite temperature due to the Monte Carlo simulation characters introduces a finite gap, with the gap being the coefficients, playing the role of $\alpha(T)$. Therefore, at zero temperature, $\Sigma_{\text{T}}$ term vanishes. The ~$\frac{1}{\omega_n}$ behavior can be examined by a few smallest frequencies.
For $\Sigma_{\text{Q}}(\omega_n)$, the quantum part contribution comes from dynamical bosonic fluctuations. At zero temperature, $\Sigma(\omega_n) = \Sigma_{\text{Q}}(\omega_n)$ is just fermionic self-energy at QCP, indicating nFL behavior.

One needs to solve the following self-consistent Eliashberg equation, 
\begin{small}
\begin{equation}
	\begin{aligned}
	&-i \Sigma(\mathbf{q},\omega_n)=\bar{g} T \sum_{m} \int \frac{\mathrm{d}^{2} \mathbf{q'}}{(2 \pi)^{2}} G(\mathbf{q'}+\mathbf{q},\Omega_m+\omega_n) D(\mathbf{q'},\Omega_m) \\
	&\Pi(\mathbf{q'},\Omega_m)=2 N_{f} \bar{g} T \sum_{n} \int \frac{\mathrm{d}^{2} \mathbf{q}}{(2 \pi)^{2}} G(\mathbf{q}+\mathbf{q'},\omega_n+\Omega_m) G(\mathbf{q},\omega_n).
	\end{aligned}
	\label{eq:B2}
\end{equation}
\end{small}
Here, $N_f=2$ is the number of fermion flavors. $\Pi(\mathbf{q},\Omega_m)$ is the bosonic self-energy solving by free fermions propagator. $D(\mathbf{q},\Omega_m)$ is the bosonic propagator, which can be extracted by calculation of dynamic susceptibility, along with $\Pi(\mathbf{q},\Omega_m)$, while $G(\mathbf{q},\omega_n)$ is the free fermionic propagators. 

On the condition when total spin along $z$-axis is conserved, e.g. Hamiltonian like Eq.~\eqref{eq:A2}, we find the leading order of $\Pi(\mathbf{q},\Omega_m) \sim \frac{|\Omega_m|}{q} $, known as the Landau damping term calculated using Eq.~\eqref{eq:B2}. Bringing the fermionic self-energy calculation, eventually we obtain the analytic formula of $\Sigma(k_{\mathbf{F}})$ at $\omega_n \ll \omega_b$ regime in Eq.~\eqref{eq:B3}.
\begin{equation}
	\Sigma_{Q}\left(\omega_{n}\right)=\omega_{\mathrm{F}}^{1 / 3}\omega_{n}^{2 / 3} \sigma\left(\omega_{n}\right)+\cdots,
	\label{eq:B3}
\end{equation}
which indicates the fermionic self energy is $\sim \omega_{n}^{2 / 3}$ and deviates from the Fermi liquid behavior.

In view of such behavior emerging near nFL region, we further consider a more general case, where total spin along $z$-axis is not conserved. The origin comes from the unexpectable dynamical structure factor measured in the uranium metallic materials, e.g. $\text{UGe}_2$ and $\text{UCoGe}$~\cite{huxley2003Magnetic,stock2011Anisotropic}. One has, $\chi(\mathbf{q},\Omega) \approx \frac{1}{\Omega^2 + q^2+\Pi(\mathbf{q,\Omega})}$, where $\Pi(\mathbf{q,\Omega})$ is the above mentioned bosonic self-energy. As we derived, we expect $\Pi(\mathbf{q,\Omega}) \sim \frac{|\Omega|}{\sqrt{(v_{\mathbf{F}}q)^2 + \Omega^2}}$, and we have $\Pi(\mathbf{q}) \sim \frac{|\Omega|}{v_{\mathbf{F}}q} = \frac{|\Omega|}{\Gamma(\mathbf{q})}$ at $|\Omega| \ll v_{\mathbf{F}} q$, thus $\Pi(\mathbf{q}=0,\Omega)=0$ due to canceling out of the localized and itinerant fermions. Actually in $\text{UGe}_2$ and $\text{UCoGe}$, experiments found $\Pi(\mathbf{q}=0,\Omega)$ remains a finite value~\cite{huxley2003Magnetic,stock2011Anisotropic}. Essentially, this is because of the change of $\Gamma(\mathbf{q})$ from $\Gamma(\mathbf{q}) \sim q$ to $\Gamma(\mathbf{q}) \sim \Gamma_0$, i.e., a finite value. It is shown in the diagramatic calculation, the (i) the self-energy diagram, (ii) the Maki-Thompson-type vertex correction diagram, and (iii) the Aslamazov-Larkin-type diagrams possess same order contributions~\cite{chubukovNonLandau2014}. We find that,  as will explain later, for conserved spin systems, e.g., Ising coupled fermions, they are actually of the same order as the other
terms. The corresponding coefficient cancels out, which leads to vanishing $\Gamma(\mathbf{q})$ at $q \rightarrow 0$. While for non-conserved spin case, e.g. rotor coupled fermions, the Aslamazov-Larkin diagrams become irrelevant at weak coupling because the balance between the damping and coupling is lost. Thus the extra power of the coupling is not cancelled out, and finally one obtains $\Gamma(\mathbf{q}) \sim \Gamma_0$.

\begin{figure}[!htp]
\centering
\includegraphics[width=\columnwidth]{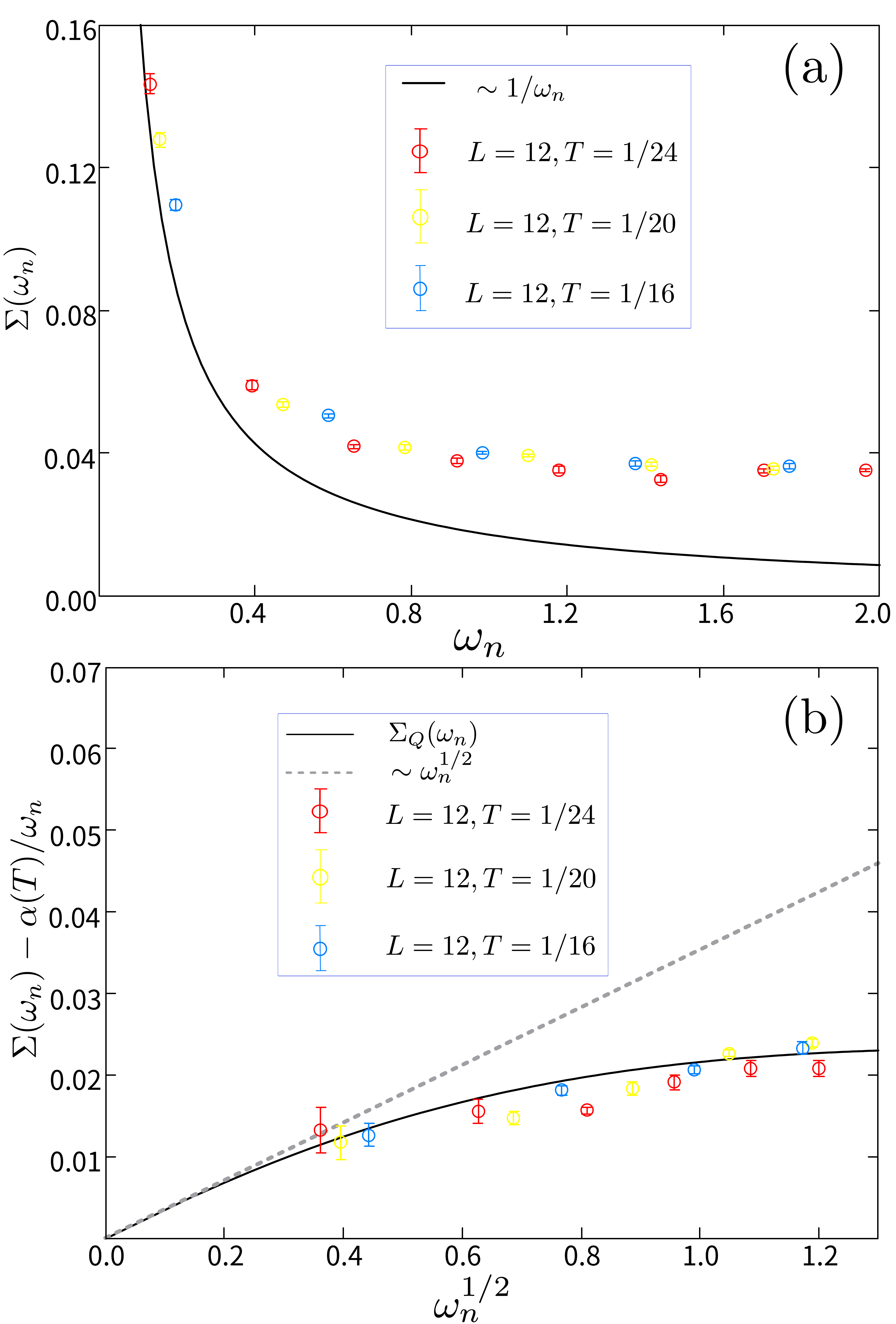}
\caption{Self-energy analysis of rotor coupled fermions model in Eq.~\eqref{eq:eqA3}. (a) shows the self energy versus Matsubara frequencies $\omega_n$ at various temperature, where clear $\frac{1}{\omega_n}$ behavior(shown in dashed line) is observed at small $\omega_n$. (b) Subtracting contributions of thermal part, $y$-axis reviews quantum part contribution. In contrast with Ising coupled fermions model, $x$-axis is set to be $\omega^{1/2}$, and dashed line is guided to the eyes as $\Sigma_Q \sim \omega^{1/2}$. Numerica data of solid dots fall on the black dashed line, which asymptotically approaches $\sim \omega^{1/2}$. The figure is adapted from Ref.~\cite{liuDynamical2022}.}
	\label{fig:fig_se2}
\end{figure}

Applying the results in the calculation of the self-consistent Eliashberg equation. The input bosonic has following form,
\begin{equation}
	D^{-1}(\mathbf{q},\Omega_m) = \frac{\chi_0}{M_0^2 + |\mathbf{q}|^2 + \Omega_m/\Gamma_0}
	\label{eq:eqB4}
\end{equation}
where $\chi_0, M_0, \Gamma_0$ are all extracted from the Monte Carlo simulation results. Note in the next subsection, we will give a detailed description of such formula. Eq.~\eqref{eq:eqB4} considered high-order diagramatic representations, which is different from first-order calculation in Eq.~\eqref{eq:B2}. Especially, the coefficients $\chi_0, M_0, \Gamma_0$ are obtained by fitting the bosonic dynamical susceptibility data in Monte Carlo simulations. Repeating similar calculation for solving fermionic self-energy as Eq.~\eqref{eq:B2}, we finally get fermionic self-energy $\Sigma_Q(\omega_n) \sim \omega_n^{1/2}$ in Fig.~\ref{fig:fig_se2}, which is also a nFL behavior. It needs to be noticed that, only in small coupling strength $K$ for rotors and fermions, e.g. $K=1$, one can separate the thermal and quantum parts of $\Sigma(\omega)$~\cite{liuDynamical2022}. 

In antiferromagnetic model, it was hard to directly reveal the scaling form of the nFL self-energies. 
	
\begin{figure}[!htp]
\centering
\includegraphics[width=0.9\linewidth]{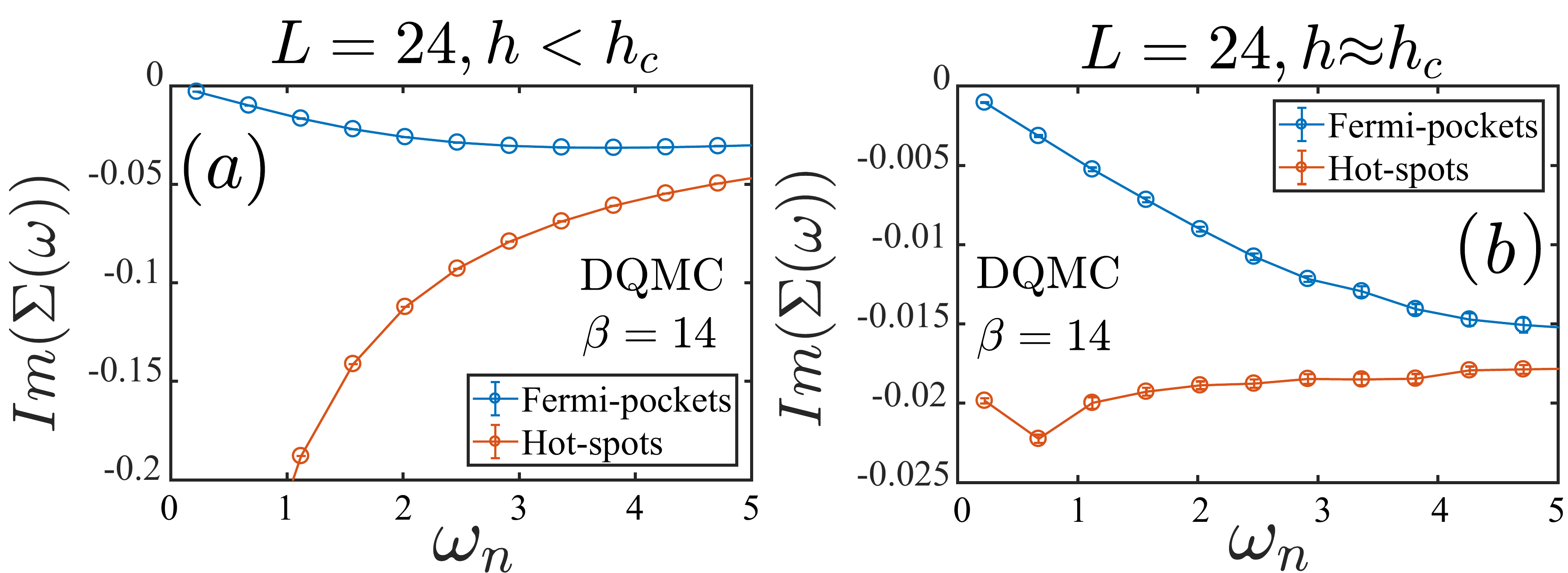}
\small
\caption{Fermionic self-energy for antiferromagnetic fluctuations coupled fermions. (a) On the Fermi pockets, the system is in a Fermi liquid state as shown by the linearly vanishing of $\text{Im}(\Sigma(\mathbf{k},\omega_n))$ at small $\omega_n$. At the hot spot, due to the Fermi surface reconstruction, an energy gap opens up, resulting in a divergent $\text{Im}(\Sigma(\mathbf{k},\omega_n))$. (b)  On the Fermi pockets, the system remains a Fermi liquid as shown by the linearly vanishing $\text{Im}(\Sigma(\mathbf{k},\omega_n))$ at small $\omega_n$. At the hot spot, $\text{Im}(\Sigma(\mathbf{k},\omega_n))$ has a small but finite value as $\omega_n$ goes to 0, which is the signature of a non-Fermi-liquid, but still different from the HMM expectation discussed in Sec.~\ref{sec:IIc} and therefore remains to be explained. The figure is adapted from Ref.~\cite{liuItinerant2019}.}
\label{fig:fig9}
\end{figure}

At $h<h_c$, trival gap leads to the divergence at small $\omega$
at hotspots. Near QCP, we find the finite value of $\text{Im} \Sigma(\omega)$, when approaching $\omega=0$. This is in contrast with the previous analysis, which separating the thermal part and the quantum part. Therefore, we conclude that the result for fermionic self-energy at antiferromagnetic QCP needs further explored. 

\label{sec:IId}
\begin{figure*}[!htp]
	\centering
	\includegraphics[width=\textwidth]{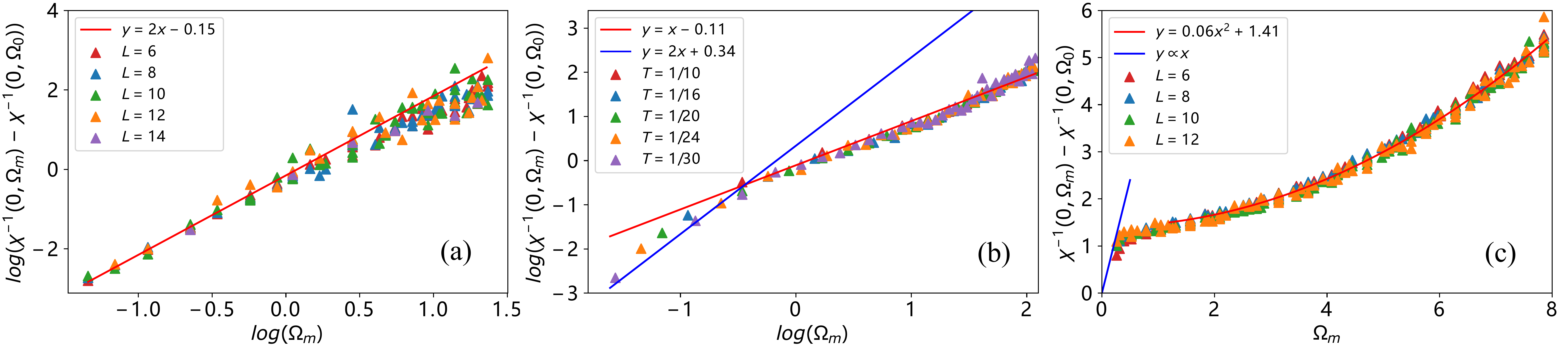}
	\small
	\caption{Bosonic dynamical susceptibilities $\chi^{-1}(q=0,\Omega_m) - \chi^{-1}(q=0,\Omega_0)$ at $U=3,6,8$, corrsponding subfigure (a), (b), (c), respectively. The subfigure (a) and (b) is plotted on log-log scale. (a) At ferromagnetic phase $U=3.0$, the lowest frequencies follows the red line guided to the eye, whose slope is of 2, indicating $z=1$ behaviors. (b) Near QCP at $U=6.0$, The fermionic parts go through nFL, pseudogap, superconducting state from higher to lower temperatures. On account of this, the bosonic dynamical susceptibilities exhibits non-landau damping behavior, transition period, and bare rotor model from higher to lower frequencies, corresponding the crossover from $z=2$(slope of 1) to $z=1$(slope of 2), descibed by the red and blue line. (c) At disorder phase $U=8.0$, the behaviors is similar to that in ferromagnetic phase with both $z=2$. The red line is the best fit of the data at $\Omega_m>0.5$, which possess square behaviors. While, a finite intercept exhibits at extrapolation of  $\Omega=0$, as the RPA calculation, where one use free fermionic propagators~\cite{abanovQuantum2003}. At small frequencies, guided by blue lines, the bosonic part is effected by the pseudogap behavior, and deviates from the red line. The figure is adapted from Ref.~\cite{jiangMonte2022}.}
	\label{fig:dynsus}
\end{figure*}

\subsection{Bosonic dynamics}
The bosonic propagators or bosonic dynamic susceptibility $\chi(\mathbf{q},\Omega_m)$ also reveal important information about the quantum critical metals. Their definitions for the Ising and rotors are,
\begin{equation}
	\begin{aligned}
		\chi\left(h, T, \mathbf{q}, \Omega_{m}\right)&=\frac{1}{L^2} \int d \tau \sum_{i j} e^{i \Omega_{m} \tau-i \mathbf{q} r_{i j}}\left\langle s_{i}^{z}(\tau) s_{j}^{z}(0)\right\rangle \\
		\chi \left(U, T, \mathbf{q}, \Omega_{m}\right)&=\frac{1}{L^2} \int d \tau \sum_{i j} e^{i \Omega_{m} \tau-i \mathbf{q} r_{i j}}\left\langle \theta_{i}(\tau) \theta_{j}(0)\right\rangle
	\end{aligned}
	\label{eq:eqB5}
\end{equation}
Generally, in such spin-fermion model, we couple free fermions to the bosons, which have their own dynamics. For the bare Ising or rotors, they both satisfy the dynamic critical exponent $z=1$, i.e., the susceptibility reads,
\begin{equation}
	\chi^{-1}(\mathbf{q},\Omega_m, r) = \frac{\chi_0}{M_r^2 + |\mathbf{q} - \mathbf{Q}|^2 + \Omega_m^2 }
	\label{eq:eqB6}
\end{equation}
where $r$ is the tuning parameter, $M_r$ represents the distance towards the QCP. $\mathbf{Q}=(0,0)$ for ferromagnetic and $(\pi,\pi)$ for antiferromagnetic models, is the ordered wave vector. When tuning on the coupling between boson and fermion, the bosonic dynamics change their behavior. Previously, HMM theory predicts the dynamic critical exponents at ferromagnetic and antiferromagnetic QCP change to $z=3$ and $z=2$, respectively. Several recent simulation results found the difference with the universality of Hertz-Millis–type exponents, indicating non-universal behavior near QCP~\cite{xuNonFermi2017,liuItinerant2018,liuItinerant2019}. Likewise, we give a detailed analysis of the results in our models. The bosonic dynamics of other cases also deserve an explanation, e.g. near QCP but covered by the superconducting fluctuations. Moreover, far from the QCP, the bosonic dynamics can be easily achieved by diagramatic calculations, such as RPA. In the subsection, we focus on the dynamic susceptibility $\chi$ near nFL state, and plenty of situation is summarized in Table.~\ref{tab:tab1}.  

\begingroup
\setlength{\tabcolsep}{10pt} 
\renewcommand{\arraystretch}{1.5} 
\begin{table*}
\caption{Summary of quantum critical bosonic self-energies for the systems in the review. $\mathbf{q}$ is the momentum measured with respect of the ordered wave vector $\mathbf{Q}=\Gamma$ for ferromagnetic and nematic QCPs and $\mathbf{Q}=\mathbf{Q_{AF}}$ for antiferromagnetic QCP. $\Omega$ is the bosonic Matsubara frequency.}
\scriptsize
\begin{tabular}{l}
	\toprule
	\multicolumn{1}{c}{FM / AFM / Nematic quantum critical scaling obtained from QMC} \\
	\midrule
	1. Ferromagnetic QCP $\;$
	$\mathbf{Q} = \Gamma$  \\
	\qquad 1a. Square lattice, $Z_2$ bosonic symmetry~\cite{xuNonFermi2017} \\
	\qquad \quad $\chi\left(T, h, \mathbf{q}, \Omega_{m}\right) \propto$ $\frac{1}{c_{t}T^{a_{t}}+c_{h}\left|h-h_{c}\right|^{\gamma}+\left(c_{q}|\mathbf{q}|^{2}+c_{\Omega} \Omega^{2}\right)^{a_{q} / 2}+\Delta(\mathbf{q}, \Omega)}$, $z=3$ and $a_{q}=2-\eta$ with $\eta=0.15(3)$, $\Delta(\mathbf{q}, \Omega)= \frac{|\Omega|}{\sqrt{(v_{\mathbf{F}}q)^2 + \Omega^2}}$\\
	\qquad 1b. Square lattice, $O(2)$ bosonic symmetry~\cite{liuDynamical2022} \\
	\qquad \quad$\chi\left(T \rightarrow 0, h \rightarrow h_c, \mathbf{q}, \Omega_{m}\right) \propto$ $\frac{1}{c_{q}|\mathbf{q}|^{2}+c_{\Omega} \Omega}$, $z=2$ for non-conserved order parameter. \\
	2. Antiferromagnetic QCP with $\mathbf{Q}=\mathbf{Q_{AF}}$\\
	\qquad 2a. Triangle lattice, $Z_2$ bosonic symmetry~\cite{liuItinerant2018} \quad $3\mathbf{Q}=\Gamma \quad$ \\
	\qquad \quad $\chi\left(T, h, \mathbf{q}, \Omega_{m}\right) \propto$ 
	$\frac{1}{\left(c_{t} T+c_{t}^{\prime} T^{2}\right)+c_{h}\left|h-h_{c}\right|^{\gamma}+c_{q}|\mathbf{q}|^{2}+\left(c_{\Omega} \Omega+c_{\Omega}^{\prime} \Omega^{2}\right)}$, $z=2$\\
	\qquad 2b. Square lattice, $O(3)$ bosonic symmetry~\cite{bauerHierarchy2020} $\quad 2 \mathbf{Q}=\Gamma \quad$ \\
	\qquad \quad $\chi\left(T \rightarrow 0, h, \mathbf{q}, \Omega_{m}\right) \propto$ 
	$\frac{1}{c_{h}\left|h-h_{c}\right|^{\gamma}+c_{q}|\mathbf{q}|^{2}+\left(c_{\Omega} \Omega\right)}$, $z=2$\\
	\qquad 2c. Square lattice, $Z_2$ bosonic symmetry~\cite{liuItinerant2019} $\quad 2 \mathbf{Q}=\Gamma \quad$ \\
	\qquad \quad $\chi\left(T, h, \mathbf{q}, \Omega_{m}\right) \propto$ $ \frac{1}{c_{t} T^{a_{t}}+c_{h}\left|h-h_{c}\right|^{\gamma}+\left(c_{q}|\mathbf{q}|^{2}+c_{\Omega} \Omega\right)^{1-\eta}+c_{\Omega}^{\prime} \Omega^{2}}$, $z=2$ and $\eta=\frac{2}{N_{h s}}=0.125\left(\right.$ with $\left.N_{h s}=16\right)$\\
	3. Nematic QCP~\cite{schattnerIsing2016} $\;$
	$\mathbf{Q} = \Gamma$  \\
	\qquad  $\chi\left(T, h, \mathbf{q}, \Omega_{m}\right) \propto$ $\frac{1}{c_t T+c_{h}\left|h-h_{c}\right|+ c_{q}|\mathbf{q}|^{2}+c_{\Omega} |\Omega| }$, $z=2$\\
	\bottomrule
\end{tabular}
\label{tab:tab1}
\end{table*}
\endgroup

Besides bare nFL state, in Sec.~\ref{sec:IIB}, we discuss the case for $K=4$ rotor model and find the bosonic susceptibility is influenced by superconducting fluctuation and the distance from the QCP. Below, we briefly discuss that even in this case, where the QCP is masked by the superconducting dome, the bosonic susceptibility can still reveal the different scaling behavior. Fig.~\ref{fig:dynsus} shows $\chi^{-1}(\mathbf{q}=0,\omega) - \chi^{-1}(\mathbf{q}=0,\omega=0)$ for different $U$. We choose $U=3.0$(ferromagnetic phase), $U=6.0$(near QCP), $U=8.0$(disordered phase), to demonstrate the joint effect of superconducting fluctuation, fermionic self-energy and non-Landau damping behavior. Considering $U$ is farm from $U_c$, first we draw attention to $U=8.0$. At bosonic Matsubara frequencies $\Omega_n > 1$, $\chi$ follows RPA calculation with respect to free fermions propagators. The behavior is reminiscent of Ising coupled fermions model, as the similar non-analytic function $\Delta(\mathbf{q},\omega)$ in Tab.~\ref{tab:tab1} 1a. At small frequencies, $\chi$ deviates from the nonzero extrapolation, since the fermions enter the pseudogap region at low temperature. Similar thing manifestes at $U=3.0$, where on log-log scale, the slope of 2(corresponding $z=1$) behavior is proper description of ferromagnetic phase. Due to the spin gap, the coupled model evolves like bare rotor model, and have no intercept in ferromagnetic phase. However, at $U=6.0$ near QCP, we separate the $\Omega$ region into three parts according to the temperature region. At large $\Omega$, corresponding $T>0.1$, the bosonic self-energies follow non-Landau damping regime and satisfies $z=2$(slope is of 1), similar to the expression of Tab.~\ref{tab:tab1} 1b. Towards low frequencies, the dominant pseudogap region drives $z=2$ to $z=1$. At lowest temperature studied, because of the superconducting gap, the behavior goes back to bare rotor model with $z=1$. 

There are still many open questions in the study of quantum critical metal. The results summarized in Tab.~\ref{tab:tab1} for dynamical critical exponent $z$, are largely consistent with the HMM prediction, for example, when the bosonic order parameters are conserved (Ising spin), one obtains the $z=3$ for ferromagnetic QCP, $z=2$ for antiferromagnetic QCP bosonic self-energies, and only when the bosonic order parameters are not conserved (rotors with O(2) symmetry), one would need higher order perturbative calculation beyond the HMM theory to obtain the $z=2$ bosonic self-energy~\cite{chubukovNonLandau2014,jiangMonte2022,liuDynamical2022}. However, there are many predictions that the antiferromagnetic bosonic fluctuations coupled fermions will eventually deviate from the HMM form and new fixed points could be found, where $z$ has strong violations from $z=2$~\cite{metlitskiQuantum2010_2,schliefExact2017}. At the present stage of the model design and computation, such deviations are yet to be unambiguously found. For example, we find in Ref.~\cite{liuItinerant2019} that the predicted rotation of the renormalization fermion velocity towards the Ising $\mathbf{Q_{AF}}$ were not as obvious as the theoretical prediction to be parallel. 

Recently there are hybrid Monte Carlo simulation results~\cite{luntsNon2022} showing at antiferromagnetic $O(3)$ QCP, there exist clear deviations of the dynamic exponent from $z=2$ to $z \approx 1.65$ at the smallest nesting angle. Furthermore, a breaking of the O(2) symmetry form of the momentum dependence down to $C_4$ happens near QCP, which also violates the HMM theory. Remarkably, one recent long review~\cite{franciscoField2022} discuss nFL implemented on the 2d AFM QCP in field-theoretic functional renormalization group formalism. The author developed the low-energy effective field theory of nFL  including all gapless modes around the Fermi surface. Through functional renormalization group flow, nFL is ascribed to the fixed point, thus its low-energy physics can be studied.  These results and statements are certainly at our interests and the interests of the community for the future research activities of the sport and pastime of model design and computation in quantum critical metals. 


\section{Yukawa-Sachdev-Ye-Kitaev Model} 
\label{sec:III}
As discussed in Secs.~\ref{sec:I} and ~\ref{sec:II}, the nFL behavior of interacting electron systems, is widely believed to be relevant
to the microscopic origin of the "strange metal" behavior in unconventional superconductors~\cite{KeimerFrom2015, LiuNematic2016,GuUnified2017,CustersThe2003,ShenStrange2019,CaoStrange2019,ShenCorrelated2019,ChenFermi2021} and many other condensed matter systems. And the nFL often
occurs via electron interactions mediated by gapless bosonic modes that render the electrons incoherent~\cite{metlitskiQuantum2010,metlitskiQuantum2010_2,metlitskiCooper2015,raghuMetallic2015,steveEnhancement2015,lawlerNonperturbativel2006,lawlerLocal2007,xuNonFermi2017,liuItinerant2018,liuItinerant2019,xuMonte2019,xuRevealing2019}. In Sec.~\ref{sec:II}, we study many 2D lattice models where we couple itinerant fermions with soft boson mode with critical fluctuations, and then study the nFL behaviors in the vicinity of the ferromagnetic and antiferromagnetic QCPs. 

Due to the lack of a natural small control parameter, the analytical solution to these models remains challenging. And in order to see nFL behavior numerically, such models always require us to turn the mass of the boson to the critical value, as seen in the examples in Sec.~\ref{sec:II}. But the system will go back to FL behaviors once away from the QCP. The position of QCP and region of nFL are model dependent and affected by the finite-size effect. In addition, we also need to deduct the non-negligible thermal contributions to the fermionic self-energy and control the strength of effective coupling~\cite{xuIdentification2020,kleinNormal2020,liuDynamical2022}, so as to see a clear signal of nFL from fermionic self-energy. These difficulties make it hard to see the scaling form of nFL, which inspired us to design other models. 

Recently, nFL behaviors in Sachdev-Ye-Kitaev (SYK) models\cite{sachdevGapless1993,kitaevTalks2015,SachdevBekenstein2015,KitaevThe2018} has garnered widespread attention. SYK model is an exactly solvable model initially proposed by Subir Sachdev and Jinwu Ye, and then modified by Alexei Kitaev. The motivation of SYK model is to come up with a model to explain the phenomena of strange metal and a simple model for quantum holography. The advantage of SYK model is that it is exactly solvable, and its exact solution is a nFL\cite{chowdhurySachdev2021}. Inspired by such exactly solvable SYK models with nFL behavior, recently a class of SYK-like models featuring random Yukawa interactions between fermions and bosons has been put forward to analyze the nFL problem~\cite{wangSolvable2019, esterlisCooper2019, schmalianEliashberg2019,wangQuantum2020,inkofQuantum2022}. 

In this Section, we will discuss our model design and computation for Yukawa-SYK model and the nFL, self-tuned quantum criticality and superconductivity therein~\cite{panYukawa2021,wangPhase2021}. It is interesting to see that the Yukawa-SYK model acquires a very unique self-tuned quantum criticality and in that the system is always inside an nFL independent of the tuning of the parameters, and the quantum fluctuations in the system also naturally give rise to the superconductivity, as we shall explain below. 

\begin{figure}[H]
	\centering
	\includegraphics[width=0.8\linewidth]{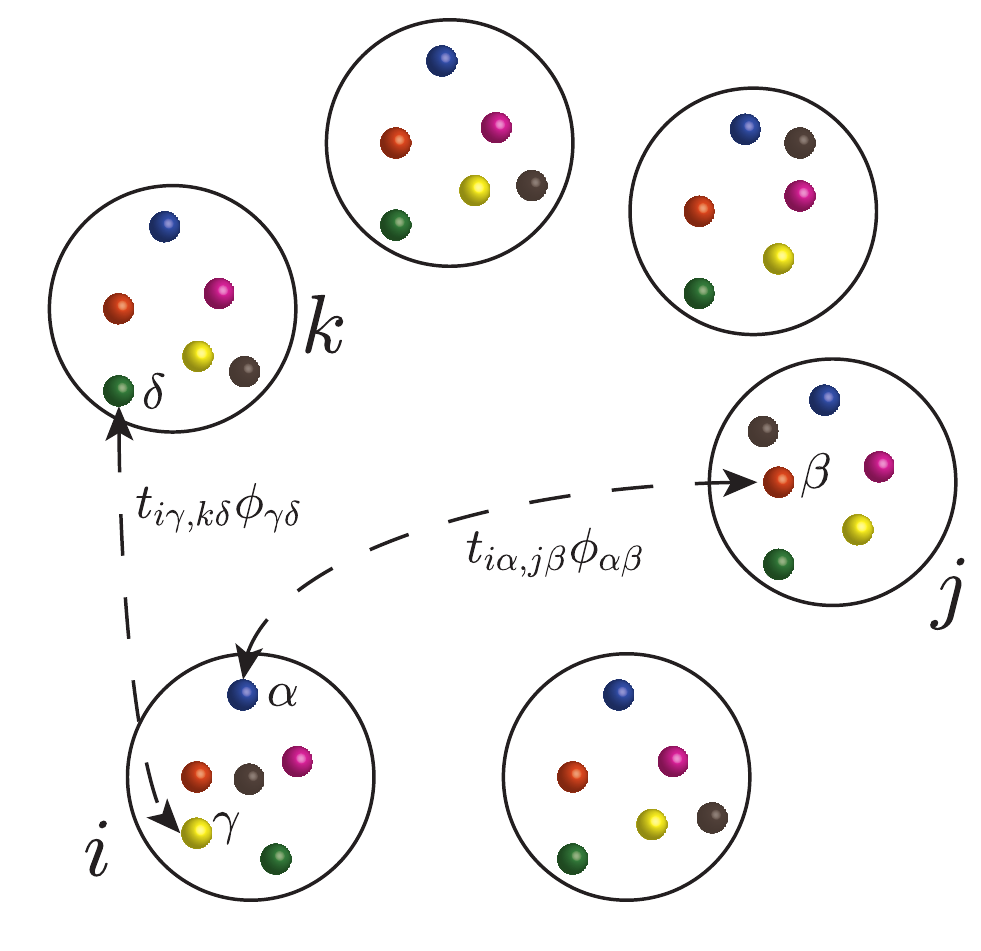}
	\caption{$M$ quantum dots labeled by $\{i,j\}$ have $N$ flavors labeled by $\{\alpha,\beta\,\gamma,\delta\,\cdots\}$ each. Fermions in different dots are coupled to bosons through a  random Yukawa coupling $t_{i\alpha,j\beta }$, where bosons are given by antisymmetric fields $\phi_{\alpha\beta}$. The figure is adapted from Ref.~\cite{panYukawa2021}.}
	\label{fig:fig11}
\end{figure}

\subsection{Spin-1/2 Yukawa-SYK model and its phase diagram}
\label{sec:IIIa}

As shown in the original literature~\cite{sachdevGapless1993,kitaevTalks2015,SachdevBekenstein2015,KitaevThe2018}, a simple form of the SYK model is $ H = -\frac{1}{4!} \sum_{ijkl} J_{ijkl} \psi_i \psi_j \psi_k \psi_l$
 , where $\psi_i$ are Majorana fermion operators which satisfy $\psi_{i}^{\dagger}=\psi_{i}$ and Clifford relation $\left\{\psi_{i}, \psi_{j}\right\}=2 \delta_{i j}$ and $\langle J_{ijkl} \rangle =0$ and $\langle J^2_{ijkl}  \rangle =1$ are the random coupling.
 
%

Compared with the SYK model that only involves interacting fermions, our Yukawa-SYK model, which is also analytically solvable in the large-$N$ limit, has a dynamical bosonic degree of freedom. As shown in Fig.~\ref{fig:fig11}, there are $M$ quantum dots and each hosting $N$ flavors of fermions. Fermions of different dots interacted with $N^2$ bosons via all-to-all random Yukawa interactions. The Hamiltonian~\cite{panYukawa2021,wangPhase2021} is:
\begin{equation}
	\begin{aligned}
		H =& \sum_{i,j=1}^{M}\sum_{\alpha,\beta=1}^{N} \sum_{m,n}^{\uparrow,\downarrow}\left(\frac{i}{\sqrt{MN}} t_{i\alpha,j\beta}\phi_{\alpha\beta}c^\dagger_{i \alpha;m}\sigma^z_{m,n}c_{j\beta ;n} 
		\right)  \\
		&+\sum_{\alpha , \beta =1}^N\left(\frac{1}{2}\pi_{\alpha\beta}^2+\frac{m_0^2}{2}\phi_{\alpha\beta}^2\right) -\mu \sum_{i=1}^{M} \sum_{\alpha=1}^{N} \sum_{m=\uparrow, \downarrow} c_{i \alpha m}^{\dagger} c_{i \alpha m}
		\end{aligned}
		\label{eq:eq10}
	\end{equation}
where $M$ and $N$ are the indices of quantum dots and fermion flavors in Fig.~\ref{fig:fig11}, respectively. As in the SYK model, the random coupling $t_{i\alpha,j\beta}$ satisfy $\langle t_{i\alpha,j\beta}\rangle=0$, $\left\langle t_{i \alpha, j \beta} t_{k \gamma, l \delta}\right\rangle=\left(\delta_{\alpha \gamma} \delta_{i k} \delta_{\beta \delta} \delta_{j l}+\delta_{\alpha \delta} \delta_{i l} \delta_{\beta \gamma} \delta_{j k}\right) \omega_{0}^{3}$ and $\pi_{\alpha \beta}$ is the canonical momentum of bosons $\phi_{\alpha \beta}$. Bosonic bare mass is $m_0$ and chemical potential is $\mu$. We set $\omega_0$ = 1 as the energy
unit. 

We find the high-rank randomness of the Yukawa coupling $t_{i\alpha,j\beta}$ in our model is important for stabilizing the nFL behavior. Also the spinful fermions, with the Pauli matrix $\sigma_z$ in Yukawa coupling term, actually help us to avoid the sign problem in the QMC simulation. Since $t_{i\alpha,j\beta}$ is symmetric matrix and $\phi_{\alpha \beta}$ is antisymmetric, the Hamiltonian is invariant under the antiunitary transformation $\mathcal{T}=i \sigma_{y} \mathcal{K}$, where $\mathcal{K}$ is the complex-conjugate operator. Or to put it simply, the matrix elements of different spins are all Hermitian to each other, so no sign problem in the evaluation of the fermion determinants~\cite{panYukawa2021,panSign2022}.

The major difference between the Yukawa-SYK model and those in Sec.~\ref{sec:II} is the random all-to-all coupling. Within the large-$N$ approximation Yukawa-SYK model is “self-tuned” to quantum criticality. The system becomes critical, independent of the bosonic bare mass $m_0$. This means we don't need to adjust the parameters close to QCP to study nFLs. Despite this benefit, there are also costs: all-to-all coupling makes the computational complexity of QMC increase  from $O(\beta N^3)$ to $O(\beta N^5 M^3)$. Since it is not onsite boson-fermion coupling, the Sherman-Morrison formula for efficiently computing the matrix inversion~\cite{blankenbeclerMonte1981,scalapinoMonte1981,hirschDiscrete1983,hirschTwo1985} does not apply in this case, which makes the computational complexity of weight calculation and updating Green's function change from $O(N^2)$ to $O(N^3 M^3)$. This is because the Yuakwa-SYK model has $N(N-1)/2$ independent bosonic fields at each time slice, instead of $N$, while the number of time slices is proportional to $\beta$. In addition, because random terms $t_{i\alpha,j\beta}$ exist, we need to perform multiple Monte Carlo processes to achieve the disorder-average, on top of the regular Markov chain for each set of $\{t_{i\alpha,j\beta}\}$.

\begin{figure}[!htp]
\centering
\includegraphics[width=0.9\columnwidth]{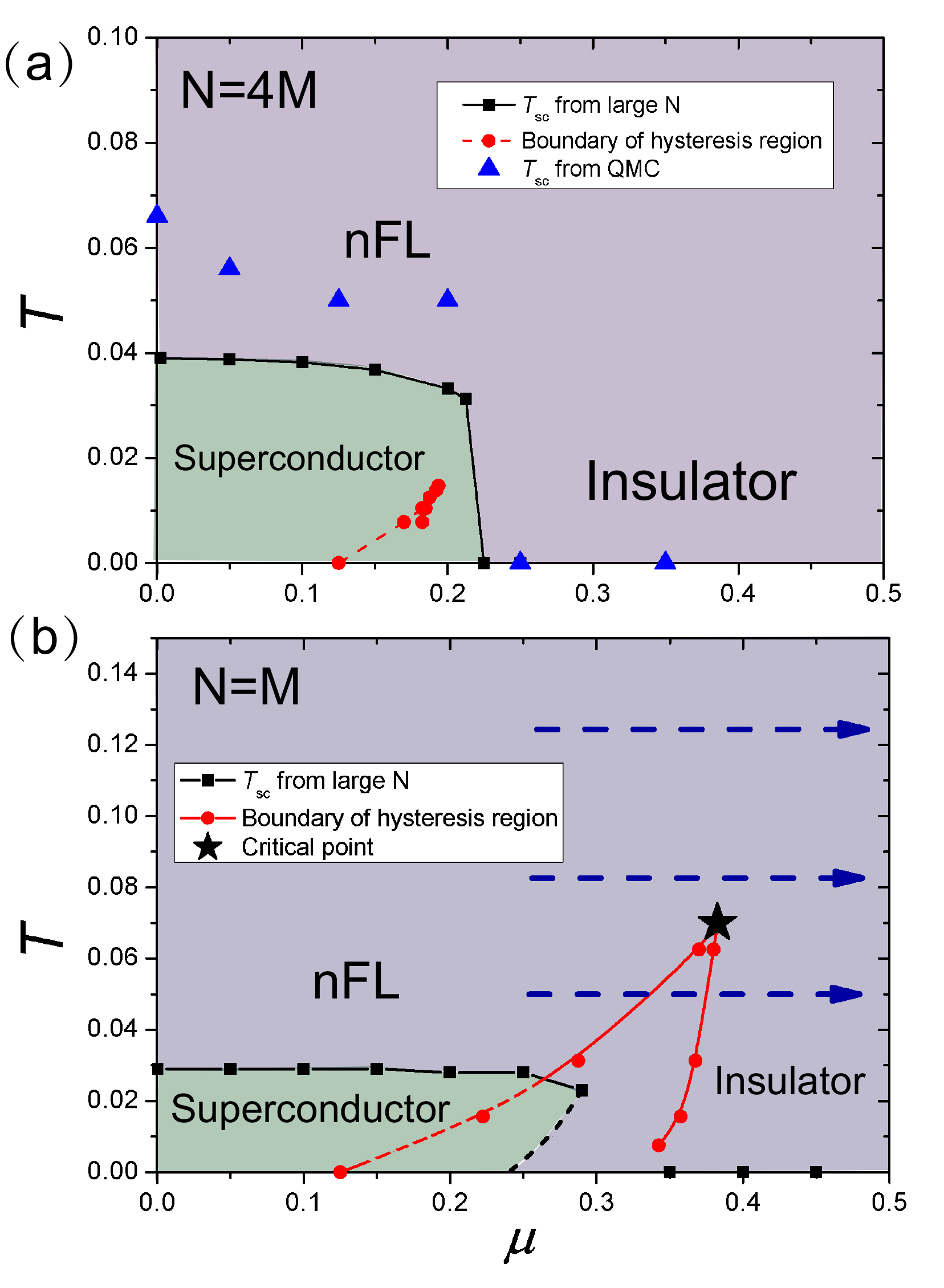}
\caption{(a) Global phase diagram of the spin-1/2 Yukawa-SYK model at $N=4M,\omega_0=1,m_0=2$. From the large-$N$ calculation, over a wide range of chemical potentials $\mu$, we can see a finite-temperature transition from nFL to SC, until the chemical potential increases to the line where a first order superconductor-to-insulator phase transition occurs. The first-order hysteresis region is denoted by the red points and lines. The thermal critical point that terminates the first order transition locates at $(\mu_c=0.194,T_c=0.015)$. And the blue triangles are the transition points from nFL to superconductor obtained from QMC at finite $N=4M$, which are consistent with the large-$N$ calculations (black squares). (b) Global phase diagram of the spin-1/2 Yukawa-SYK model at $N=M,\omega_0=1,m_0=2$ from the large-$N$ calculation. Similarly, large-$N$ calculations will give the first-order hysteresis region, while the dashed-line portion of this boundary means it is renormalized by superconductivity phase. The QMC $n-\mu$ parameter scans in Fig.~\ref{fig:fig13} are along the blue dashed paths. The thermal critical point at $(\mu_c=0.3825,T_c=0.07)$ is denoted by the black star. The figure is adapted from Ref.~\cite{wangPhase2021}.}
\label{fig:fig12}
\end{figure}

  Through numerical simulation of QMC and large-$N$ calculation, we obtained the global phase diagram of Yukawa-SYK model, as shown in the Fig.~\ref{fig:fig12}. By solving the linear Eliashberg equation using the large-N result of the Green’s functions and considering the pairing susceptibility datas in QMC simulations, a finite temperature phase transition from nFL to superconductivity can be seen in some intervals of the chemical potential $\mu$. And we can see that by solving the Schwinger-Dyson equation: the first-order quantum phase transition extends to low
  temperature and terminates at a (thermal) critical point, which
  is common in lots of metal-insulator transitions in correlated materials~\cite{imadaMetal1998,limeletteUniversality2003}. And the reason for choosing two different sets of parameters is that the superconducting dome completely preempts the would-be nFL/insulator transition for $N=4 M, \omega_{0}=1, m_{0}=2$ (Fig.~\ref{fig:fig12} (a)), while for $N=M, \omega_{0}=1, m_{0}=2$, the first-order phase transition is stronger and the corresponding thermal critical point occurs outside of the superconducting phase (Fig.~\ref{fig:fig12} (b)).
  	
\subsection{Normal-state results at $N, M\rightarrow \infty$}
One can slove the Yukawa-SYK model at the $N, M\to \infty$ limit analytically~\cite{wangSolvable2019,esterlisCooper2019, schmalianEliashberg2019,wangQuantum2020}, by following the original treatment of the SYK model~\cite{sachdevGapless1993,kitaevTalks2015,SachdevBekenstein2015,KitaevThe2018}. In this limit we have the Schwinger-Dyson(SD) equations:
	\begin{align}
		\Pi (i\Omega_m) =& \frac{4M}{N}\omega_0^3T\sum_{n}
		G_f(i\omega_n-i\Omega_m/2) G_f(i\omega_n+i\Omega_m/2)\nonumber\\
		\Sigma(i\omega_n) =& -\omega_0^3
		T\sum_{m}
		G_b(i\Omega_m) G_f(i\omega_n-i\Omega_m),
		\label{eq:eq11}
	\end{align}
where $\Sigma$ and $\Pi$ are fermionic and bosonic self-energies. The fermionic Green's functions is $G_f(i\omega_n) =\left[i\omega_n+\mu+\Sigma(i\omega_n)\right]^{-1}$ while $G_b(i\Omega_m) =  \left[\Omega_m^2+ \Pi(i\Omega_m)+m_0^2\right ]^{-1}$ are bosonic Green's functions. Since we are considering large $N,M$ limit, only the ratio $M/N$ matters in Eq.~\eqref{eq:eq11}.

\begin{figure}[!htp]
	\centering
	\includegraphics[width=\columnwidth]{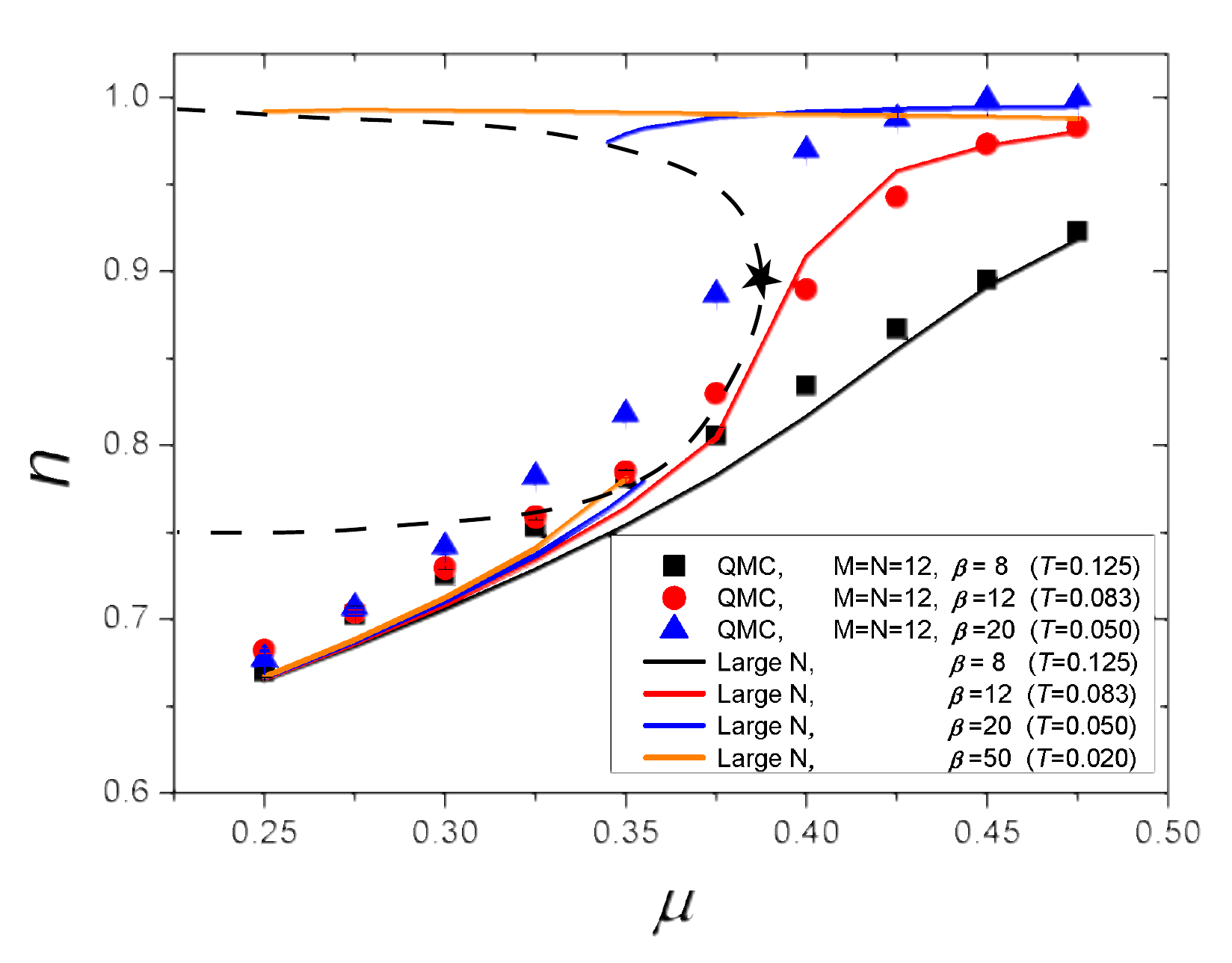}
	\caption{Filling $n$ versus $\mu$ for certain $T$ in the vicinity of the thermal critical point in Fig.~\ref{fig:fig12} (b) for $N=M, \omega_0=1, m_0=2$. Both QMC and large-$N$ results are shown in the figure. At higher temperature $T$, one sees the $n(\mu)$ curves both from large-$N$ and QMC are smooth, while $T=0.125, \mu=0.375$ (marked by the star), is a thermal critical point. At temperature $T=0.083$, which is close to the thermal critical point,  a sharp turn of $n(\mu)$ signifying the divergence of the compressibility $dn/d\mu$ can be seen. At lower $T$, there is a range of $\mu$---the hysteresis region---where the filling is double-valued. The upper branch represents the insulating behavior and the lower branch represents the nFL behavior. A first order transition connects the two branch at a chemical potential given by a Maxwell construction. The dashed line delimits the region where solutions are unstable. The figure is adapted from Ref.~\cite{wangPhase2021}.}
	\label{fig:fig13}
\end{figure}

After iterative solution of Eq.~\eqref{eq:eq11} (introducing the “stabilizers”
for each step of the iteration~\cite{wangPhase2021}), we obtain the hysteresis region which separates nFL from insulator (red lines in Fig.~\ref{fig:fig12}). And we can get filling $n$ since we have imaginary-time fermionic Green’s function from QMC and solution of Eq.~\eqref{eq:eq11}:
\begin{equation}
n=\frac{1}{2}+\sum_{\omega_{n}} G_{f}\left(i \omega_{n}\right).
\end{equation}
Fig.~\ref{fig:fig13} shows the exact $(n,\mu)$ curve in different parameters. The coexistence of two solutions for certain $(\mu, T)$ indicates a metastable state. Similar to the water-vapor transition, the $n-\mu$ curve connecting the two branches is a straight line determined by the Maxwell construction. The two types of solutions become smoothly connected at a chemical potential $\mu_c$ above a certain temperature $T_c$. Here $(T_c,\mu_c)$ is a thermal critical point of the system, for that the compressibility $dn/d\mu$ diverges.

The parameter transformation range in Fig.~\ref{fig:fig13} corresponds to the blue dashed line in the phase diagram, Fig.~\ref{fig:fig12} (b). The filling $n$ is doubly valued in the hysteresis zone(a range of $\mu$) at lower $T$. The nFL behavior is represented by the lower branch, while the insulating behavior is represented by the upper branch; a first-order transition connects the two at a chemical potential $\mu_c=0.3825$, which is denoted by the black star. The dashed line marks the boundary between stable and unstable solutions.
	 
\subsection{nFL Green's functions}
\label{sec:IIIc}
In Refs.~\cite{wangSolvable2019,esterlisCooper2019}, some large-$N$ results have been found that when $T=0,\mu=0$, for $m_{0} \sim \omega_{0}$ and $\omega, \Omega \ll \omega_{0}$, the fermionic and bosonic self-energies are given by:
\begin{equation}
\begin{aligned}
&\Sigma(\omega)=-G_{f}(\omega)^{-1}=i c \operatorname{sgn}(\omega)|\omega|^{x} \omega_{0}^{1-x} \\
&\Pi(\Omega)=G_{b}(\Omega)^{-1}=-m_{0}^{2}+c^{-2} \alpha(x)|\Omega|^{1-2 x} \omega_{0}^{1+2 x}
\end{aligned}\label{eq:eq13}
\end{equation}
where $c$ is a nonuniversal constant, the function $\alpha(x)$ is:
\begin{equation}
\alpha(x)=-\frac{\Gamma^{2}(-x)}{4 \pi \Gamma(-2 x)}
\end{equation}
and $0<x<1 / 2$ is determined by
\begin{equation}
\frac{4 M}{N}=\frac{1 / x-2}{1+\sec (\pi x)}
\label{eq:eq15}
\end{equation}

In the time domain, after a Fourier transform we have:
\begin{equation}
\begin{aligned}
\Pi(\tau, \tilde{\tau}) & \propto|\tau-\tilde{\tau}|^{-(2-2 x)} \\
G_{b}(\tau, \tilde{\tau}) & \propto|\tau-\tilde{\tau}|^{-2 x} \\
\Sigma(\tau, \tilde{\tau}) & \propto|\tau-\tilde{\tau}|^{-(1+x)} \operatorname{sgn}(\tau-\tilde{\tau}) \\
G_{f}(\tau, \tilde{\tau}) & \propto|\tau-\tilde{\tau}|^{x-1} \operatorname{sgn}(\tau-\tilde{\tau})
\label{eq:eq16}
\end{aligned}
\end{equation}
Here it's worth noting that the Fourier transform for $\Pi(\Omega)$ shall be carried out carefully, this is because the positive power-law $|\Omega|^{1-2 x}$ does not have a Fourier transform in the common sense as the Fourier integral is UV divergent for $0<$ $x<1 / 2$. This divergence is canceled by the Fourier transform of the $m_{0}^{2}$ term in $\Pi(\Omega)$ with the UV information.

In our Yukawa-SYK model, at a finite but low temperature $T=1 / \beta \ll$ $\omega_{F}= \omega_{0}^{3} / m_{0}^{2}$ , the long-time correlated behavior persists~\cite{sachdevGapless1993,kitaevTalks2015,SachdevBekenstein2015,KitaevThe2018,wangSolvable2019,esterlisCooper2019}. By applying a reparametrization symmetry transformation~\cite{sachdevGapless1993,kitaevTalks2015,SachdevBekenstein2015,KitaevThe2018} $\tau \rightarrow f(\tau)=\tan (\pi \tau / \beta)$, we know that at low temperatures and long-time limit:
\begin{align}
	G_f(\tau, 0) \propto & \left(\frac{\pi}{\beta\sin(\pi\tau/\beta)}\right)^{1-x}\nonumber\\
	G_b(\tau,0) \propto& \left(\frac{\pi}{\beta\sin(\pi\tau/\beta)}\right)^{2x}.		\label{eq:eq17}
\end{align}
where  $G_{f}(\tau, 0)=$ $\frac{1}{(M N)^{2}} \sum_{i, j=1}^{M} \sum_{\alpha, \beta=1}^{N}\left\langle c_{i, \alpha, \sigma}(\tau) c_{j, \beta, \sigma}^{\dagger}(0)\right\rangle $ and $ G_{b}(\tau, 0)=$ $\frac{1}{N(N-1)} \sum_{\alpha, \beta=1, \alpha \neq \beta}^{N}\left\langle\phi_{\alpha \beta}(\tau) \phi_{\alpha \beta}(0)\right\rangle$. The Eq.~\eqref{eq:eq15} gave us the value of $x$, which comes form self-consistent Schwinger-Dyson equations to leading order in $O(1/MN)$.

 It was already well known that in SYK model, the fermion Green's function is incoherent~\cite{sachdevGapless1993}: $G(\tau) \sim \tau^{-1 / 2}$ at large $\tau$, while in FL the quasiparticles have the coherent Green's function: $G(\tau) \sim 1 / \tau$. And when $ T>0$, the SYK Green's function has conformal invariance as follow: $G \sim\left(T / \sin \left(\pi k_{B} T \tau / \hbar\right)\right)^{1 / 2}$~\cite{parcollet1999non}.

 Back to our spin-1/2 Yukawa-SYK model, the exponent $x$ is between 0 (same as in a noninteracting disordered electron system) and $1/2$ (same as in the SYK model). The power-law behavior of bosonic and fermionic Green's functions both in the Matsubara frequency and imaginary time, can be used to identify the nFL behavior.
 
  The question for the QMC simulation of the Yukawa-SYK model is to explore whether at small and finite $M$ and $N$, where the perturbative analysis no longer works, the system could still exhibit the nFL behavior as in Eqs.~\eqref{eq:eq13} and ~\eqref{eq:eq17}, and whether there will be emergent symmetry breaking phases such as superconductivity generated by the quantum fluctuations in the nFL. These questions, as we shall see below, have been answered affirmatively by our model design and computation. 

\begin{figure}[!htp]
	\centering
	\includegraphics[width=0.5\textwidth]{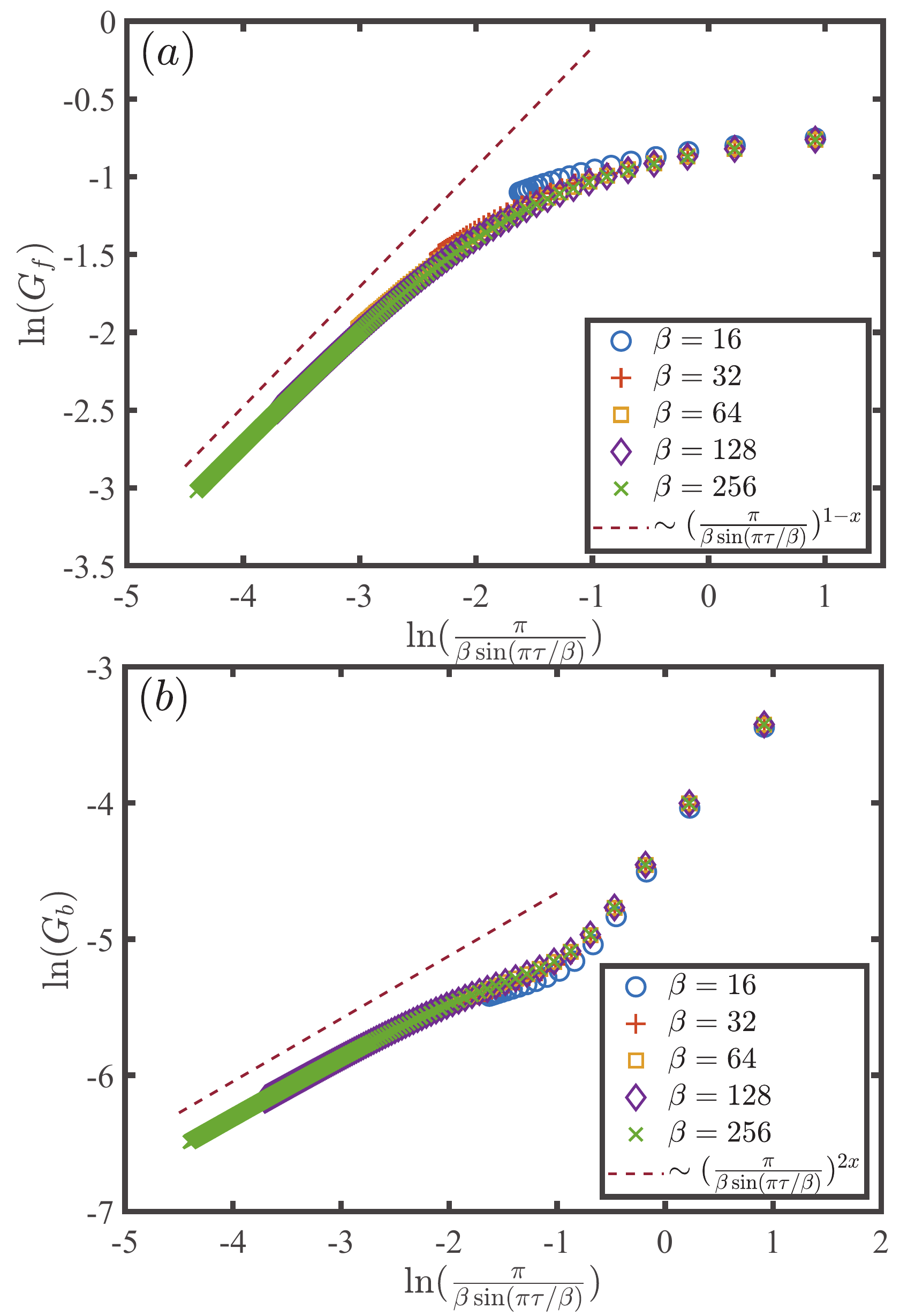}
	\caption{Theoretical result of $G_f$ and $G_b$ at $N=4M \rightarrow \infty,\;\omega_0 = 1, m_0 = 2,\mu=0$. When $\beta$ is very large, the solution of the SD equation agrees very well with the analytic form in Eqs.~\eqref{eq:eq17}. The figure is adapted from Ref.~\cite{panYukawa2021}. }
	\label{fig:fig14}
\end{figure}

According to Eqs.~\eqref{eq:eq17}, we can look for the nFL fermionic and bosonic Green's functions directly in the imaginary time decay. In Fig.~\ref{fig:fig14}, we present the log-log plot of the behavior of $G_{f}$ and $G_{b}$ at $N=4 M$, $\omega_{0}=1, m_{0}=2,\mu=0$, and different temperatures from iterative theoretical calculation. Here for $4 M=N$, one finds $x \approx 0.231$. The solution of Eq.~\eqref{eq:eq17} matches well in the long-time limit with the approximate result obtained using time-reparametrization symmetry, exhibiting self-tuned criticality and nFL behaviors. The system behaves as a quantum critical NFL, with a pairing instability at $T_c < \omega_F$.

\begin{figure}[! htp]
	\centering
	\includegraphics[width=0.5\textwidth]{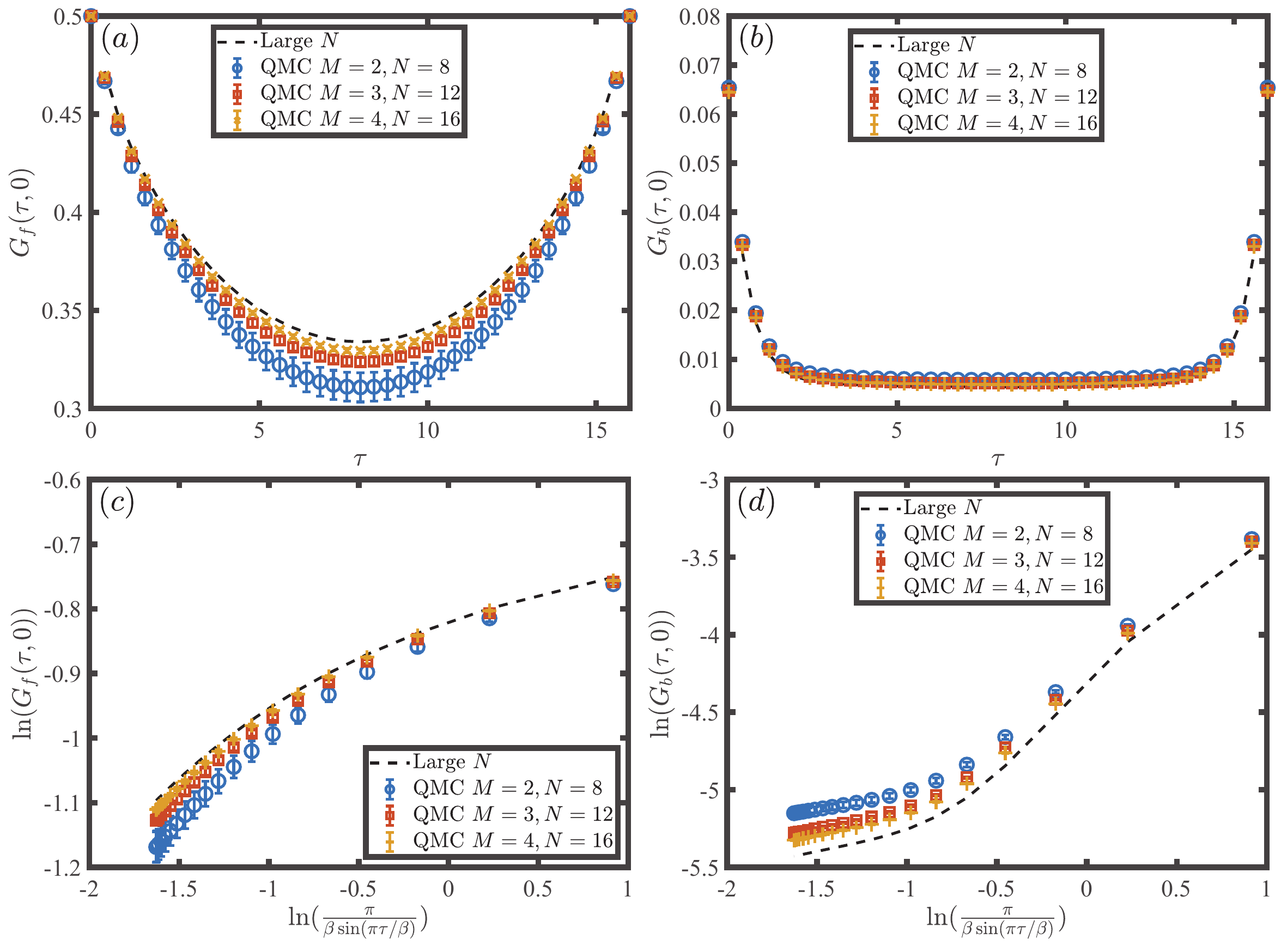}
	\caption{QMC results at $N=4M$, $m_0=2,\omega_0=1,\mu=0$ and $\beta=16$ for $M=2,3,4$. (a)(b)  Green's function $G_{f}(\tau,0)$ and $G_{b}(\tau,0)$ versus $\tau$ in the range of $\tau\in[0,\beta]$. Blue, red and yellow dots are DQMC data and the black dashed line is the large-$N$ result. (c) and (d) The same as above, but in a special log-log scale 
		The convergence towards the large-$N$ results as $(M,N)$ increase is obvious. The figure is adapted from Ref.~\cite{panYukawa2021}. } 
	\label{fig:fig15}
\end{figure}

The QMC results with the same parameters are shown in Fig.~\ref{fig:fig15} , which also show the nFL behaviors that match with Eqs.~\eqref{eq:eq17}. In Fig.~\ref{fig:fig15}, with the increase of system size $L$, the numerical results of QMC approach the analytic solution of large $N$, plotted in linear and log-log scale $\ln(G_{b,f})$ versus $\ln(\pi /[\beta \sin (\pi \tau / \beta)])$. All of the QMC data are obtained by averaging over 20 disorder realizations in $\left\{ t_{i\alpha,j\beta}\right\}$, and the errorbar is the variance of different realizations. Here the system is not in a spin glass but self-average, which is different from the low-rank model (See Eq.~\ref{eq:eq22}) discussed later. Self-average means that one does not need to perform many disorder realizations.

\begin{figure}[!htp]
	\centering
	\includegraphics[width=\columnwidth]{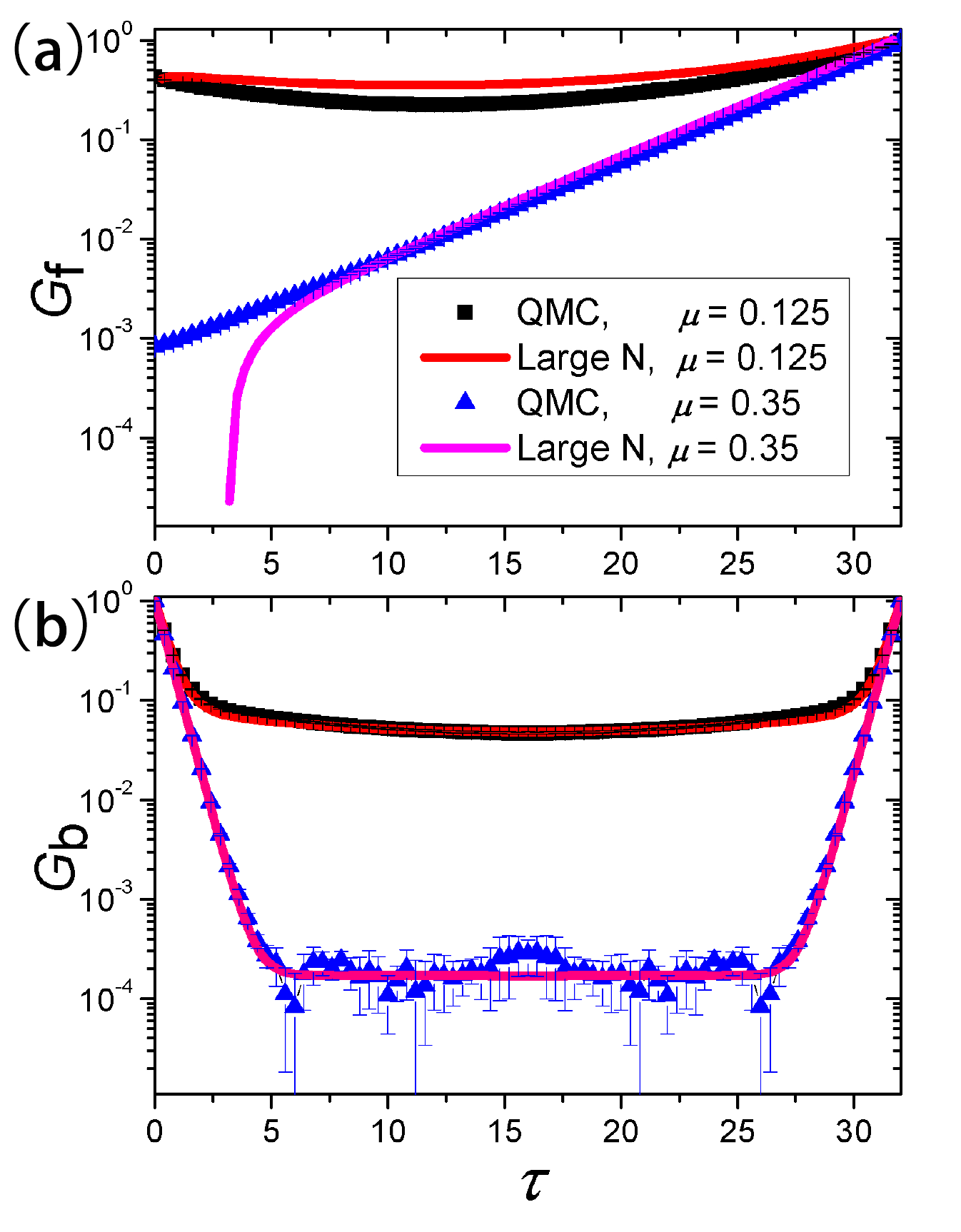}
	\caption{The QMC fermionic Green's functions (a) and the bosonic Green's functions (b) with different $\mu$. $M=4$, $N=16$, $\beta=32$, $\omega_0=1$, $m_0=2$. For convenience both the $G_f$ and $G_b$ have been normalized to 1 at $\tau=\beta$. The system become gapped with the increase of the chemical potential. Noted that the sharp downturn of the large-$N$ result in panel (a) is an artifact of keeping finite frequency points. The figure is adapted from Ref.~\cite{wangPhase2021}.}
	\label{fig:fig16}
\end{figure}

In addition, when we consider $\mu \neq 0$ case, the agreement is also excellent. It is shown in Fig.~\ref{fig:fig16}.  Since there is no longer particle-hole symmetry, fermionic Green's function $G_{f}$ is not symmetric with respect to $\tau=\beta / 2$, and we normalize the data with respect to $G_{f}(\tau=\beta)$ here. When $\mu= 0.125$, power-law decay in imaginary time of $G_{f}$ and $G_{b}$ is  seen, which is similar to that in Ref.~\cite{panYukawa2021}. At larger doping $\mu=0.35$, both $G_{f}$ and $G_{b}$ decay exponentially, consistent with insulating behavior discussed before. And considering finite $N, M$ the system does not develop superconductivity at $\mu=0.35$, since it is in the gapped insulator phase. So it is sensible to compare it with Green's functions at large $N$ for the normal state. 

\subsection{Self-tuned quantum criticality}
\label{sec:IIId}
Self-tuned quantum criticality means the
system becomes critical due to the strong mutual feedback between the bosonic and fermionic sectors, independent of the bosonic bare mass $m_0$. From Eq.~\eqref{eq:eq13} we have:
\begin{equation}
m_{0}^{2}-\Pi(\Omega=0)=0
\end{equation}
which means the boson is critical/gapless no matter what the bosonic bare mass $m_0$ is, the system renormalizes it to zero via interaction effects. This has been proved to be true in Refs.~\cite{wangSolvable2019,esterlisCooper2019} at the large-$N$ limit.

\begin{figure*}
\centering
\includegraphics[width=\linewidth]{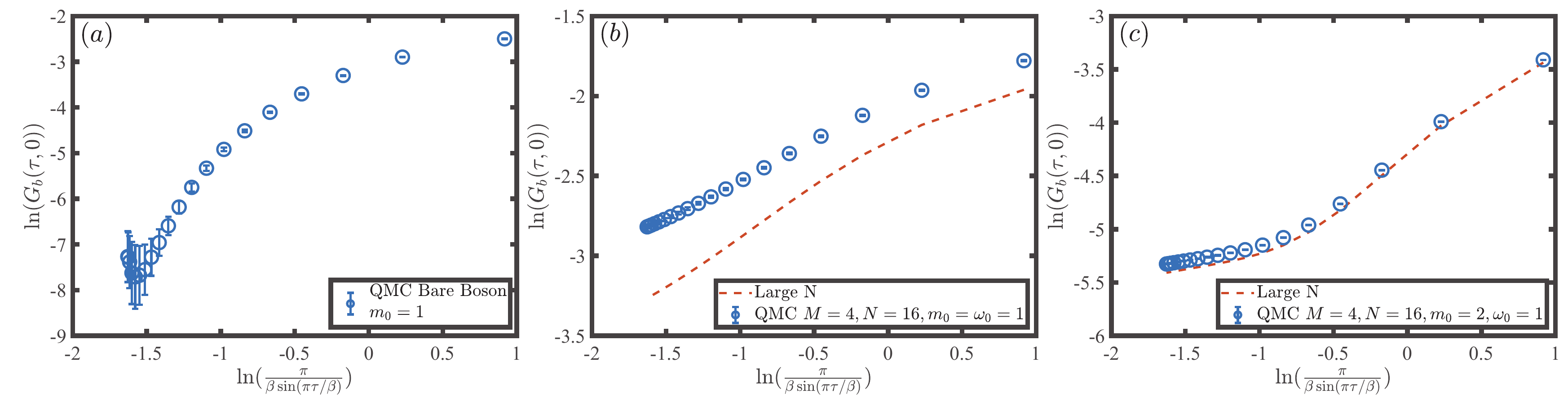}
\caption{Self-tuned quantum criticality with different boson masses. Blue dots are DQMC data and the red dashed lines are large-$N$ result (following Eq.~\eqref{eq:eq17}). It is clear to see power law decay of $G_b$ at low-temperatures and long-time limit in log-log plot. These results reveal the self-tuned quantum criticality in our system. The figure is adapted from Ref.~\cite{panYukawa2021}.}
\label{fig:fig17}
\end{figure*}

Fig.~\ref{fig:fig17} (a) shows the bare bosonic Green's function generated from the bosonic part in Eq.~\eqref{eq:eq10}, with the mass $m_{0}=1$ and $\beta=16$. Then the Green's function clearly exhibits exponential decay in imaginary time to $G_{b}(\tau=\beta / 2,0) \approx 0$. However,  as shown in Fig.~\ref{fig:fig17} (b) and (c), once coupled with fermions in our model, the bosonic Green's functions become critical. These results reveal the self-tuned quantum criticality in our system. The QMC data in Fig.~\ref{fig:fig17} (c) are very close to the theoretical result. The self-tuned quantum criticality is consistent with analytical predictions at large $N$.

\subsection{Superconductivity emerging from nFL}
\label{sec:IIIe}
We can see that in the phase diagram shown in Fig.~\ref{fig:fig12}(a),  $N=4M,\omega_0=1,m_0=2$, in low temperatures the systems are in superconducting phases in a range of chemical potential $\mu$. And when $\mu$ increases, the $T_{sc}$
is moderately reduced until a sudden drop at larger $\mu$. The QMC results and large-N calculation gave us the critical $\mu_c = 0.25$ at zero temperature, which coincides with each other. Details will be shown below.

The leading pairing instability mediated by the critical boson mode is toward a spin-singlet, intra-dot, and intra-flavor channel~\cite{panYukawa2021}: $\sum_{i,\alpha}\langle c^\dagger_{i\alpha\uparrow}c^\dagger_{i\alpha\downarrow}\rangle$ and the system has an instability and a low-temperature pairing
phase. The Eliashberg equation is given by:
\begin{equation}
\begin{aligned}
\Phi\left(\omega_{n}\right)=& \omega_{0}^{3} T \sum_{\Omega_{m}} G_{b}\left(i \Omega_{m}\right) G_{f}\left[i\left(\omega_{n}+\Omega_{m}\right)\right]  \\
& \times G_{f}\left[-i\left(\omega_{n}+\Omega_{m}\right)\right]  \Phi\left(\omega_{n}+\Omega_{m}\right)
\end{aligned}
\label{eq:eq19}
\end{equation}

For $\mu=0$, the pairing problem was analyzed in Ref.~\cite{panYukawa2021}, since there is particle-hole symmetry. Fermionic Green's function $G_f(\tau)$ is symmetric with respect to $\beta/2$.  Now the Eliashberg equation is given by: $ \Phi\left(\omega_{n}\right)= \omega_{0}^{3} T \sum_{\Omega_{m}} G_{b}\left(i \Omega_{m}\right) |G_{f}\left[i\left(\omega_{n}+\Omega_{m}\right)\right] |^2 \, \Phi\left(\omega_{n}+\Omega_{m}\right)$

However, for $\mu \neq 0$, now the mismatch between $G\left(\pm i \omega_{n}\right)$ leads to a reduced pairing tendency. Now we can consider the weak-coupling limit $\omega_{0} \ll m_{0}$, and determine the value of $\mu_{s c}$ beyond which pairing vanishes. At zero temperature, the Schwinger-Dyson equations gave an insulating solution approximated \cite{wangQuantum2020} by $\Sigma(i \omega)=-\omega_{F} / 2$, where $\omega_{F} \equiv \omega_{0}^{3} / m_{0}^{2}$ and $\Pi(i \Omega)=0$ as long as $\mu>\omega_{F} / 2$. In this regime the pairing equation becomes:
\begin{equation}
\Phi(\omega)=\omega_{0}^{3} \int \frac{d \Omega}{2 \pi} \frac{1}{(\Omega-\omega)^{2}+m_{0}^{2}} \frac{1}{\Omega^{2}+\left(\mu-\omega_{F} / 2\right)^{2}} \Phi(\Omega)
\end{equation} 
and at very weak coupling, with the ansatz $\Phi(\omega)=$ const~\cite{panYukawa2021}, there is a pairing transition at $\mu_{s c}=\omega_{F}\left(\equiv \omega_{0}^{3} / m_{0}^{2}\right)$.

If we view the Eliashberg equation Eq.~\ref{eq:eq19} as a matrix equation/an eigenvalue problem, we can solve for the critical temperatures $T_c$. As the temperature lowers, the eigenvalues of the kernel increase, and $T_c$ corresponds to the temperature at which the largest eigenvalue approaches 1.
 
\begin{figure*}[!htp]
\includegraphics[width=\textwidth]{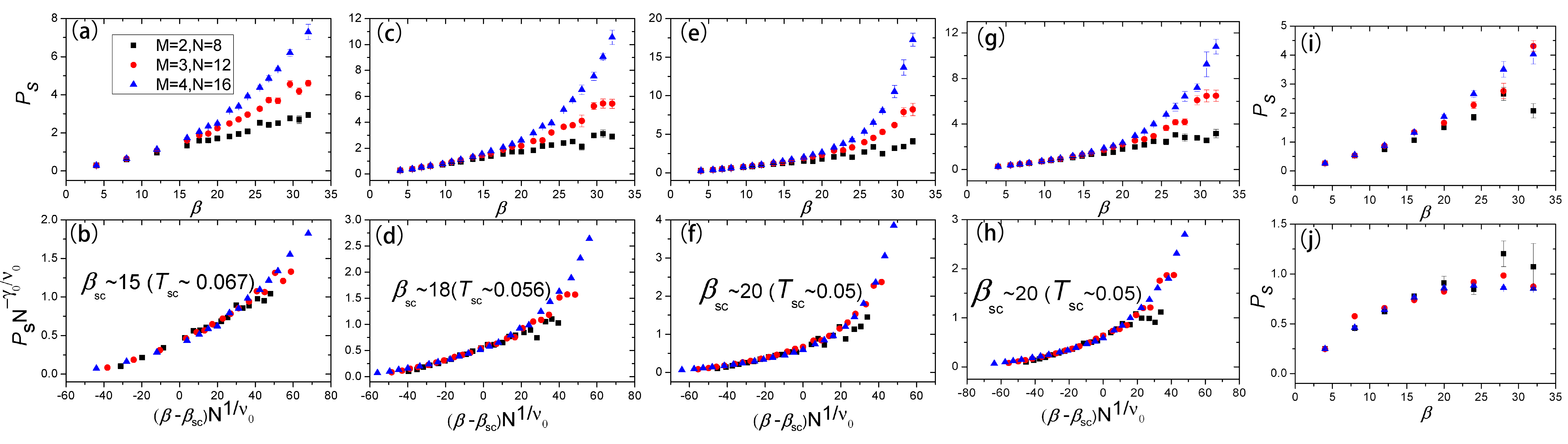}
\caption{Pair susceptibility $P_s$ measured at different chemical potential $\mu$. The obtained $T_c$ ($\beta_c$) are denoted by the blue triangles in Fig.~\ref{fig:fig12} (a). From the temperature dependence of the $P_s$ with different system sizes $(M,N)$ the authors in Ref~\cite{wangPhase2021} perform the data collapse using mean-field exponents $\gamma_0 = 1$, $\nu_0 = 2$ and the transition temperatures $T_c$ ($\beta_c$) are obtained accordingly. The parameters are: $N=4M$, $\omega_0=1$, $m_0=2$. $\mu=0$ in (a) and (b); $\mu=0.05$ in (c) and (d); $\mu=0.125$ in (e) and (f); $\mu=0.2$ in (g) and (h); $\mu=0.25$ in (i); $\mu=0.35$ in (j). The superconducting transition temperature reduces as $\mu$ increases. For $\mu=0.25$ (i) and $\mu=0.35$ (j) the pairing susceptibilities are not divergent at larger $N$, and the system enter a gapped insulator phase. The figure is adapted from Ref.~\cite{wangPhase2021}.} 
	\label{fig:fig18}
\end{figure*}

In QMC simulation, we can study superconducting phase transitions by measuring pairing correlation. The pair susceptibility is defined as $P_{s}=\int_{0}^{\beta} d \tau\left\langle\Delta(\tau) \Delta^{\dagger}(0)\right\rangle$, where $\Delta$ is the pairing field:
\begin{equation}
\Delta^{\dagger}=\frac{1}{\sqrt{M N}} \sum_{i=1}^{M} \sum_{\alpha=1}^{N} c_{i, \alpha, \uparrow}^{\dagger} c_{i, \alpha, \downarrow}^{\dagger}.
\end{equation}
At finite $N$, the pairing susceptibility $P_{s}$ does not diverge and can be represented as $P_{s}^{(N)} \sim N^{a} f\left[N^{1 / v}\left(T-T_{s c}\right)\right]$. Here $N$ (and $M$ for a fixed ratio) plays the role of the system size\cite{ChenFermi2021,isakovInterplay2003,paivaCritical2004,costaPhonon2018} . One thing to be careful about here is, there is not notion of space for the Yukawa-SYK system, the role of correlation length is replaced by a correlation "cluster size" $n \sim$ $\left(T-T_{s c}\right)^{-v}$ and hence the functional dependence of $f(x)$. In the large-$N$ limit, all fluctuation effects are suppressed by $1 / N$, and such a phase transition is mean-field-like~\cite{wangSolvable2019,esterlisCooper2019, schmalianEliashberg2019,wangQuantum2020,inkofQuantum2022}. This means that for a fixed $T-T_{c}$, the exponent $v=2$ follows the analog of Josephson's identity \cite{wangPhase2021} and that $f(x) \sim 1 / x$. Further requiring that in the large-$N$ (thermodynamic) limit the susceptibility diverges independent of $N$, we obtain $a=$ 1/2. The data collapse shown in Fig.~\ref{fig:fig18} gave us the critical temperature $\beta_c$.
 
One can see in Fig.~\ref{fig:fig18} (i) and (j), the pairing susceptibility $P_{s}$ no longer diverges with large $N$ when $\mu>0.25$, which is consistent with the solution of Eliashberg equation Eq.~\eqref{eq:eq19}, $\mu_{s c}=\omega_{0}^{3} / m_{0}^{2}=0.25$. And when $\mu<0.25$, data collapse gave us the $T_{sc}(\beta_{sc})$, which are ploted in Fig.~\ref{fig:fig18}(a-h) and as the phase boundary in Fig.~\ref{fig:fig12}(a).

\subsection{Planckian metals, spin glass and open questions}

 In the original SYK model, people use the replica trick and then take the replica diagonal ansatz to formally obtain analytical results of the nFL behavior. The Replica trick is $\overline{\ln Z}=\lim _{M \rightarrow 0} \frac{\ln \overline{Z^{M}}}{M}$ and replica diagonal means replacing the quenched disorder with annealed disorder. The justification for it is that replica nondiagonal processes are usually suppressed by $1/N$ \cite{guLocal2017}. However, since it is not clear whether the summation of the subdominant processes, each small in $1/N$, is convergent, the validity of this ansatz is not obvious. 
 
On the other hand, we have the following understanding: if the system breaks replica symmetry, the true ground state is then a spin glass. For example, in some bosonic SYK models~\cite{georgesQuantum2001,fuNumerical2016}, there is replica symmetry breaking, and in fact it has been shown recently that similar situations occur for all random interacting
 bosonic models~\cite{baldwinQuenched2020,tulipmanStrongly2020,tulipmanStrongly2021}. While for the fermionic SYK model, it seems that a glass phase is absent and the nFL state persists down to $T = 0$ ~\cite{georgesQuantum2001,fuNumerical2016}. Since our Yukawa-SYK model involves
 both fermions and bosons, whether there is spin glass is a question.

To properly address this question, we can construct a model in a similar form of Eq.~\eqref{eq:eq10} but the random coupling is now of a lower rank,
\begin{equation}
\begin{aligned}
H =& \sum_{i,j=1}^{M}\sum_{\alpha,\beta=1}^{N} \sum_{m,n}^{\uparrow,\downarrow}\left(\frac{i}{\sqrt{MN}} t_{i,j}\phi_{\alpha\beta}c^\dagger_{i \alpha;m}\sigma^z_{m,n}c_{j\beta ;n} 
\right) \\
&+\sum_{\alpha , \beta =1}^N\left(\frac{1}{2}\pi_{\alpha\beta}^2+\frac{m_0^2}{2}\phi_{\alpha\beta}^2\right),
\end{aligned}
\label{eq:eq22}
\end{equation}
we note now $\left\langle t_{i j}\right\rangle=0,\left\langle t_{i j} t_{k l}\right\rangle=\left(\delta_{i k} \delta_{j l}+\delta_{i l} \delta_{j k}\right) \omega_{0}^{3}$.

\begin{figure}
	\centering
	\includegraphics[width=\linewidth]{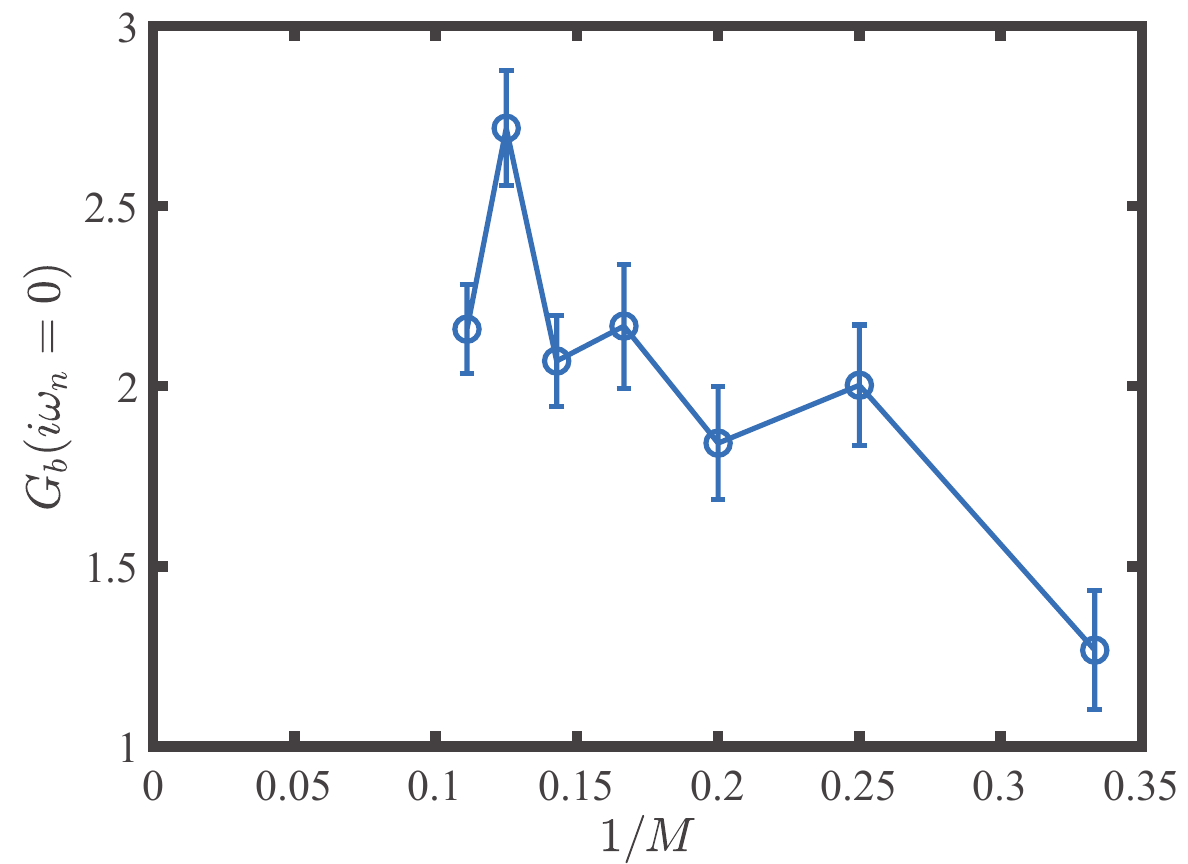}		
	\caption{We show the static component (with $\omega_n=0$) for bosonic Green's function  $G_b(\omega_n)$. This component can be viewed as an Edwards-Anderson order parameter of the spin glass phase in the large-$N$ limit, t. As $N$ increases, the static component, as well as its variance for different disorder realizations, increases with the increase of $N$ at $M=N, \beta=16, m_0=\omega_0=1$, indicating a spin glass behavior. The figure is adapted from Ref.~\cite{panYukawa2021}.}
	\label{fig:fig19}
\end{figure}

In Fig.~\ref{fig:fig19}, we show the static component (with $\Omega_m=0$ ) for bosonic Green's function $G_{b}\left(\Omega_{m}\right)$, obtained from the QMC simulation of Eq.~\eqref{eq:eq22}. The component can be regarded as an Edwards-Anderson order parameter of the spin-glass phase~\cite{fuNumerical2016} in the large-$N$ limit. As $N$ increases, the static component, along with its variance for different disorder realizations, increases with the increase in $N$ at $M=N, \beta=16$, and $m_{0}=\omega_{0}=1$, which is indicative of spin-glass behavior. 

\begin{figure}
		\centering
		\includegraphics[width=\linewidth]{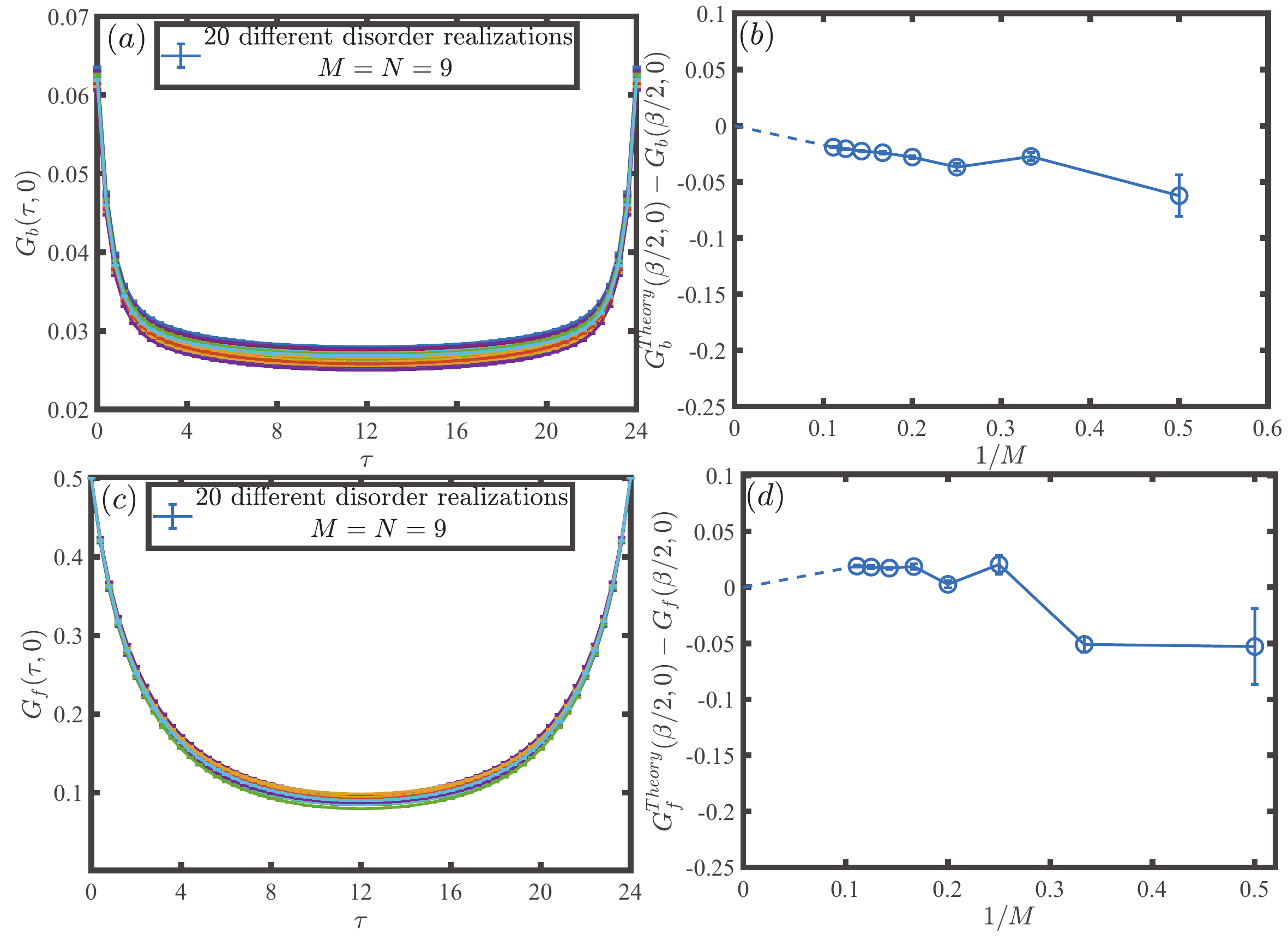}
		\caption{QMC bosonic and fermionic Green's functions $G_b$ and $G_f$, which are from 20 different disorder realizations with $M=N=9$, $\beta=24$ , $m_0=2$ and $\omega_0=1$, are shown in (a) and (c). And (b) and (d) present the difference between theoretical (large-$N$) results $G^{Theory}(\beta/2,0)$ and QMC numerical simulation data $G(\beta/2,0)$. The figure is adapted from Ref.~\cite{panYukawa2021}.}
		\label{fig:fig20}
	\end{figure}

In contrast, we can see that in Fig.~\ref{fig:fig20}, similar data of Eq.~\eqref{eq:eq10} show that our Yukawa-SYK model is not spin glass, since results of $G_f$ and $G_b$ show self-average of different realizations and the difference between the QMC obtain Green's functions and the large-$N$ solutions decreases with the increase of $M,N$.

Besides the Yukawa-SYK QMC studies presented in Secs.~\ref{sec:IIIc}, \ref{sec:IIId} and \ref{sec:IIIe}, there is also new QMC research on complex SYK model~\cite{kangSign2022}, where it was found one can design the similar model like our Yukawa-SYK with only fermions without the sign problem, and it is shown that the self-tuned and nFL behaviors also persist. 

Overall, there are still many interesting open questions in the study of SYK and SYK-like models. For example, planckian metal\cite{patelTheory2019,chowdhurySachdev2021} has attracted a lot of attention recently. The $T$-linear resistivity behaviors above the superconducting critical temperatures near optimal doping in the cuprate high-temperature superconductors is a long-standing puzzle. And recently the $T$-linear resistivity has been quantitatively characterized by the Drude formula $\rho=\frac{m^{*}}{n e^{2}} \frac{1}{\tau_{\mathrm{tr}}}$, with a "Planckian" metal transport scattering time $\tau_{\text {tr }}$ obeying $\frac{1}{\tau_{\mathrm{tr}}}=f \frac{k_{B} T}{\hbar}$~\cite{patelTheory2019}. It is found that $f \approx 1$ in many correlated electron systems such as cuprate, twisted bilayer graphene, and ultracold atoms, then it  is dependent only on Planck's constant and the absolute temperature in units of energy. Such linear $T$ dependence of resistivity has been observed in different models like SYK-like model theoretically as well as in experiments~\cite{chowdhurySachdev2021,hartnollPlanckian2021}. Some solvable models~\cite{patelTheory2019} in large $N$ limit with $N$ flavor fermions and random interactions, which are similar to SYK model have been designed to describe Planckian metal. Some bosonic SYK models also have showed Planckian dissipation and Corresponding transport characteristics\cite{tulipmanStrongly2020,tulipmanStrongly2021}. And in a lattice of coupled SYK models, the Hubbard model in single-site dynamical mean-field theory~\cite{chaSlope2020} and near QCP of the spin-1/2 model with random interaction~\cite{chaLinear2020,dumitrescuPlanckian2022}, people also observe the linear $T$ resistivity. Experimentally, it has been seen in various semiconducting materials\cite{ahnPlanckian2022} and some hole-doped cuprate\cite{grissonnancheLinear2021}. It will be of great interests that one can design more SYK-like models with nFL, superconductivity and low computation complexity such that one can compute the planckian metal behaviors, in a controlled manner, therein.

\section{Quantum Moir\'e lattice models and their computation}
\label{sec:IV}
 Quantum Moir\'e systems such as TBG and TMD systems, bestowed with the quantum geometry of wavefunctions -- manifested in the distribution of Berry curvature in the flat bands -- and strong long-range Coulomb electron interactions, exhibit rich phase diagram of correlated insulating and superconducting phases thanks to the high tunability by twisting angles, gating and tailored design of the dielectric environment. The extremely active research field of 2D quantum Moir\'e materials is developing very fast~\cite{tramblyLocalization2010,tramblyNumerical2012,bistritzerMoire2011,lopesContinuum2012,lopesGraphene2007,caoUnconventional2018,shenCorrelated2020,xieSpectroscopic2019,KhalafCharged2021,KevinStrongly2020,pierceUnconventional2021,caoCorrelated2018,liaoValence2019,liaoCorrelated2021,luSuperconductors2019,moriyamaObservation2019,chenTunable2020,rozhkovElectronic2016,ChatterjeeSkyrmion2020,kerelskyMaximized2019,rozenEntropic2021,tomarkenElectronic2019,soejimaEfficient2020,liuSpectroscopy2021,KhalafSoftmodes2020,ZondinerCascade2020,saitoPomeranchuk2021,GhiottoCriticality2021,SchindlerTrion2022,WangTMD2020,Parkchern2021,liaoCorrelation2021,anInteraction2020,huangGiant2020}. 
 
The theoretical efforts, from model design to the computation solutions in the quantum Moir\'e lattice models are also moving forward rapidly. The real space effective lattice models gave rise to the proper description of the ground state insulating phases with various symmetry breakings patterns and topological nature have been obtained both from the mean-field type analysis and quantum Monte Carlo and tensor-network many-body computations~\cite{koshinoMaximally2018,xuKekule2018,kangSymmetry2018,kangStrong2019,liaoValence2019,zhangCorrelated2020,liaoCorrelated2021,liaoCorrelation2021,chenRealization2021,linExciton2022}.

Later on, it was understood that due to the quantum metric in the flat-band wavefunction and the 2D long-range Coulomb interactions, the proper description of the quantum Moir\'e lattice model shall be constructed in continuum in momentum space. And it is from here the exact solutions at the chiral limit~\cite{bernevig2020tbg5,schindlerTrions2022} and mean-field analyses of the ground state phase digram~\cite{bultinckGround2020,liuPseudo2019,liuTheories2021,zhangCorrelated2020} and the momentum-space quantum Monte Carlo algrithm~\cite{zhangMomentum2021,daiQuantum2022,hofmannFermionic2022} to solve these systems and their intricate Monte Carlo sign structure~\cite{ouyangProjection2021,zhangSign2021,panSign2022} have been gradually and successfully revealed. 

Many interesting properties of the quantum Moir\'e lattice model have been discovered ever since and people gradually find many evidence~\cite{chenRealization2021,linExciton2022} pointing to the fundamental difference between the correlated flat-band lattice model and those of the conventional correlated electron lattice models systems, such as Hubbard-type models, in that, the interplay of topological wavefunction and long-range Coulomb interaction not only give rise to complex ground state phase diagrams with possibly different pairing mechanism and symmetry breaking patterns~\cite{zhangSuperconductivity2021}, but also hint at the difference in their thermodynamic and dynamic responses~\cite{panDynamical2022}.

Since the field of quantum Moir\'e materials is developing very fast, it is nevertheless not the purpose of this section to present an exhaustive survey of all the efforts in the understanding of the rich physics of quantum Moir\'e materials, but mainly focus on the model design and computation solutions inspired by the fast development of the field, and in a mutually constructive manner, demonstrate how our attitude of a sport and a pastime of doing so, actually either solve the present mysteries or point out the new directions that await to be explored.

\begin{figure}[!t]
\includegraphics[width=\columnwidth]{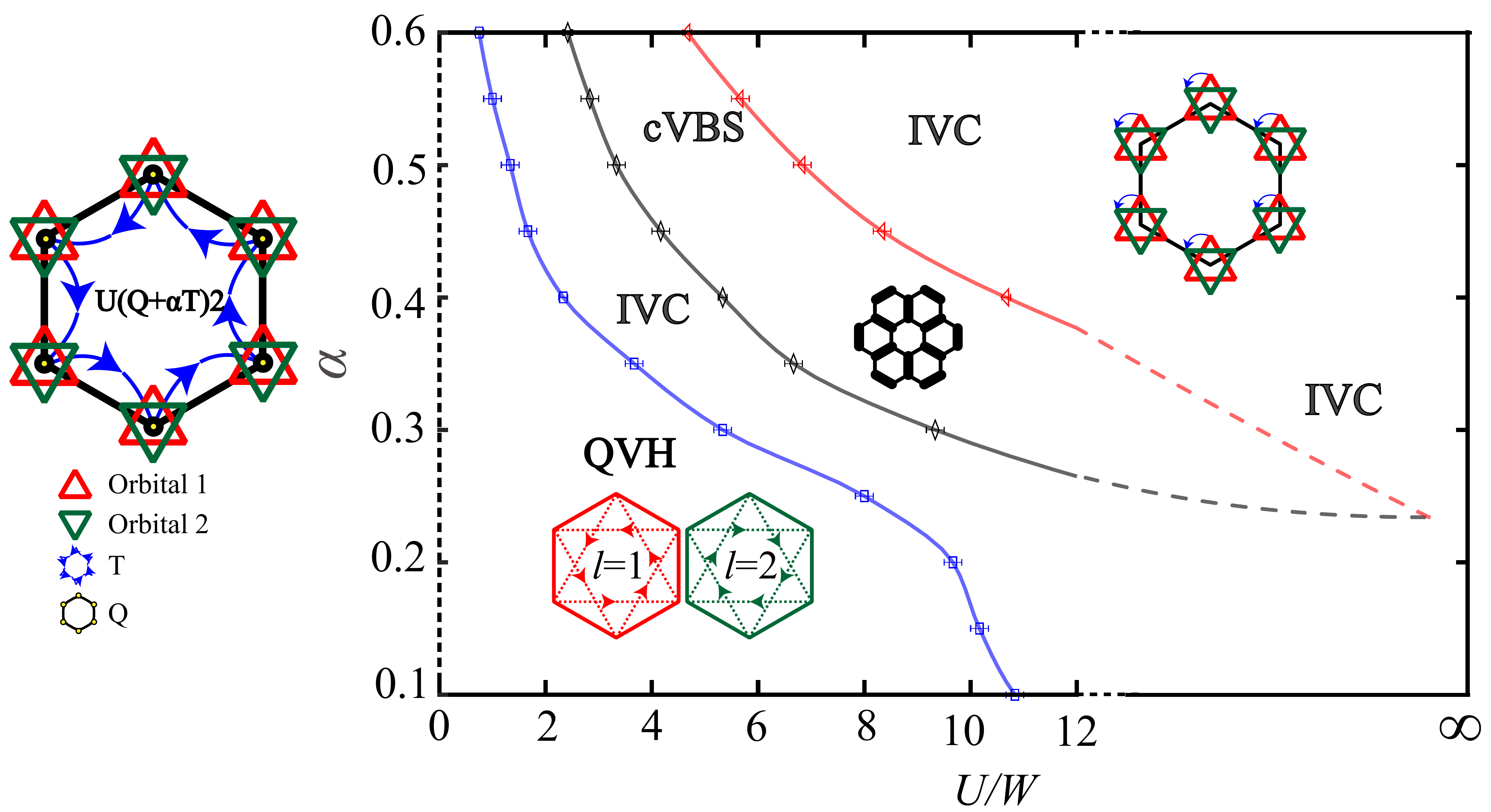}
\caption{(Left) The effective model with two valleys (red and green triangles) and spins and spin-valley SU(4) symmetry. The interactions act on every hexagon and consist of the cluster charge term $Q$ and the assisted-hopping interaction term $T$. (Right) Ground-state phase diagram, spanned by the $U/W$ and $\alpha$ axes, obtained from QMC simulations. The $y$-axis at $U=0$ (dashed line) stands for the Dirac semimetal phase. At small $U$, the ground state is a QVH (quantum Valley Hall) phase characterized by emergent imaginary next-nearest-neighbor hopping with complex conjugation at the valley index, as illustrated by the red and green dashed hoppings with opposite directions. Upon further increasing $U$, an IVC (inter-valley-coherent) insulating state is found, which breaks the SU(4) symmetry at every lattice site by removing the valley symmetry. Since	it preserves the lattice translational symmetry, it is ferromagnetic-like. The cVBS insulator, which appears after the IVC phase, breaks the lattice translational symmetry and preserves the on-site SU(4) symmetry. The IVC phase is stable at strong coupling limit. This figure is adapted from Ref.~\cite{liaoCorrelation2021}.} 
\label{fig:fig21}
\end{figure}

\subsection{Real-space lattice models}

To properly take the long-range Coulomb interactions into consideration in the Moir\'e system and explain various correlated insulating phases at the integer fillings such as the quantum anomalous Hall and intervalley-coherent insulators, one would need to start with the continuum model at momentum space, i.e., the Bistritzer-MacDonald (BM)~\cite{tramblyLocalization2010,tramblyNumerical2012,bistritzerMoire2011} type of model and project the Coulomb interactions onto the flat-bands while at the same time including the renormalization effects of the remote bands. The difficulty of doing so lies in the fact that there are no generic quantum many-body approaches that handle the long-range interactions well, except Hartree-Fock-type of mean-field calculation~\cite{zhangCorrelated2020,liuPseudo2019,liuTheories2021} and ED for very small cluster~\cite{xieTwisted2021}, and this is the reason that inspired us to successfully developed the momentum-space quantum Monte Carlo method for this purpose~\cite{zhangMomentum2021,daiQuantum2022}, which we will particularly focus on in later sections.

But before going directly to the continuum model, one can still construct real-space lattice effective model at the Moir\'e scale which captures the extended interactions beyond the Hubbard-type local interaction~\cite{kangSymmetry2018,koshinoMaximally2018,kangStrong2019,poOrigin2018,bernevig2020tbg3}, and develop controlled quantum many-body numerical methods such as QMC, DMRG and tensor-network approaches to solve them and determine the phase diagram. This is indeed the successful path taken by many (including us) and different topological phases and intervalley-coherent phases have been identified~\cite{xuKekule2018,liaoCorrelated2021,liaoCorrelation2021,liaoValence2019,chenRealization2021,linExciton2022}. 

As shown in Fig.~\ref{fig:fig21}, we design a QMC method that can simulate the real-space lattice model with extended charge and assisted  hopping interactions, with the Hamiltonian,
\begin{equation}
H =  -t\sum_{\langle ij \rangle l \sigma}\left(c^{\dagger}_{il\sigma}c^{\phantom{\dagger}}_{jl\sigma}+\rm{h.c}. \right) + U\sum_{\varhexagon}(Q_{\varhexagon}+\alpha T_{\varhexagon}-4)^2
\label{eq:eq23}
\end{equation}
here the Wannier orbitals, denoted by the operators $c_{il\sigma}$, live on the sites $i$ of the Moir\'e superlattice and are labeled by spin $\sigma=\uparrow,\downarrow$ and two orbital degrees of freedom $l=1,2$ (roughly corresponding to the two valleys). $U$ sets the overall strength of the Coulomb interaction. The two interaction terms in Eq.~(\ref{eq:eq23}), illustrated in Fig.~\ref{fig:fig21} (left), consist of the cluster charge $Q_{\varhexagon} \equiv \sum_{j\in \varhexagon}\frac{n_j}{3}$, with $n_j = \sum_{l\sigma}c^\dagger_{jl\sigma}c^{\phantom{\dagger}}_{jl\sigma}$, and the cluster assisted hopping 
$T_{\varhexagon} \equiv \sum_{j,\sigma} \left(i c_{j+1,1\sigma}^\dagger c^{\phantom{\dagger}}_{j,1\sigma} - i c_{j+1,2\sigma}^\dagger c^{\phantom{\dagger}}_{j,2\sigma} + h.c. \right)$. The index $j=1,\ldots,6$ sums over all six sites of the elemental hexagon in the honeycomb lattice. The hopping term is the nearest-neighbor band dispersion displays Dirac points at charge neutrality. And since we are interested in the strong-coupling regime, we try to keep the full non-trivial interaction term and simplify the tight-binding Hamiltonian in order to circumvent the sign problem at charge neutrality (four electrons per hexagon once averaging over the lattice). 

The QMC simulation results are summarized in the Fig.~\ref{fig:fig21} (right). We find that even very small interaction values trigger a transition from the non-interacting Dirac semi-metal phase to an insulating state. Upon increasing $U$, the nature of the insulator changes from a non-symmetry-breaking topological QVH phase, to an onsite SU(4) symmetry-breaking (intervalley-coherence) IVC insulator, to a translational symmetry-breaking columnar valence bond solid (cVBS) phase, and then finally back to a reentrant IVC state. This rich phase diagram is a consequence of the interplay between two different types of interaction terms: a cluster-charge repulsion $Q_{\varhexagon}$ and a non-local assisted-hopping interaction $T_{\varhexagon}$. The former is analogous to the standard Hubbard repulsion and, as such, is expected to promote either SU(4) antiferromagnetic order or valence-bond order in the strong-coupling regime. The latter, on the other hand, arises from the topological properties of the flat bands in TBG. When combined with $Q_{\varhexagon}$, it gives rise not only to SU(4) ferromagnetic-like order, but also to correlated insulating phases with topological properties, such as the QVH phase. 

While the precise value of $U/W$ in TBG is not known, a widely used estimate is that this ratio is of order $1$~\cite{kangStrong2019}. Referring to our phase diagram in Fig.~\ref{fig:fig21} (right), this means that certainly the QVH phase and possibly the IVC phase can be realized at charge neutrality, provided that $\alpha$ is not too small. Experimental probes do report a gap at charge neutrality Refs.~\cite{luSuperconductors2019,xieSpectroscopic2019} and the main manifestation of the QVH phase would be the appearance of gapless edge states, whereas in the case of the IVC state, it would be the emergence of a $\mathbf{q}=0$ order with onsite coupling between the two different valleys.

Furthermore, we also initiated the DMRG simulation at the filling factor of 3/4 of the same real-space model with extended interactions and found the quantum anomalous Hall (QAH) insulator~\cite{chenRealization2021,linExciton2022} which is consistent with experimental findings at the same filling. These results are shown in Figs.~\ref{fig:fig22} and ~\ref{fig:fig23}. The interaction-only Hamiltonian is now,
\begin{equation} 
H = U\sum_{\varhexagon}({Q_{\varhexagon}} + \alpha T_{\varhexagon} - 1)^2,
\label{eq:eq24}
\end{equation}
where $U = 1$ sets the energy unit {($\sim$ 40 meV in realistic system~\cite{chenRealization2021}), the cluster charge term $Q_{\varhexagon} \equiv \frac{1}{3}\sum_{l\in\varhexagon}c^\dagger_l c^{\phantom{\dagger}}_l$ counts the numbers of electrons in each hexagon, and $T_{\varhexagon} \equiv \sum_{l\in\varhexagon} [(-1)^l c^\dagger_l c^{\phantom{\dagger}}_{l+1} +\text{h.c.}]$ represents the assisted hopping term with alternating sign, as discussed in Fig.~\ref{fig:fig21}(left).

\begin{figure}[!t]
	\includegraphics[angle=0,width=1\linewidth]{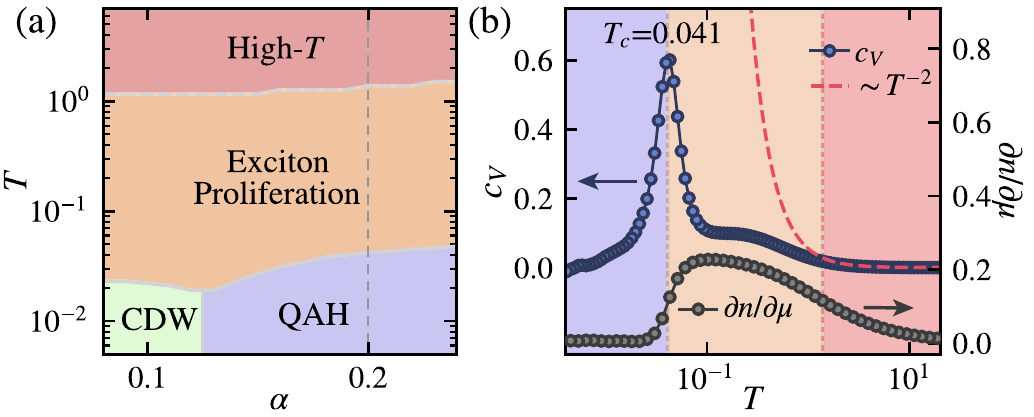}
	\caption{ (a) Finite-temperature phase diagram of the KV model, with
		the red regime being the high-$T$ disorder phase, the intermediate-$T$
		orange one where exciton proliferates, the green one being the CDW
		phase, and the purple one being the QAH phase. 
		The vertical dash line denotes the $\alpha=0.2$ case analysed in (b) panel. 
		(b) For the case of $\alpha=0.2$, the specific heat $c_V$ data is shown versus $T$, 
		where the peak at $T_c=0.041$ signifies the thermal phase transition. 
		The right axis of panel (b) shows the compressibility $\partial n/\partial\mu$ vs. $T$.
This figure is adapted from Ref.~\cite{linExciton2022}.}
	\label{fig:fig22}
\end{figure}
	
A quantum anomalous Hall (QAH) topological Mott insulator (TMI) with spontaneous time-reversal symmetry breaking and nonzero Chern number and a charge-density-wave (CDW) insulator have been discovered in the DMRG computation of Eq.~\eqref{eq:eq24}~\cite{chenRealization2021}. We further employ the state-of-the-art thermal tensor network and the perturbative field-theoretical approaches to obtain the finite-$T$ phase diagram and the dynamical properties of the TBG model~\cite{linExciton2022}. As shown in Fig.~\ref{fig:fig22}(a), the phase diagram includes the quantum anomalous Hall and charge density wave phases at low $T$, and an Ising transition separating them from the high-$T$ symmetric phases. Due to the proliferation of excitons -- particle-hole bound states -- the transitions take place at a significantly reduced temperature than the mean-field estimation. The exciton phase is accompanied with distinctive experimental signatures such as enhanced charge compressibilities and optical conductivities close to the transition. These results explain the smearing of the many-electron state topology by proliferating excitons and open the avenue for controlled many-body investigations on finite-temperature states in the TBG and other quantum Moir\'e systems.

\begin{figure}[!htp]
\includegraphics[angle=0,width=1\linewidth]{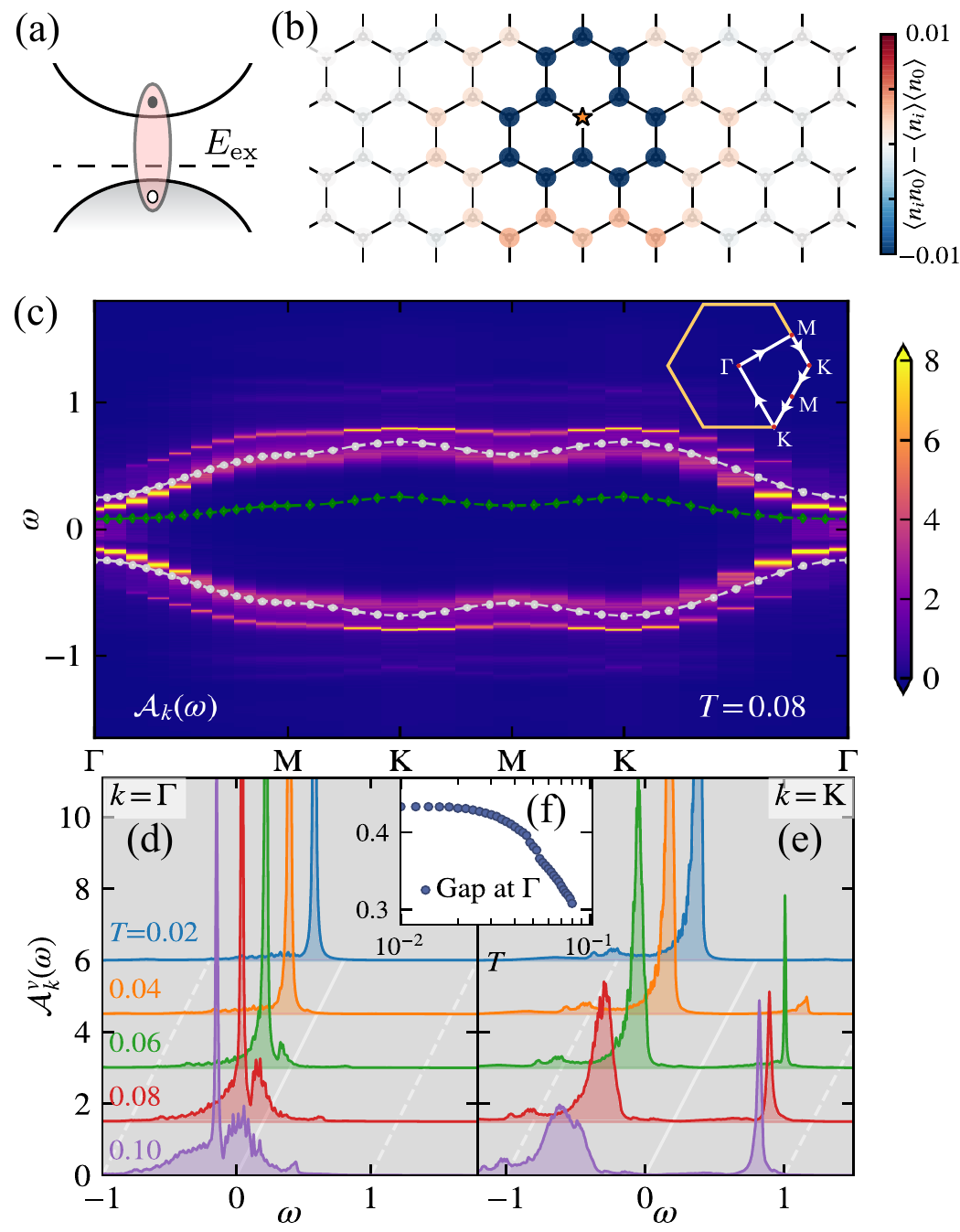}
\caption{(a) illustrates the formation of
exciton between two quasi-particles from the
valence and conduction bands. (b) At intermediate
temperature $T=0.244$ inside the exciton proliferation regime, the charge correlations between a central site (the asterisk) and other sites are shown here. (c) Quasi-particle spectral function $\mathcal{A}_k(\omega)$ as the sum of valence and conduction band spectrum, for $\alpha=0.2$ at $T=0.08$. Compared to the mean-field single-particle dispersion, which consist of upper conduction and lower valence bands (white dots), the spectral weight distribution gets broadened. The exciton band (denoted by green diamonds) lies within the single-particle gap. The corresponding path in the Brillouin zone is drawn in the inset diagram. Panels (d,e) show the valence electron spectral weights $\mathcal{A}_k^v(\omega)$ at $k=\Gamma$ and $k=\mathrm{K}$ for various temperatures. Panel (f) shows the gap at $k=\Gamma$ and $k=\mathrm{K}$ as a function of temperature $T$. This figure is adapted from Ref.~\cite{linExciton2022}.}
\label{fig:fig23}
\end{figure}

It has been found that the CDW and QAH phase are separated by a first-order transition extending from the transition point $\alpha \simeq 0.12$ \cite{chenRealization2021}. Then we turn to consider the finite-temperature Ising transition, such as at $\alpha=0.2$. In Fig.~\ref{fig:fig22}(b), the specific heat $c_V$ has a peak at $T_{c} \simeq 0.041$, which is the evidence of a second-order phase transition from QAH to excition proliferation region at this temperature. As for the compressibility $\partial n / \partial \mu$, its curve has a peak at higher temperatures above $T_c$. Inside the exciton regime it also keeps an enhanced value. The above facts show the exciton (bosonic bound state) proliferation above $T_c$. When $T$ increases, a crossover between the intermediate-$T$ and high-$T$ regimes will be seen. Then the $c_V$ curve scales as the high-$T$ limit of $T^{-2}$.

Fig.~\ref{fig:fig23} (a) and (b) show the schematic plot of the exciton excitations and its real-space presence from the density-density correlation functions for YC4$\times$12 DMRG cylinder at $T=0.244$ (above $T_c$) and $\alpha=0.2$ inside the exciton proliferation phase. As shown in Fig.~\ref{fig:fig23} (b),
if we place a hole at the center of the lattice, bunching and antibunching modulation behaviors of electrons are present. It's the evidence of existence of particle-hole bound states—excitons—in the system. 

When $T=0.08$, a
representative temperature, the  renormalized spectral function quasi-particle $A_{\mathbf{k}}(\omega)=A_{\mathbf{k}}^{c}(\omega)+A_{\mathbf{k}}^{ v}(\omega)$ gave us the information of the correlated band in Fig.~\ref{fig:fig23} (c). Here $c,v$ are for electrons in conduction and valence bands. Then we know that the exciton (green diamonds line) has a
much lower energy $E_{\mathrm{ex}} \sim 0.08$ than the mean-field band gap (white dots line), as shown in Fig.~\ref{fig:fig23} (c). In panel (d) and (e), the spectral
functions of electrons in the valence band $A_{\mathbf{k}}^{ v}(\omega)$ at $k=\Gamma$ and $k=\mathrm{K}$ of different temperatures are shown. From the above data, we can see how the energy gap varies with temperature, as shown in Fig.~\ref{fig:fig23} (f). It is clear that above the $T_c \sim 0.04$ but still below the mean-field QAH band gap at $T\sim 0.1$, the spectral function is already gapless due to the exciton collective excitations.

\subsection{Momentum-space models and quantum Monte Carlo method}

In order to consider the long-range Coulomb interaction and the Wannier obstruction problem, recently there have been a lot of studies to simulate TBG and other Moir\'e lattice models in momentum space~\cite{leeA2021,zhangMomentum2021,hofmannFermionic2022,panDynamical2022}. In our previous work, we start from the continuous BM model~\cite{Trambly2010,Trambly2012,Bistritzer_TBG,ROZHKOV20161,Santos2007,lopesContinuum2012,YiZhang2020} and long-range single-gate Coulomb interaction, and integrate out the states on the remote bands to obtain the projected Hamiltonian as~\cite{song2020tbg2,bernevig2020tbg3}:
	\begin{equation}
		H_{\mathrm{int}}=\frac{1}{2 \Omega} \sum_{\mathbf{q}, \mathbf{G},|\mathbf{q}+\mathbf{G}| \neq 0} V(\mathbf{q}+\mathbf{G}) \delta \rho_{\mathbf{q}+\mathbf{G}} \delta \rho_{-\mathbf{q}-\mathbf{G}}
			\label{eq:eq25}
	\end{equation}
	and
	\begin{equation}
	\begin{aligned}
		\delta \rho_{\bq+\bG}=&\sum_{\bk, m_{1}, m_{2},\tau,s} \lambda_{m_1,m_2,\tau}(\bk,\bk+\bq+\bG) \\
		&\left(d_{\bk, m_{1},\tau,s}^{\dagger} d_{\bk+\bq, m_{2},\tau,s}  -\frac{1}{2} \delta_{q,0}\delta_{m_1,m_2}\right) 
		\end{aligned}
		\label{eq:eq26}
	\end{equation}
where $V(\mathbf{q})=\frac{e^{2}}{4 \pi \varepsilon} \int d^{2} \mathbf{r}\left(\frac{1}{\mathbf{r}}-\frac{1}{\sqrt{\mathbf{r}^{2}+d^{2}}}\right) e^{i \mathbf{q} \cdot \mathbf{r}}=\frac{e^{2}}{2 \varepsilon} \frac{1}{q}\left(1-e^{-q d}\right)$ is the long-ranged single gate Coulomb interaction, with $d/2$ is the distance between graphene layer and single gate and $\epsilon$ is the dielectric constant. Here the Moiré lattice vector $L_{M}=\frac{a_{0}}{2 \sin \left(\frac{\theta}{2}\right)}$, the area of system $\Omega=N_{\mathbf{k}} \frac{\sqrt{3}}{2} L_{M}^{2}$ with $N_{\mathbf{k}}$ the number of $\mathbf{k}$ points in $\mathrm{mBZ}$ (here we have $N_{\mathbf{k}}=36, 81$ for $6 \times 6$ and $9\times 9$ meshes). We use $\varepsilon=7 \varepsilon_{0}$, $d=40$ nm,$ \mathrm{mBZ}$ reciprocal lattice vector $|\mathbf{G}|=\frac{4 \pi}{\sqrt{3} L_{M}}$ in our simulation. And $\theta=1.08^{\circ}$, $a_0=0.246$nm.

\begin{figure*}
	\includegraphics[width=\textwidth]{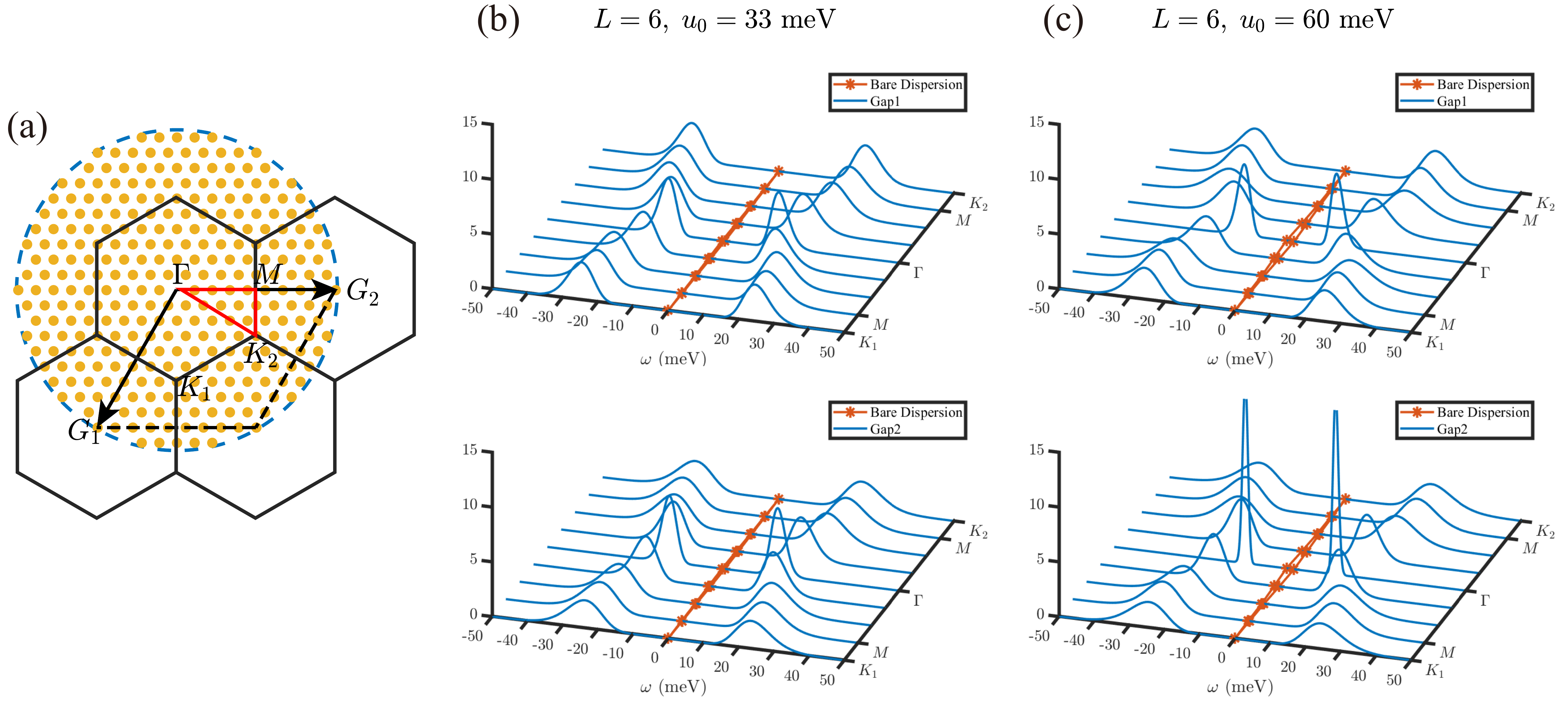}
	\caption{ (a) The moir\'e Brillouin zones (mBZ) at one valley. The high-symmetry path $\Gamma-M-K_1(K_2)-\Gamma$ are marked by the red solid line. $\bG_{1,2}$ are the reciprocal lattice vectors of the mBZ. All of the yellow dots mark possible momentum transfer in QMC simulations, $\bq+\bG$, and the blue dashed circle is the momentum space cut-off. Here we show a $9\times 9$ mesh in the mBZ, while there are 300 allowed momentum transfers. In (b) and (c), blue lines show the single particle spectra of $L=6,\; T =0.667$ meV, $u_0=33 $ meV and $60$ meV. They are obtained from the momentum space QMC with analytic continuation. The red stars and lines indicate the bare dispersions of $H_0$. Here $u_0=60$ meV is a realistic case \cite{YiZhang2020,bernevig2020tbg1,song2020tbg2,bernevig2020tbg3,tarnopolskyOrigin2019} which leads to a bandwidth of 1.08 meV. And $u_0=33$ meV is a case between the realistic one and the chiral limit. This figure is adapted from Refs.~\cite{zhangMomentum2021,panDynamical2022}.}
	\label{fig:fig24}
\end{figure*}

As for the form factor $\lambda_{m_{1}, m_{2}, \tau}(\mathbf{k}, \mathbf{k}+\mathbf{q}+\mathbf{G})$, it is defined as $\lambda_{m_{1}, m_{2}, \tau}(\mathbf{k}, \mathbf{k}+\mathbf{q}+\mathbf{G}) \equiv\left\langle u_{\boldsymbol{k}, m_{1}, \tau}\right|$ $\left.u_{k+\mathbf{q}+\mathbf{G}, m_{2}, \tau}\right\rangle$. Here $\left|u_{k, m, \tau, s}\right\rangle$ is the Bloch eigenstate of BM Model $H_{B M}^{\tau}(\mathbf{k})$. The operator $d_{\boldsymbol{k}, m . \tau, s}^{\dagger}$ is the corresponding creation operator, with $m, s, \tau$ the band, spin, and valley indices. 

The BM model in plane wave basis~\cite{tramblyLocalization2010,tramblyNumerical2012,bistritzerMoire2011,rozhkovElectronic2016,lopesGraphene2007,lopesContinuum2012,zhangCorrelated2020} is: $H_{B M}^{\tau}(\mathbf{k})=\sum_{\mathbf{k}^{\prime}} H_{B M \mathbf{k}, \mathbf{k}^{\prime}}^{\tau} \quad e^{-i \mathbf{k} \cdot \mathbf{r}} e^{i \mathbf{k}^{\prime} / \mathbf{r}} \quad$ where \begin{equation}
		\begin{aligned}
			H^{\tau}_{BM,\mathbf{k},\mathbf{k {}^\prime}}=\delta_{\mathbf{k},\mathbf{k {}^\prime}}&\left(\begin{array}{cc} 
				-\hbar v_F ({\bk}-\bK_1^{\tau}) \cdot \pmb{\sigma}^{\tau}  &  U_0  \\
				U_0^\dagger  & -\hbar v_F ({\bk}-\bK_2^{\tau}) \cdot \pmb{\sigma}^{\tau}
			\end{array}\right) \\ 
			+&\left(\begin{array}{cc} 
				0  &  U_1^{\tau} \delta_{\mathbf{k},\mathbf{k {}^\prime}-\tau \bG_1 }  \\ 
				U_1^{\tau \dagger} \delta_{\mathbf{k},\mathbf{k {}^\prime}+\tau \bG_1 }  & 0
			\end{array}\right)\\ 
			+&\left(\begin{array}{cc} 
				0  &  U_2^{\tau} \delta_{\mathbf{k},\mathbf{k {}^\prime}-\tau(\bG_1+\bG_2) }  \\ 
				U_2^{\tau \dagger} \delta_{\mathbf{k},\mathbf{k {}^\prime}+\tau(\bG_1+\bG_2) }  & 0
			\end{array}\right)
		\end{aligned}
		\label{eq:eq1}
	\end{equation}
	 and $ U_{0}=\left(\begin{array}{ll}u_{0} & u_{1} \\ u_{1} & u_{0}\end{array}\right), U_{1}^{\tau}=\left(\begin{array}{cc}u_{0} & u_{1} e^{-\tau \frac{2 \pi}{3} i} \\ u_{1} e^{\tau \frac{2 \pi}{3} i} & u_{0}\end{array}\right)$ , $U_{2}^{\tau}=\left(\begin{array}{cc}u_{0} & u_{1} e^{\tau \frac{2 \pi}{3} i} \\ u_{1} e^{-\tau \frac{2 \pi}{3} i} & u_{0}\end{array}\right)$, with  $u_{0}$ and $u_{1}$ are the intra-sublattice and inter-sublattice interlayer tunneling amplitudes. $\tau=\pm$ is the valley index and $\boldsymbol{\sigma}^{\tau}=\left(\tau \sigma_{x}, \sigma_{y}\right)$. $\mathbf{K}_{1}^{\tau}$ and $\mathbf{K}_{2}^{\tau}$ are the corresponding Dirac points of the bottom and top layers of graphene, and $\mathbf{k} \in \mathrm{mBZ}$ and $\mathbf{G}_{1}$ and $\mathbf{G}_{2}$ are the reciprocal vectors of $\mathrm{mBZ}$. We set $\hbar v_{F} / a_{0}=2.37745 \mathrm{eV}$, with $a_0=0.246$ nm and $u_{1}=0.11 \mathrm{eV}$. The twist angle $\theta=1.08^{\circ}$ is the first magic angle, which can result in flat band in chiral limit($u_0$=0 eV).

Adding back to the flat-band dispersion, the Hamiltonian in Moir\'e lattice model in continuum reads,
\begin{equation}\begin{aligned}
H&=H_{0}+H_{\mathrm{int}}\\
H_{0} &=\sum_{m=\pm 1} \sum_{\mathbf{k} \tau s} \epsilon_{m, \tau}(\mathbf{k}) d_{\mathbf{k}, m, \tau, s}^{\dagger} d_{\mathbf{k}, m, \tau, s}
\label{eq:eq28}
\end{aligned}
\end{equation}
and $\epsilon_{m, \tau}(\mathbf{k})$ is the bare dispersion\cite{vafekRenormalization2020}, which is eigenvalue of $H_{BM}$. We note that in Refs.~\cite{bultinckGround2020,parkerStrain2021,kwanKekule2021,hofmannFermionic2022}, the mean-field contribution of the remote band interaction from the flat band is removed, which is different from our choice.

We develop the momentum space QMC approaches to solve the Hamiltonian in Eq.~\eqref{eq:eq28}, the method is free of the sign problem at charge-neutrality point(CNP) with the $C_2 P$ and $C_2 T$ symmetries, and we further proved that the flat-band system acquires a unique {\it Sign bound theory} \cite{zhangSign2021} , some of them, like chiral limit single valley and single spin TBG at CNP, acquires a polynomial sign problem rather than exponential, this effective means that even with the polynomial sign, the computational complexity of the lattice models such as Eq.~\eqref{eq:eq28} is polynomial instead of exponential and we can solve them without hitting the "exponential well"~\cite{zhangMomentum2021,zhangSign2021,panSign2022}. Some of the other filling cases, like $\nu=-1$, can also be simulated in polynomial time based on the {\it Sign bound theory}.

With the method developed, we can now compute the detailed static, dynamic and thermodynamic propers of the Moir\'e lattice models in continuum. As shown in Fig.~\ref{fig:fig24} (b) and (c), we demonstrate the QMC + stochastic analytic continuation (SAC)~\cite{Sandvik2016,HShao2017,sunDynamical2018,maDynamical2018,yan2021topological,zhou2020amplitude,hu2020evidence}, obtained single-particle gaps at the charge neutrality of the TBG with both $u_0$ and $u_1$ finite at low temperature $T=0.667$meV, i.e., the realistic material parameters~\cite{YiZhang2020,bernevig2020tbg1,song2020tbg2,bernevig2020tbg3,tarnopolskyOrigin2019}, we also showed the bare dispersion of the BM model with parameters which give rise to a narrow bandwidth of 1.08 meV. One can clearly see that the Coulomb interaction opens an interaction gap at the scale of $\sim 10$ meV, consistent with the experimental observations.

\subsection{Dynamical and thermodynamic signatures of the flat-band correlations}
\begin{figure*}[tb]
\centering
\includegraphics[width=\linewidth]{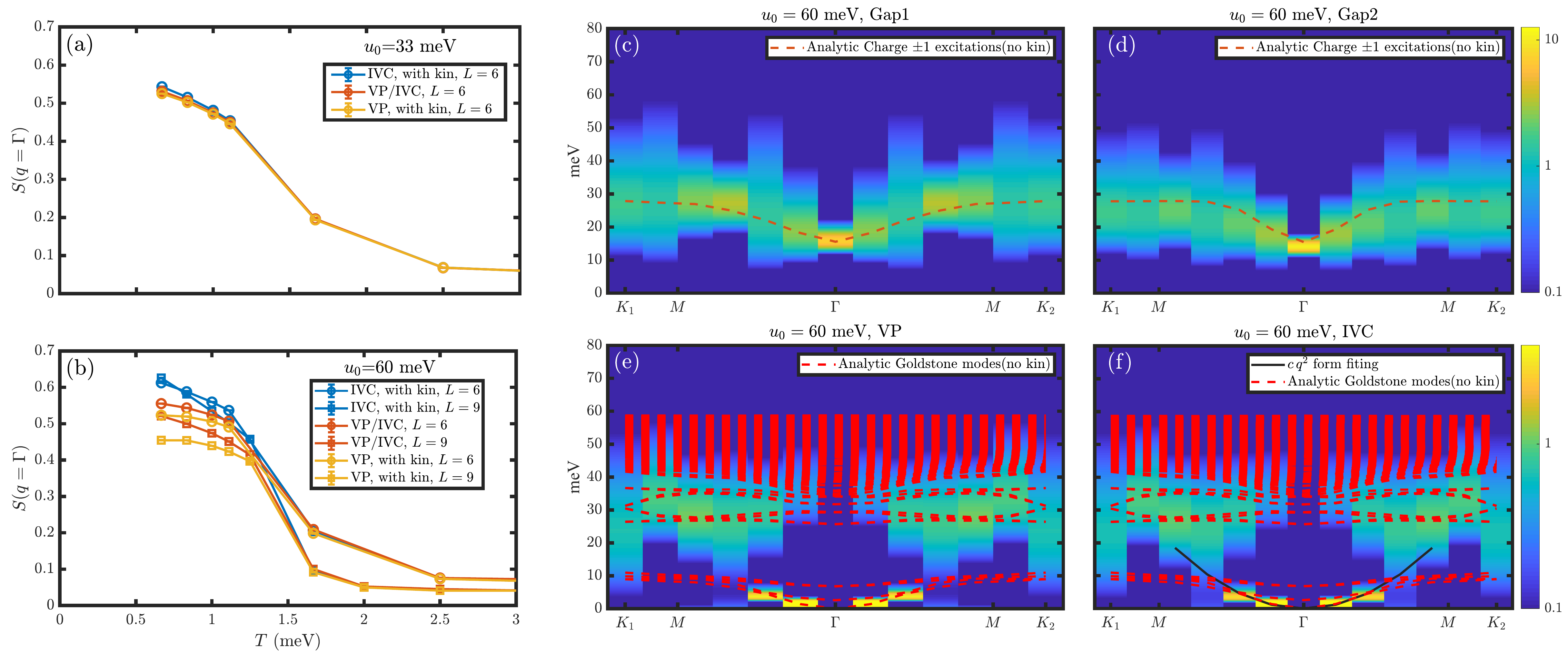}
\caption{ (a) and (b). Order parameter correlation functions, $S(\mathbf{q}=\Gamma,\tau=0)$, for VP and IVC at $u_0=33,60$ meV and $L=6,9$, as a function of temperature. When the kinetic energy is taken into account ("with kin"), which breaks the symmetry, the degeneracy of VP/IVC is lifted. While ignoring the kinetic energy will recover the degeneracy due to an emergent $SU(2)$ [$U(4)$] symmetry. It is interesting to see the competition of the two orders, with the IVC slightly stronger.
(c) and (d). Single-particle spectra at $T =0.667$ meV, $u_0=60$ meV and $L=9$, which shows an insulating gap $\sim 10$ meV (consistent with that in Fig.~\ref{fig:fig24} (b) and (c)). The dashed lines are the exact solution of the single-particle dispersion at the flat-band limit following Ref.~\cite{bernevig2020tbg5}. (e) and (f). Dynamical spectra of VP and IVC with the same parameters. We obtain sharp and ferromagnetic-like valley waves in both channels near $\mathbf{q}=\Gamma$. The black line gave the $q^2$ fitting. At the energy scale of twice the single-particle gap, $\sim 20$ meV, valley waves are over-damped into the particle-hole continuum. Again the dashed lines are the analytic computation of the Goldstone mode at the flat-band limit following Ref.~\cite{bernevig2020tbg5}. They are consistent with QMC results(with kinetic energy) only at low energy and near $\Gamma$ point. This figure is adapted from Ref.~\cite{panDynamical2022}.}
	\label{fig:fig25}
\end{figure*}

Although the ground state of the Moir\'e lattice models at integer fillings are exactly solvable~\cite{Bultinck2020,hofmannFermionic2022,bernevig2020tbg5,vafekLattice2021} if there is no kinetic energy term $H_0$, the dynamical and thermodynamical information originated from the long-range interactions and the interplay of quantum and thermal fluctuations cannot be accessed analytically, and it is here our momentum-space QMC, working in path-integral formalism, could provide these valuable information and make connection and great extend the analytic understandings~\cite{panDynamical2022}.	

For example, to properly study the symmetry breaking phases in the charge-neutrality point of TBG, we can define the valley polarization (VP) and intervalley-coherence (IVC) order parameters as $\mathcal{O}_{a}(\boldsymbol{q},\tau)  \equiv \sum_{\boldsymbol{k}} d_{\boldsymbol{k}+\boldsymbol{q}}^{\dagger}(\tau) M_a d_{\boldsymbol{k}}(\tau)$, with $M_a= \tau_z \eta_0$ ($\eta_0$ for band index) for VP and $M_a= \tau_x \eta_y$ or $\tau_y \eta_y$ for the IVC states~\cite{Bultinck2020,KhalafSoftmodes2020,liuTheories2021,bernevig2020tbg5,hofmannFermionic2022}. These three order parameters  generate a $SU(2)$ symmetry group at $q=0$. And the nonzero value of them means  the spontaneously symmetry breaking of the $SU(2)$ symmetry, which will result in gapless Goldstone modes. 

When we add the kinetic energy term, the degeneracy of IVC/VP will be broken, since an IVC state breaks the continuous $U(1)$ symmetry while a VP state breaks  the discrete $Z_2$ symmetry. The difference will results different behaviors of dynamics fluctuations in VP and IVC states. In the QMC simulation, we can compute the correlation of VP/IVC order parameters and investigate their difference. The correlation functions of order parameters are defined as:
\begin{equation}
S_{a}(\boldsymbol{q},\tau) \equiv \frac{1}{L^{4}}\left\langle\mathcal{O}_{a}(-\boldsymbol{q},\tau) \mathcal{O}_{a}(\boldsymbol{q},0)\right\rangle,
\label{eq:eq29}
\end{equation} 
and their results, $S(\mathbf{q}=\Gamma, \tau=0)$ of VP/IVC for various temperatures, are shown in Fig.~\ref{fig:fig25} (a) and (b). One can indeed see the degeneracy of VP/IVC without the kinetic energy term and when the kinetic energy is taken into account (date curves “with kin” in Fig.~\ref{fig:fig25} (a), (b)), which breaks the symmetry, this degeneracy is lifted. From the finite size behaviors with $L=6$ and $9$, we can conclude that in the presence of flat-band kinetic term, the ground state is more likely to be IVC state, since the IVC correlation function increases with system size $L$ as temperature reduces, while that of VP reduces with $L$.

\begin{figure*}[htp!]
\includegraphics[width=1.0\textwidth]{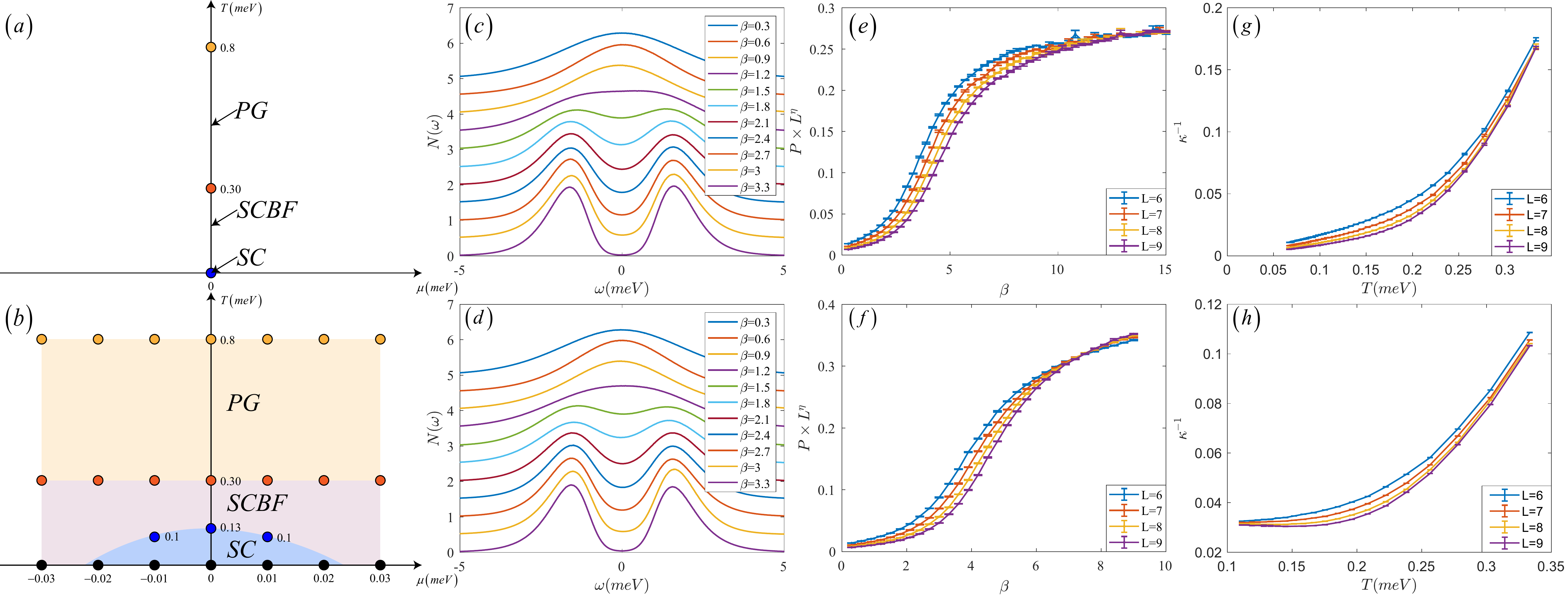}
\caption{In this figure, top row corresponds to the results in flat-band limit and bottom row is that away from the flat-band limit. The bandwidths of non-interacting band structures are set to $0$ and $0.8$ meV, respectively. (a-b) Phase diagrams. In the flat-band limit (a), a finite chemical potential drives the system into a trivial insulator with either empty ($\mu\ll 0$) or complete filled ($\mu \gg 0$) bands. In Fig.~(b), the yellow region is the pseudogap (PG) while purple region is super-compressible bosonic fluid (SCBF) phase. The blue region marks the superconducting (SC) dome. (c-d) Density of states for a $9\times9$ system at $\mu=0$.  The Cooper gap survives until it  turns into a pseudogap above $T\sim0.3$. (e-f)  Data cross of $P\times L^{\eta}$ versus $\beta=1/T$ with BKT anomalous dimension $\eta=1/4$. The crossing point of different system sizes ($L=6,7,8,9$) gave the superconducting transition temperature: $T_c\sim0$ in (e), which is consist with that the SU(2) symmetry requires the cross point at $\beta\to\infty$, and $\sim 0.13$ in (f). (g-h) Inverse of compressibility $\kappa$ for $L=6,7,8,9$ at $\mu=0$. At low temperature, the compressibility $\kappa$ diverged in (g), while it converges in superconducting phase as shown in (h). This figure is adapted from Ref.~\cite{zhangSuperconductivity2021}.}
	\label{fig:fig26}
\end{figure*}

We can further compute the dynamic correlations of the IVC and VP fermion bilinears, and their spectra with the system size of $9\times9$ for the realistic case with kinetic energy at $u_0=60$ meV at low temperature $T=0.667$ meV,  are shown in Fig.~\ref{fig:fig25} (e) and (f), with Fig.~\ref{fig:fig25} (c) and (d) the associated single-particle spectra. When there is no kinetic energy term, the single-particle dispersion and Goldstone modes can be obtained exactly by calculating the commutator~\cite{bernevig2020tbg5}, corresponding to the red dotted line in Fig.~\ref{fig:fig25} (c-f).  Below the single-particle gap $\sim 10$ meV, there are sharp Goldstone-like modes in VP and IVP spectra near $\Gamma$ point, with $\omega \propto c\,q^2$. They are in strong analog to $SU(2)$ ferromagnetic Goldstone modes. It means an approximate $SU(2)$ symmetry survives in our model, which is consistent with our small bare dispersion. This is also supported by the fact that the IVC and VP spectra are almost identical and are strikingly similar to the flat-band limit~\cite{bernevig2020tbg5,feldnerDynamical2011}. We fit the quadratic dispersion and find $c=31.32\pm 0.03 \;\mathrm{meV}/k_{\theta}^2$, (where $k_{\theta}=8 \pi \sin (\theta / 2) /\left(3 a_{0}\right)$ and the lattice constant of the monolayer graphene $a_{0}=0.246 \mathrm{~nm}$). What needs to be noted is that the sharp quadratic Goldstone-like modes means an approximate $SU(2)$ symmetry survives, but it is an approximate $SU(2)$ symmetry. At very small $q$ and $\omega$, there's no such symmetry and no dominant ferromagnetic Goldstone modes, and then we will see a linear dispersion $\omega \propto q$, corresponding to the SU(2) symmetry breaking and magnon-like excitation~\cite{KhalafSoftmodes2020}. Due to the small kinetic energy term we used, this SU(2) symmetry breaking is really weak and the linear dispersion is not visible in our QMC data.

In addition, we can see damping of collective modes above the energy scale of $\sim 20$ meV. Such results beyond mean-field type of calculations have the following possible origins: (1) scattering between collective modes and (2) damping due to the fermion particle-hole continuum. When the energy is larger than twice the fermion gap, the second damping channel arises, which is responsible for the over-damped features at energy above $20$ meV shown in Fig.~\ref{fig:fig25} (e) and (f). A similar phenomenon, the damping of ferromagnetic spin excitations,  has been seen in the graphene nanoribbons. In that case, the spin waves become over-damped in the particle-hole continuum while the flat band gives rise to the ferromagnetic long-range order~\cite{feldnerDynamical2011,golorQuantum2013,golorQuantum2014}

\subsection{Symmetry, Pairing mechanism and Superconductivity}
The TBG model in Eq.~\eqref{eq:eq25} has been proved to be sign-problem-free at charge-neutrality point and acquires polynomial sign bound at integer fillings~\cite{zhangMomentum2021,zhangSign2021,panSign2022}. However, away from integer fillings, the QMC simulation still suffers from the exponential sign problem. But the experimental observation of the superconductivity in TBG~\cite{caoUnconventional2018,shenCorrelated2020} and TMD~\cite{anInteraction2020} systems are all away from the integer fillings. To be able to still study the pairing of the flat-band continuum models, we adjust the interaction term in Eq.~\eqref{eq:eq26} to:
\begin{equation}
\begin{aligned}
\delta \rho_{\mathbf{q}+\mathbf{G}}=& \sum_{\mathbf{k}, m, n}\left[\lambda_{m, n, \tau}(\mathbf{k}, \mathbf{k}+\mathbf{q}+\mathbf{G}) d_{\mathbf{k}, m, \tau}^{\dagger} d_{\mathbf{k}+\mathbf{q}, n, \tau}\right.\\
&\left.-\lambda_{m, n,-\tau}(\mathbf{k}, \mathbf{k}+\mathbf{q}+\mathbf{G}) d_{\mathbf{k}, m,-\tau}^{\dagger} d_{\mathbf{k}+\mathbf{q}, n,-\tau}\right]
\end{aligned}
\end{equation}
which accounts to an inter-valley attraction and intra-valley repulsion in this new model, while the original model in Eq.~\eqref{eq:eq25} is repulsive both inter- and intra-valley.  With such a simple change, our new model is sign-problem-free at any fillings~\cite{zhangSuperconductivity2021}. Besides, for simplicity, here we only consider one flat-band per valley, and choose parameters according to twisted homobilayer transition metal
dichalcogenides (TMD)\cite{liSpontaneous2021,suMassive2021,zhangSuperconductivity2021}. 

We find for the new model, at the flat band limit, the single-particle correlation function are still fully gapped, the same as in TBG model at the charge neutrality point. But the difference is that now this gap is the Cooper pair gap, instead of an insulating gap. For $T \ll V$, all electrons are
paired into Cooper pairs, which means the system is a fluid of
Cooper pairs without unpaired fermions. In addition, the bosonic fluid has a high compressibility, which diverges at zero temperature limit. We label this state as SCBF (for
super-compressible bosonic fluid) in our phase diagrams
Fig.~\ref{fig:fig26} (a) without the kinetic term and (b) with the kinetic term.

We then compute the correlation function of the Cooper pairs $\left\langle\Delta(t) \Delta^{\dagger}(0)\right\rangle$ and the corresponding susceptibility
\begin{equation}
\chi_{S C}=\frac{1}{2 L^{4}}\left\langle\Delta \Delta^{\dagger}+\Delta^{\dagger} \Delta\right\rangle
\end{equation}  to determine the superconducting phase transition temperature, where $\Delta=\sum_{\mathbf{k}, m} d_{-\mathbf{k}, m,-\tau} d_{\mathbf{k}, m, \tau}$, note here $t$ is the imaginary time and $\tau$ is used as the valley index. 

The QMC obtained diagrams are shown in Fig.~\ref{fig:fig26} (a) and (b). From high to low temperature are PG phase, super-compressible bosonic fluid (SCBF) phase, and the dome of a valley-singlet superconductor. Fig.~\ref{fig:fig26}(c) and (d) show the density of states, which are obtained from stochastic analytic continuation (SAC) of the QMC data, these result exhibit the process of change of the phases in the temperature axis: at $T < 0.3$, a full Cooper gap is observed and the system is in the superconducting phase; for $0.3 < T < 0.8$, fermion spectral weight start to emerge inside the gap, i.e., a PG. In Fig.~\ref{fig:fig26} (e) and (f), we use the BKT scaling function $\chi_{S C}=L^{-\eta} f\left(L \cdot \exp \left(-\frac{A}{\left(T-T_{c}\right)^{1 / 2}}\right)\right)$ with $\eta=1/4$ to find the critical temperature, the same analyses have been done in Figs.~\ref{fig:fig_ps} and ~\ref{fig:fig18} in other context. The critical temperature $T_c\sim 0.13$ in Fig.~\ref{fig:fig26}(f), while in (e) no cross at finite temperature since there is the $SU(2)$ symmetry in the model without kinetic term. 

Finally, in Fig.~\ref{fig:fig26}(g) and (h), we plot the inverse of compressibility $\kappa$ for $L=6,7,8,9$ at $\mu=0$, where $\kappa=\frac{1}{n_{0} L^2} \frac{d \langle \hat{N}_p\rangle}{d \mu}=\beta \frac{\left\langle\hat{N}_{p}^{2}\right\rangle-\left\langle\hat{N}_{p}\right\rangle^{2}}{n_{0} L^2}$. Here $n_{0}$ is the particle density and  $\hat{N}_p=\sum_{\mathbf{k}, m, \tau} d_{\mathbf{k}, m, \tau}^{\dagger} d_{\mathbf{k}, m, \tau}$ is the particle number
operator of flat bands.  We see the large compressibility above the superconducting dome. In Fig.~\ref{fig:fig26}(h) it then converges at low temperature . As for the flat-band case in Fig.~\ref{fig:fig26}(g), the compressibility diverged at low temperature at SCBF phase, which is consistent with the observation at superconducting dopings in TBG experiment\cite{andreiGraphene2020}. And in this model, if there is no kinetic energy term, we have the exact solutions: $\kappa=\frac{2 \beta}{n_{0} L^2}\left\langle\Delta \Delta^{\dagger}\right\rangle$ with $\left\langle\Delta \Delta^{\dagger}\right\rangle \propto L^{4}$. Then the compressibility diverges as $\kappa \propto \beta L^2$. It is also consistent with the QMC datas shown in Fig.~\ref{fig:fig26}(g).

\section{Discussions}
\label{sec:V}
The three categories of systems, the quantum critical metals, the Yukawa-SYK models and the quantum Moir\'e lattice models, are all at the heart of the frontier quantum many-body research of novel and highly-entangled matter. Their intrinsic difficulties, such as the precise treatment of the quantum critical fluctuations, the all-to-all couplings in Yukawa-SYK models and the fruitful interplay between the quantum geometry in the flat-band wave function and the long-range Coulomb interaction in the Moir\'e materials, are all the quintessential characteristics of modern strongly correlated electron problems and it is in their gradually analytic and numeric solutions lie the future of the condensed matter and quantum material research.

Complicated as they are, these problems are also extremely interesting, and it is here in this review we try to present our efforts in the spirit of "A Sport and A Pastime"~\cite{salterA1967,dowieA1988,hallBeautiful2017} in the model design and computation -- to some extend analytic -- solutions in these systems. We have completed the quantum critical scaling behaviors of the FM and AFM Ising and XY quantum critical metals to identify the nFL self-energies and bosonic susceptibilities with exponents (summarized in Tab.~\ref{tab:tab1}), we proved that in the Yukawa-SYK models the large-$N$ results from the original SYK still hold down to spin-1/2 case~\cite{panYukawa2021,wangPhase2021}, and last but not least, we have succeeded in designing both the real-space Moir\'e-scale effective lattice model and the momentum-space continuum model for the quantum Moir\'e materials and have employed the QMC, DMRG and thermal tensor-network~\cite{liaoValence2019,liaoCorrelated2021,liaoCorrelation2021,chenRealization2021,linExciton2022} and more importantly invented the momentum-space QMC approach to solve them~\cite{zhangMomentum2021,zhangSuperconductivity2021,panDynamical2022}. The rich and systematic results presented in this review are the proof that even when facing the three categories of difficult yet exciting systems, the model design and computation can indeed demystify the puzzling experimental results, provide solid understanding for theoretical paradigms and make progresses both in fundamental theory and the associated computation techniques.

It is well-known that not only all the quantum many-body problems are difficult, with different lattice geometry, interaction form, computation complexity, etc, but a particular quantum many-body system is difficult in its own way, as have been avidly shown in the three categories of models in this review. Yet we still would like to convey a message that with the spirit of a sport and a pastime, these difficult problems can still be solved with the principal organ of scientific activities -- the inspired curiosity, imagination and persistent efforts -- that is, as long as we are compelled, by the rich and beautiful and everlasting discoveries in quantum many-body systems, to invent and discover, in this context the numerical methodologies and theoretical understandings in the pursuit of the model design and computation, the even richer understanding and brighter frontiers in quantum many-body systems are yet awaiting us to explore.

\section*{Acknowledgement}
We are in great debt to the former group members Xiao Yan Xu, Zihong Liu, Yuzhi Liu, Wei Wang, Yuan Da Liao and Chuang Chen for having the pleasure and wonderful experience to work together on the works mentioned. We thank Avraham Klein, Yuxuan Wang, Kai Sun and Andrey Chubukov for close collaborations on the topics of quantum critical metals and Yukawa-SYK models over the years. We acknowledge Xiyue Li, Yi Zhang, Clara Brei\o, Jian Kang, Oskar Vafek, Brian Anderson, Rafael Fernandes, Tao Shi, Wei Li, Xu Zhang, Bin-Bin Chen, Hongyu Lu, Heqiu Li and Kai Sun for stimulating collaborations on the topics related with quantum Mori\'e material models. 
 
GPP, WLJ and ZYM
acknowledge support from the Research Grants Council of Hong Kong SAR of China (Grant Nos. 17303019,
17301420, 17301721 and AoE/P-701/20), the K. C. Wong
Education Foundation (Grant No. GJTD-2020-01), and
the Seed Funding “Quantum-Inspired explainable-AI” at
the HKU-TCL Joint Research Centre for Artificial Intelligence.
We thank the Computational
Initiative at the Faculty of Science and HPC2021 system under the Information Technology Services at the University of Hong Kong,
and the Tianhe platforms at the National Supercomputer
Centers for their technical support and generous allocation of CPU time.

\newpage

\bibliographystyle{iopart-num.bst}
\bibliography{review}

\end{document}